\documentclass[11pt]{article}
\pdfoutput=1
\usepackage{jheppubmod}

\usepackage{amsmath, amssymb, comment, braket, mathtools, psfrag, array, amsthm, graphicx, subcaption, epstopdf, scrextend, color, epsfig }

\usepackage{hyperref} 
\hypersetup{
     colorlinks=true, 
     linkcolor=blue,  
     citecolor=red,  
     filecolor=cyan,
     urlcolor=blue,
     linktoc=section    
          }

\usepackage[parfill]{parskip}
\def\pb[#1,#2]{\{#1, #2\}}
\def\deb[#1,#2]{[#1,#2]_{\text{D.B.}}}

\def\tO{\widetilde{\cal O}}

\def\tphi{\widetilde{\phi}}

\def\Or[#1]{{\text{O}}\left({#1}\right)}
\def\dotl[#1,#2]{\left\langle #1,\, #2 \right\rangle}
\def\dotlb[#1,#2]{\left\langle #1,\, #2 \right\rangle}
\def\dotlm[#1,#2]{\left[ #1,\, #2 \right]}
\def\dotp[#1,#2]{(\vect{#1} \cdot\vect{#2})}
\def\aff[#1,#2]{\hat{#1}(#2)}

\def\n4sym{{\cal N}=4 SYM}
\def\>{\rangle}
\def\<{\langle}
\def\weight[#1,#2,#3]{\{(#1),#2,#3\}}
\def\ads[#1]{$\text{AdS}_{#1}$}

\def\tarelr[#1]{\widetilde{a}^{\text{rel}}_{R#1}}
\def\Oright[#1]{{\cal O}_{R#1}}
\def\Oleft[#1]{{\cal O}_{L#1}}
\def\aleft[#1]{a_{L#1}}
\def\arelr[#1]{a^{\text{rel}}_{R#1}}

\newcommand{\tfd}{|{\rm TFD}\rangle}

\newcommand{\be}{\begin{equation}}
\newcommand{\ee}{\end{equation}}
\newcommand{\ba}{\begin{align}}
\newcommand{\ea}{\end{align}}
\newcommand{\bs}{\begin{split}}
\def\sess\end{split}

\newcommand{\vect}[1]{{#1}}

\def\tO{\widetilde{\cal O}}
\def\rs{|\Psi\rangle}
\def\ls{\langle \Psi|}

\def\rsz{|\Psi_{0}\rangle}
\def\lsz{\langle \Psi_0|}

\def\Tr{{\rm Tr}}
\def\tO{\widetilde{\cal O}}
\def\rs{|\Psi\rangle}
\def\ls{\langle \Psi|}

\def\htfd{H_{\rm TFD}}
 
 \author[a]{Jan de Boer}
 \author[b,c]{, Rik van Breukelen}
 \author[a]{, Sagar F. Lokhande}
 \author[d]{, Kyriakos Papadodimas}
 \author[a]{and Erik Verlinde}
 \emailAdd{j.deboer@uva.nl}
 \emailAdd{rik.van.breukelen@cern.ch}
 \emailAdd{sagar.f.lokhande@gmail.com}
 \emailAdd{kyriakos@ictp.it}
 \emailAdd{e.p.verlinde@uva.nl}
 \affiliation[a]{\vspace{2mm} 
 Institute for Theoretical Physics and Delta Institute for Theoretical Physics, University of Amsterdam, Science Park 904, 1098 XH Amsterdam, The Netherlands 
 \vspace{2mm}}
 \affiliation[b]{Theoretical Physics Department, CERN, CH-1211 Geneva 23, Switzerland 
 \vspace{2mm}}
 \affiliation[c]{Geneva University, 24 quai Ernest-Ansermet, CH-1214 Geneva 4, Switzerland
 \vspace{2mm}}
 \affiliation[d]{International Centre for Theoretical Physics,
Strada Costiera 11, 34151 Trieste, Italy
\vspace{2mm}}

\keywords{Black Holes, Information Paradox}

\abstract{We investigate the possibility that the geometry dual to a typical AdS black hole microstate corresponds to the extended AdS-Schwarzschild geometry, including a region spacelike to the exterior. We argue that this region can be described by the mirror operators, a set of state-dependent operators in the dual CFT. We probe the geometry of a typical state by considering state-dependent deformations of the CFT Hamiltonian, which have an interpretation as a one-sided analogue of the Gao-Jafferis-Wall traversable wormhole protocol for typical states. We argue that the validity of the conjectured bulk geometry requires that out-of-time-order correlators of simple CFT operators on typical pure states must exhibit the same chaotic effects as thermal correlators at scrambling time. This condition is related to the question of whether the product of operators separated by scrambling time obey the Eigenstate Thermalization Hypothesis. We investigate some of these statements in the SYK model and discuss similarities with state-dependent perturbations of pure states in the SYK model previously considered by Kourkoulou and Maldacena. Finally, we discuss how the mirror operators can be used to implement an analogue of the Hayden-Preskill protocol.}

\preprint{\\\hspace*{\fill}

\vspace{-30pt}
}

\begin{document}
\title{Probing typical black hole microstates}

\maketitle

\section{Introduction}

The black hole information paradox is a long-standing open problem, which is related to the smoothness of the black hole horizon \cite{Mathur:2009hf, Almheiri:2012rt}. The AdS/CFT correspondence provides an ideal setting to investigate the issue of smoothness. Large typical black holes in AdS are expected to be dual to typical high-energy pure states in the dual CFT. These typical black holes are approximately in equilibrium and hence do not evaporate. Even then, it is challenging to reconcile the smoothness of the horizon with unitarity of the dual CFT \cite{Almheiri:2013hfa, Marolf:2013dba, Bousso:2013wia}. In this paper, we make some inroads into investigating the geometry of such a typical black hole microstate. 

\begin{figure}[!h]
\begin{center}\includegraphics[width=.23\textwidth]{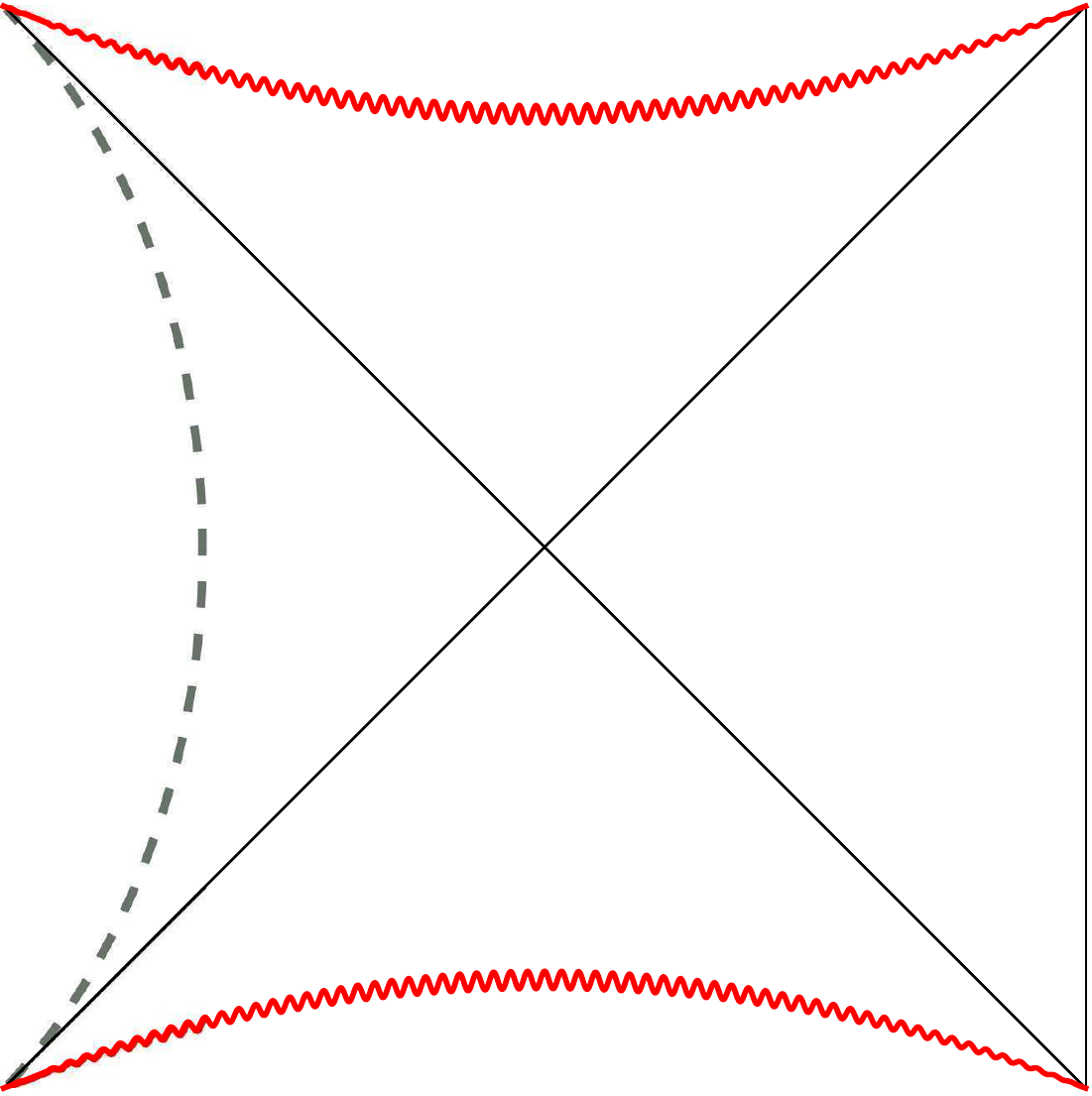}
\caption{A proposal for the  geometry dual to a typical black hole microstate.}
\label{figureintronew}
\end{center}
\end{figure}

Owing to robust arguments in the AdS/CFT framework, it is widely believed that at large $N$ the geometry of a typical black hole microstate contains at least the region exterior to the black hole horizon, which is described by the AdS-Schwarzchild metric. The question then is: do there exist any other regions in the geometry dual to a typical black hole microstate? It seems reasonable that any proposed answer to this question needs to satisfy two constraints: (1) the geometry in the exterior should be that of the AdS-Schwarzschild black hole, (2) the geometry should manifest the approximate time-translation-invariance of the typical pure state in the CFT, through the existence of an approximate timelike Killing isometry. We discuss the time-translation-invariance of typical pure states in Section \ref{ssec:typical-BH}.

In \cite{Almheiri:2012rt, Almheiri:2013hfa, Marolf:2013dba}, it was suggested that the geometry of a typical black hole microstate contains only the exterior region, which gets terminated at the horizon by a firewall. However, for large typical black holes, the curvature near the horizon is low. Thus, this proposed solution demands a dramatic modification of general relativity and effective field theory in regions of low curvature. 

In this paper, we will explore the possibility that the bulk geometry of a typical pure AdS black hole microstate contains part of the extended AdS-Schwarzchild diagram, as shown in figure \ref{figureintronew}. Since the dual of this geometry is a typical pure state in a single CFT, the Penrose diagram cannot be extended arbitrarily to the left and there is no ``left'' CFT. The dotted line in figure \ref{figureintronew} denotes a surface beyond which the geometry is not operationally meaningful. We will discuss the interpretation and other features of this geometry in later sections. 
 
The proposal that the geometry dual to a typical microstate includes parts of the black hole, white hole and left regions, as depicted in figure \ref{figureintronew}, is suggested by the existence of CFT operators which have the right properties to represent these regions. These are the {\it mirror operators}, denoted by $\widetilde{\cal O}$, a set of {\it state-dependent}  operators identified by an analogue of the Tomita-Takesaki construction applied to the algebra of single-trace operators ${\cal O}$ in the CFT. At large $N$, the operators $\widetilde{\cal O}$ commute with usual single trace operators and they are entangled with them. They are the natural candidates to describe the left region of the extended black hole geometry. The black hole interior and white hole region would then be reconstructed by a combination of ${\cal O}$ and $\widetilde{\cal O}$.

Naively, the left region would be inaccessible from the CFT, at the level of effective field theory. However, starting with the work of Gao, Jafferis and Wall \cite{Gao:2016bin} and further work \cite{Maldacena:2017axo, Kourkoulou:2017zaj}, a new approach has been identified for probing the space-time beyond the horizon, including the left region. This new approach, which was formulated in the framework of the two-sided eternal black hole,  is based on the observation of \cite{Gao:2016bin} that in the case of the two-sided eternal black hole there are perturbations of the boundary CFTs of the form $\delta H = {\cal O}_L {\cal O}_R$, which can create negative energy shockwaves which can violate the average null energy condition and allow particles to traverse the horizon. This effect is related to the quantum chaotic behavior of out-of-time-ordered correlators (OTOC) at scrambling time in the boundary CFT \cite{Shenker:2013pqa, Maldacena:2015waa}.

\begin{figure}[!t]
\begin{center}\includegraphics[width=.23\textwidth]{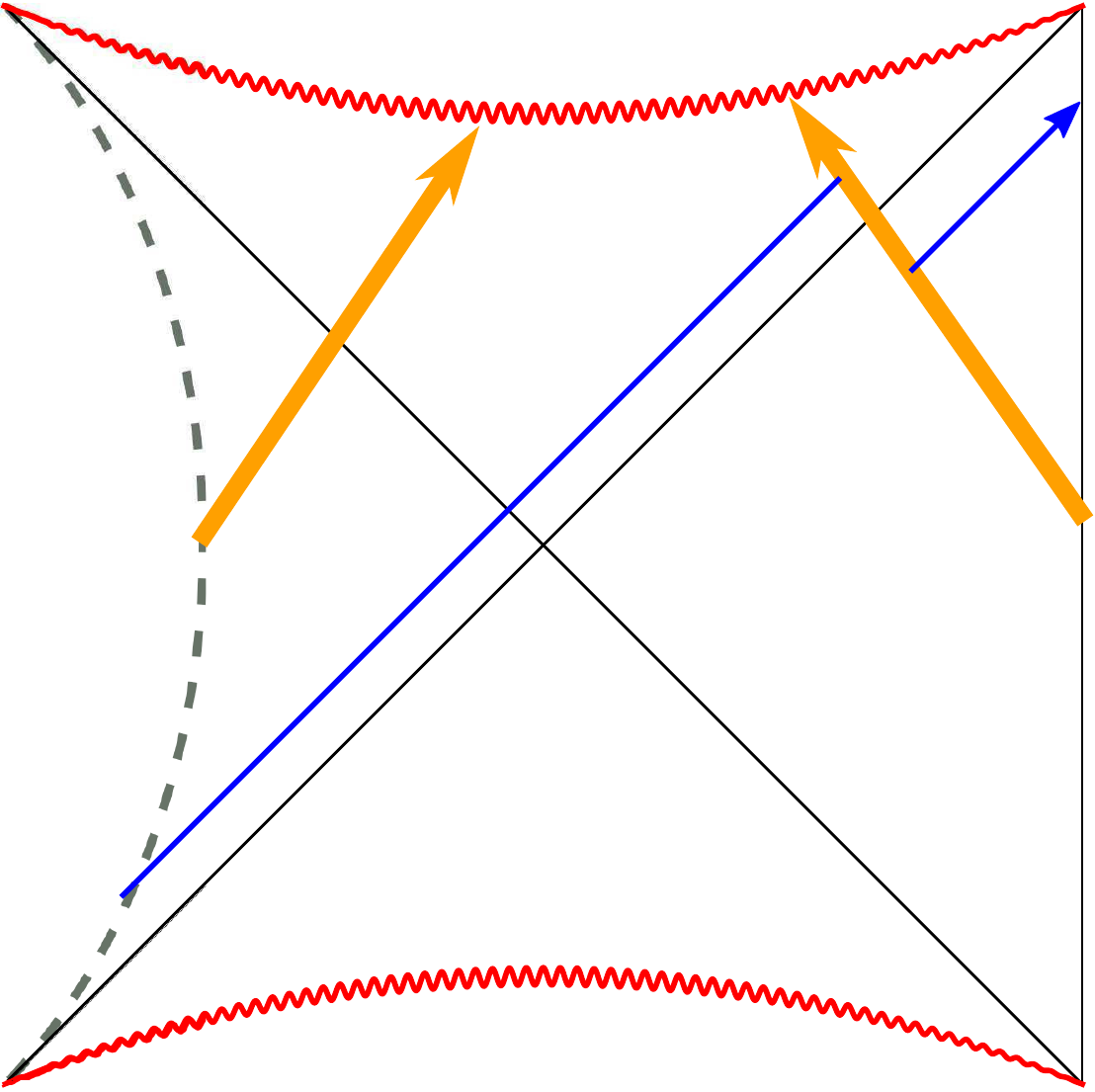}
\caption{Perturbations of $H_{\rm CFT}$ by state-dependent operators can create negative energy shockwaves (yellow), which allow a probe from the left region (blue) to get detected by simple operators in the CFT.}
\label{figintrotwo}
\end{center}
\end{figure}

In this paper, we provide evidence for the conjectured geometry of figure \ref{figureintronew} for the one-sided black hole, by perturbing the CFT Hamiltonian by the state-dependent operators $\widetilde{\cal O}$, in the schematic form $\delta H = {\cal O} \widetilde{\cal O}$. These perturbations allow particles that are localized in the left region of the geometry dual to a pure microstate, to traverse the black hole region and emerge in the right region and get directly detected by single-trace CFT operators.  This is schematically shown in figure \ref{figintrotwo}. We emphasize that the use of state-dependent operators from the point of view of the boundary CFT falls within the standard framework of quantum mechanics\footnote{We can imagine that the boundary observer has prepared many identical systems. By performing measurements in many of these copies he can determine the exact microstate. Then, the observer can prepare an experimental device that acts with the operators $\tO$ relevant for that microstate and apply them to one of the identically prepared copies which has not been  previously measured.} and is logically independent from the question of how the infalling observer can use these operators.

Using these state-dependent perturbations by mirror operators, we argue that the consistency of the space-time geometry proposed in this paper, and shown in figure \ref{figureintronew}, requires as a {\it necessary} condition that CFT correlators of {\it ordinary} CFT operators should obey the following property: {\it the effects of quantum chaos, which become important in out of time order correlators (OTOC) at scrambling time, should be the same --- to leading order at large $N$ --- in typical pure states as in the thermal ensemble.} We conjecture that this is true in large $N$ holographic CFTs and provide some indirect evidence. Notice that this conjecture about the OTOC in pure states is a statement which is independent of the bulk interpretation, and in principle it can be either verified or falsified by CFT methods. A verification of this CFT conjecture would be a  necessary condition for the validity of the geometry of figure \ref{figureintronew}.

Finally,  we argue that the mirror operators $\tO$ can be used to implement an analogue of the Hayden-Preskill protocol \cite{Hayden:2007cs}, in its formulation given in \cite{Maldacena:2017axo}. Information thrown into an AdS black hole, which was originally in a typical state, can be recovered by deforming the CFT Hamiltonian by ${\cal O} \tO$  and then measuring a mirror operator after scrambling time. An analogue of this protocol can be applied to black holes in flat space
after Page time. Then the mirror operators are mostly supported on the early  Hawking radiation, which forms the larger fraction of the total Hilbert space. Interestingly, the protocol then becomes an analogue of the Hayden-Preskill protocol. The complicated nature and state-dependence of operators $\tO$ is consistent with the fact that for the application of the Hayden-Preskill decoding protocol, the observer must have knowledge of the initial black hole microstate and apply a state-dependent decoding procedure.

The plan of the paper is as follows: in section \ref{sec:mainconjecture} we provide details about the conjectured geometry of a typical black hole microstate. In section \ref{sec:doubletrace} we describe how time-dependent perturbations of the CFT Hamiltonian using state-dependent operators allows us to probe the interior. In section \ref{sec:SYK} we formulate and investigate some of our general statements in the SYK model. In section \ref{sec:conjecture} we discuss the technical conjecture about the chaotic OTOC correlators in pure states and provide some evidence for its validity. In section \ref{hp_protocol} we discuss the connection of our experiments with the Hayden-Preskill protocol. 

Part of the results of this paper were reported in a shorter note \cite{deBoer:2018ibj}. Other recent works which investigate the region behind the horizon of special, \textit{atypical} pure states include \cite{Almheiri:2018ijj, Almheiri:2018xdw, Jefferson:2018ksk, Cooper:2018cmb, Brustein:2018fkr}.

\section{On the Interior Geometry of a Typical State}
\label{sec:mainconjecture}

In this section we present a conjecture for the bulk dual of a typical black hole microstate in the framework of AdS/CFT correspondence. We review the construction of {\it mirror operators}, CFT operators that may describe the region behind the horizon. We also discuss how time-dependent perturbations of the CFT Hamiltonian by mirror operators can create excitations behind the horizon.

\subsection{Typical Black Hole Microstates and the ``Mirror Region''}
\label{ssec:typical-BH}

\begin{figure}[!h]
\begin{center}
\includegraphics[width=.8\textwidth]{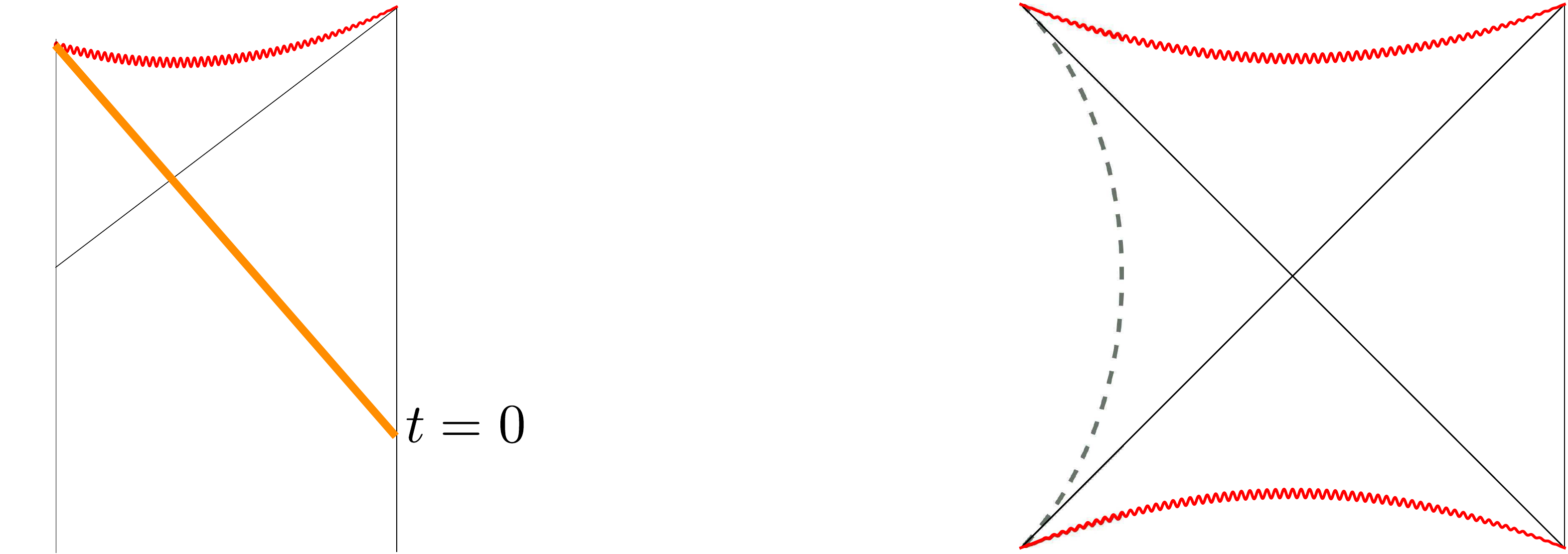}
\caption{Left: AdS black hole formed by collapse.  Right: conjectured bulk geometry of a typical black hole microstate in a single CFT. The left region makes sense up to a cut-off region (dotted lines) and there is no left CFT.}
\label{collapsetypical}
\end{center}
\end{figure}

The Penrose diagram of an AdS black hole formed by collapse is shown in figure \ref{collapsetypical}.
In this paper we are not interested in black holes formed by collapse, but rather in understanding the geometry dual to a {\it typical} black hole microstate in the CFT. This is defined as a pure state which is a random superposition of energy eigenstates
\be
|\Psi_0\rangle = \!\!\!\!\!\!\!\!\!\sum_{E_i \in (E_0,E_0+\delta E)} \!\!\!\!\!\!\!\!\!\!\!c_i |E_i\rangle ,
\label{deftypical}
\ee
selected from a narrow energy band. Here $c_i$ are random complex numbers, constrained to obey $\sum_i |c_i|^2=1$ , and selected with the uniform Haar measure. We take
\begin{equation}
\label{bandoftyp}
 \delta E \sim R_{AdS}^{-1} O(N^0) \, , \qquad E_0 \sim R_{AdS}^{-1} O(N^2) \, .
\end{equation}
Understanding the geometry of a typical black hole microstate is an important question, as --- by definition--- typical states represent the majority of black hole microstates of given energy.  In addition, understanding the geometry of a typical state is  important for the AdS/CFT version
of the firewall paradox \cite{Almheiri:2013hfa, Marolf:2013dba, Papadodimas:2015jra}. As mentioned above, the reader should keep in mind the difference between typical states and states formed by collapse. The number of states which can be formed by ``reasonable'' gravitational collapse is much smaller than those predicted by the Bekenstein-Hawking entropy, see for example an early discussion in \cite{tHooft:1993dmi}. Also, notice that strictly speaking the class of typical states defined above are not  exactly the same as the late-time configuration of a collapsing black hole. For example, the standard inequality ${1\over 2} \left|{d \langle A\rangle \over dt}\right| \leq \Delta A \Delta E$, where $\Delta A,\Delta E$ denote the variance of $A$ and the energy respectively, 	 implies that any state which undergoes gravitational collapse over a time-scale of the order of the AdS radius (hence there are observables $A$ for which ${d \langle A\rangle \over dt} \sim O(N^0)$) and which initially has a semi-classical description, i.e. $\Delta A \sim O(N^{-1})$, must have an energy variance\footnote{See appendix A of \cite{Papadodimas:2017qit} for more details.} which is $\Delta E \sim R_{AdS}^{-1} O(N)$. Such states are somewhat different
from the typical states with narrow-energy band \eqref{bandoftyp} that we consider here\footnote{It is interesting to better understand how collapsing black holes approximate certain classes of typical states at late times, and to clarify the role of complexity in studying the late time limit, see for example \cite{Susskind:2018pmk}.}.
  
Typical states in the CFT look almost time-independent when probed by simple observables, which do not explicitly depend on time, since
\begin{align}
\begin{split}
\lsz {d A \over dt} \rsz & =\Tr[\rho_m {dA \over dt} ]+O(e^{-S})\\ &= i\Tr[\rho_m [H,A]] +O(e^{-S})=O(e^{-S}),
\end{split}
\end{align}
where we introduced the microcanonical density matrix $\rho_m$ relevant for the window $(E_0,E_0+\delta E)$ and we used the approximation of a typical microstate to the microcanonical expectation value reviewed in subsection \ref{subsec:puremicro}\footnote{Notice that to prove this we do not need to use the Eigenstate Thermalization Hypothesis (ETH) \cite{Srednicki1999approach}.}. In the last equality we dropped the trace using $[\rho_m,H]=0$.  This suggests that the dual geometry  to a typical pure state $\ket{\Psi_0}$ should be characterized by an approximate Killing isometry, which is timelike in the exterior region. A natural expectation within the AdS/CFT framework is that part of the dual geometry contains the exterior of a static black hole in AdS. For the benefit of the reader we summarise the relevant arguments in Appendix \ref{exterior}. A natural question then is, does there exist an extension of this geometry behind the black hole horizon? 

If the future horizon is smooth, then the dual geometry should contain at least part of the black hole interior. Since the ensemble of typical states is time-reversal invariant, we will conjecture that the dual geometry should also contain part of the white hole region. Finally, if the dual space-time contains parts of all these three regions, it is natural to assume that it should also contain part of the left asymptotic region. This leads us to the conjectured diagram in figure \ref{collapsetypical} for the dual geometry of a typical state. A typical state is in equilibrium so nothing is happening in it. Thus one may wonder what is the meaning of the statement that the dual geometry contains these regions. The operational meaning of this statement is that a class of perturbations of the boundary CFT can be described by low-energy effective field theory perturbations of the conjectured geometry. In other words,  under a class of deformations the typical state responds ``as if it had a smooth interior'', partly extending into the left  region. Notice that there is no left CFT, but rather the geometry is effectively inaccessible (and operationally meaningless) beyond the region indicated by the dotted lines, whose nature and location we will discuss later in section \ref{leftcutoff}. 

The left region of the conjectured geometry of a typical microstate is described by the ``mirror operators'', which we discuss below. The existence of a region, which is causally disconnected from the black hole exterior, is a consequence of the fact that the algebra ${\cal A}$ describing low-energy effective field theory experiments in the exterior has a nontrivial commutant ${\cal A}'$. Moreover, this commutant is entangled with ${\cal A}$. The fact that the geometry in the left region should be a ``mirrored copy'' of the right region, at least up to some cutoff on the far left, follows from the algebraic construction of the mirror operators \eqref{tomita3}. In that construction we will notice that the commutant ${\cal A}'$ is in some sense isomorphic to the original algebra. Combining together the small algebra ${\cal A}$ and its commutant ${\cal A}'$ we get the black and white hole regions. We can think of the conjectured geometry dual to a typical state as a wormhole connecting the exterior of the black hole and the left  interior region, which represents the space of the mirror operators. The operators ${\cal O}$ and $\tO$ are entangled in a similar way as the two sides of the thermofield state, and the emergence of the wormhole is reminiscent of the ER/EPR proposal \cite{Maldacena:2013xja}. The meaning of this proposed geometry is that we can use effective field theory on it to compute CFT correlators. These correlators can be localized within a finite time domain on the boundary, of the order of few scrambling times. In particular, the domain of validity of the conjectured diagram does not need to capture experiments extended over arbitrarily long time-scales. On the other hand, the finite time domain mentioned can be centered around any time, thus allowing us to access arbitrary regions in the proposed geometry.

\begin{figure}
\hspace{100pt}\includegraphics[width=.3\textwidth]{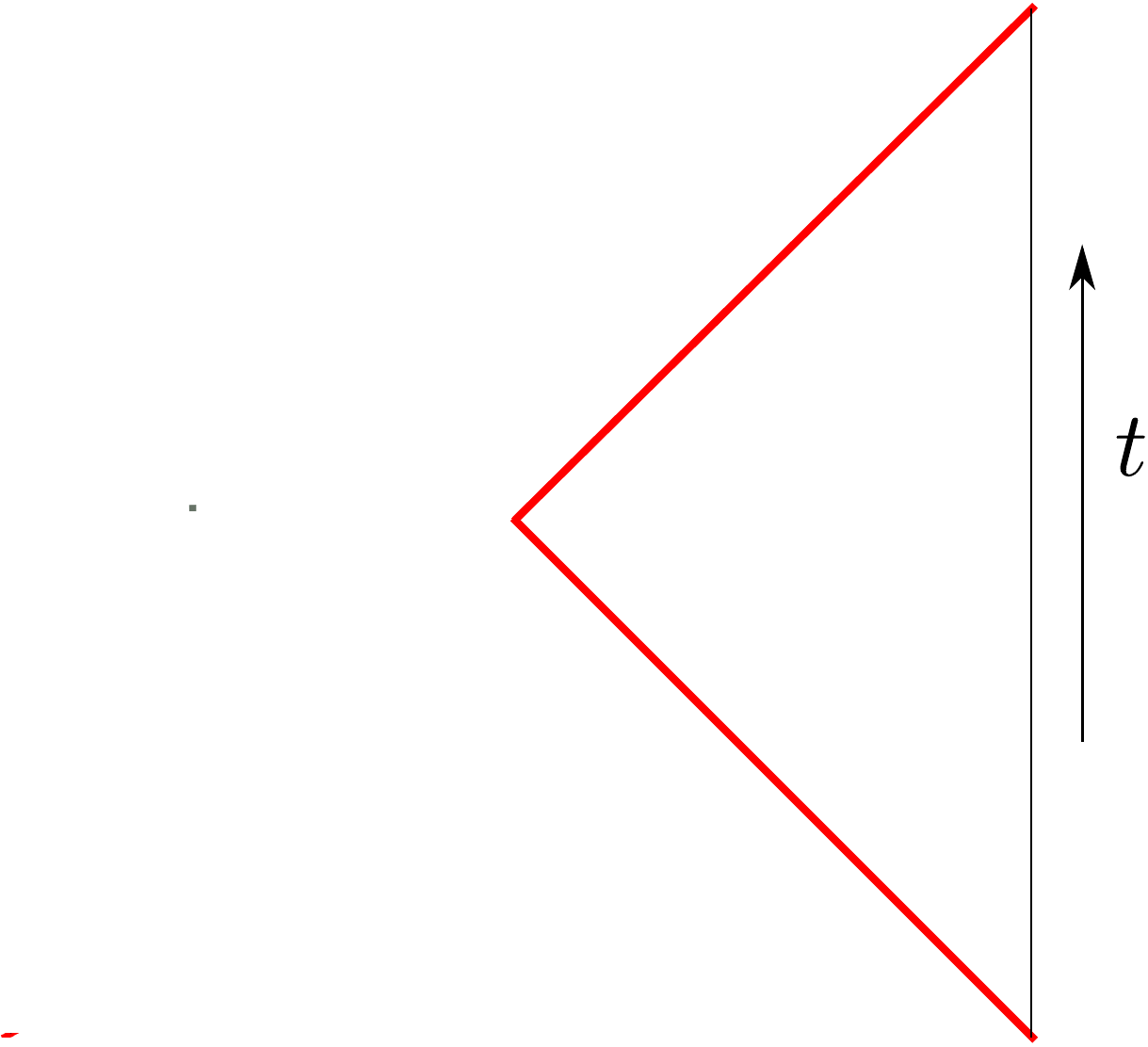}
\caption{An alternative to our proposal: typical states have a firewall on the horizon. This would be consistent with the Killing isometry, but would correspond to modifications of general relativity at low curvatures.}
\label{firewalltypical}
\end{figure}

It is important to consider what could be a possible alternative to the proposed geometry of figure \ref{collapsetypical}. It is natural that the geometry will have to be consistent with the Killing isometry, at least in some approximate sense. One extreme possibility consistent with this symmetry is that the spacetime terminates on the past and future horizons, by a firewall or other object as indicated in figure \ref{firewalltypical}.
This would violate our expectations from general relativity in a regime of low curvatures. If we want to avoid this scenario and if we want to preserve the smoothness of the horizon then we need to extend the geometry behind the horizon, up to some cutoff  which must be consistent with the Killing isometry.

\subsection{The Mirror Operators}
\label{subsec:defmirror}

The conjecture that a typical black hole state should be associated to a geometry, which has a smooth interior and moreover contains  part of the left region follows from the construction of the ``mirror operators'', which was introduced in \cite{Papadodimas:2012aq, Papadodimas:2013b, Papadodimas:2013}. Similar conclusions from a somewhat different perspective were reached in \cite{Verlinde:2012cy, Verlinde:2013vja, Verlinde:2013qya}. The construction of the mirror operators starts by defining a small algebra of observables ${\cal A}$, which correspond to simple experiments described in effective field theory in the bulk. In a large $N$ gauge theory ${\cal A}$ can be thought of as generated by products of a small number of single trace operators of low conformal dimension. Further details about the definition and limitation of the small algebra can be found in the references mentioned above. Here we only emphasize that technically the set ${\cal A}$ is not a proper algebra, since we define it as the set of small products. However, in the large $N$ limit this limitation is not important for what follows, so we will continue to refer to ${\cal A}$ as an algebra.

Given a typical black hole microstate $\rsz$, we define the ``small Hilbert space'', also called code-subspace, as
\begin{equation}
{\cal H}_{\rsz} = {\rm span} {\cal A} \rsz .
\end{equation}
This subspace of the full CFT Hilbert space, contains all states that one can get starting from $\rsz$ and acting on it with a small number of bulk operators. Hence, this subspace is the one relevant for describing effective field theory in the bulk\footnote{The code subspace knows about $1/N$ interactions which can be described within effective field theory.}. An important algebraic property is that operators of the algebra ${\cal A}$ cannot annihilate the state $\rsz$. This can be understood, for example, by noticing that for $A\in {\cal A}$ we have
\begin{equation}
|| A \rsz||^2 = \lsz A^\dagger A \rsz = Z^{-1}\Tr[e^{-\beta H} A^\dagger A] +O(1/S) ,
\end{equation}
which is positive if we ignore the subleading corrections. Here we used the approximation of a typical pure state by the thermal ensemble, which is expected to be very good at least at leading order in large N.  

The fact that ${\cal A}$ contains no annihilation operators for $\rsz$ means that for the representation of the algebra ${\cal A}$ on the code subspace ${\cal H}_{\rsz}$, the state $\rsz$ is a {\it cyclic} and {\it separating} vector. As suggested by
the Tomita-Takesaki theorem, see for example \cite{Haag:1992hx} for a review, this implies that the representation of the algebra ${\cal A}$ on the subspace ${\cal H}_{\rsz}$ is reducible and the algebra has a non-trivial commutant ${\cal A}'$ acting on ${\cal H}_{\rsz}$. The elements of ${\cal A}'$ are operators which commute with all elements of the algebra ${\cal A}$, and geometrically in the bulk correspond to local fields in a region which must be causally disconnected by the black hole exterior. It is natural to identify the region corresponding to ${\cal A}'$ with the left asymptotic region.

The commutant ${\cal A}'$ can be concretely identified by an analogue of the Tomita-Takesaki construction
\begin{equation}
 \label{tomita1}
 S A \rsz = A^\dagger \rsz ,
 \end{equation}
\begin{equation}
 \label{tomita2}
\Delta = S^\dagger S ,\quad\quad S=J \Delta^{1/2} ,
 \end{equation}
\begin{equation}
 \label{tomita3}
 \widetilde{\cal O} = J {\cal O} J .
 \end{equation}
where $J$ is an anti-unitary operator called \textit{modular conjugation} and $S$ is a general anti-linear map. Moreover, using large $N$ factorization and the KMS condition relevant for equilibrium states, it is possible to show \cite{Papadodimas:2013} that at large $N$ the CFT Hamiltonian acts on the code subspace similar to the (full) modular Hamiltonian 
\be
\label{modularh}
\Delta = \exp[-\beta(H-E_0)] + O(1/N) ,
\ee
where we assume that the energy of $\rsz$ is highly peaked around $E_0$.

To be more concrete, we will construct the small algebra ${\cal A}$ starting with single-trace operators in frequency space and later we discuss their Fourier transform to the time coordinate. The algebra ${\cal A}$ is defined as
\begin{equation}
{\cal A} = {\rm span} \left\{{\cal O}_{\omega_1}, {\cal O}_{\omega_1}{\cal O}_{\omega_2},...., {\cal O}_{\omega_1}....{\cal O}_{\omega_n}\right\} ,
\end{equation}
where $n \ll N$. Of course this linear set is not a proper algebra as we demand $n\ll N$, for example we are not allowed to multiply together $N$ single-trace operators ${\cal O}_{\omega_i}$. However, if $n$ is very large (but much smaller than $N$) this set behaves approximately as an algebra for correlation functions involving a small number of operators. 
Moreover, the fact that ${\cal A}$ is not a proper algebra is important for the realization of the idea of black hole complementarity, as has been discussed in detail in \cite{Papadodimas:2012aq, Papadodimas:2013}. Nevertheless, we will continue referring to ${\cal A}$ as the ``small algebra''.

Now we clarify the nature of the Fourier modes generating the small algebra ${\cal A}$. We first consider the exact Fourier modes of operators, defined as
\be
\label{firstfourier}
{\cal O}^{\rm exact}_\omega \equiv \int_{-\infty}^{+\infty} dt\,\,e^{i \omega t} {\cal O}(t) .
\ee
Usually $\omega$ takes values in $(-\infty,\infty)$. However, for the construction above we need to restrict the range of $\omega$ in some ways: 

i) Since the spectrum of the dual CFT is assumed to be discrete\footnote{For example consider the ${\cal N}=4$ SYM on ${\mathbb S}^3 \times {\rm time}$.}, then for generic choice of real $\omega$ there will be no pair of states such that $E_i-E_j=\omega$. Hence, if $\omega$ is a generic real number, the operator ${\cal O}^{\rm exact}_\omega$ defined by \eqref{firstfourier}  will be zero. To avoid this we
can bin together sets of frequencies in bins of size $\delta \omega$, which can be very small given that the typical energy level gap is $\beta^{-1} e^{-S}$. Therefore, we define the coarse-grained frequency operators 
\be
\label{coarsefourier}
{\cal O}_\omega \equiv {1\over \sqrt{\delta \omega}}\int_{\omega}^{\omega +\delta\omega} {\cal O}^{\rm exact}_{\omega'} d\omega' ,
\ee where now the set of allowed $\omega$'s is discretized with step $\delta\omega$. In \eqref{coarsefourier} we have divided by $\sqrt{\delta \omega}$ in order to have an operator whose correlators are stable under small changes of the bin size $\delta \omega$. We will denote these coarse-grained Fourier modes simply as ${\cal O}_\omega$, without explicitly showing the choice of step $\delta \omega$, which is not important for most calculations. Alternatively to the binning procedure, we can think of the ${\cal O}_{\omega}$ in a distributional sense, where we always use these operators inside integrals over $\omega$.

ii) We need to impose an upper cutoff in the allowed frequencies $|\omega|\leq \omega_*$. The reason is that the mirror operators are meaningful when the small algebra cannot annihilate the state. In a thermal state we find that $\langle {\cal O}_\omega^\dagger {\cal O}_\omega\rangle \propto e^{-\beta \omega}$. 
For large $\omega$ this is extremely close to zero, implying that the operator ${\cal O}_\omega$ almost annihilates the state. 	It is possible, and sufficient for our purposes, to take $\omega_*$ arbitrarily large but $N$-independent. We will discuss this and the possibility of scaling $\omega_*$ with $N$ in subsection \ref{leftcutoff}. 

iii) If we want to describe the part of the black hole interior relevant for experiments initiated around some time $t_0$ in the CFT, then for the definition of the mirror operators
we need to consider CFT operators which are localized only within a time band $t_0 \pm T_{\rm max}$, where $T_{\rm max}$ is at least as large as several times the scrambling time. This means that the Fourier modes should be defined with respect to this IR-cutoff in time, which
in turns effectively leads to a discretization of frequencies of order  ${1 \over T_{\max}}$. If $T_{\rm max}$ is large enough, this does not affect correlators significantly, see also the relevant discussion in \cite{Papadodimas:2012aq}. Since our experiments are contained inside a time band given by $T_{\rm max}$ we do not probe the infinite past and infinite future of the state. 

From \eqref{tomita3},\eqref{modularh} follows that at large $N$ the mirror operators are defined by the equations\footnote{One might worry that conjugating
${\cal O}_{\omega}^\dagger$ with $e^{\pm{ \beta H \over 2}}$ would result in a complicated operator, not necessarily obeying ETH. However, notice that the Fourier modes ${\cal O}^{\rm exact}_\omega$ \eqref{firstfourier} obey precisely $[H,({\cal O}^{\rm exact}_\omega)^\dagger] = \omega ({\cal O}^{\rm exact}_{\omega})^\dagger$ and therefore $e^{-{\beta H \over 2}} ({\cal O}^{\rm exact}_\omega)^\dagger  e^{{\beta H \over 2}}= e^{-{\beta \omega \over 2}} ({\cal O}^{\rm exact}_\omega)^\dagger$. After binning \eqref{coarsefourier} we get an operator that obeys the ETH, assuming that ${\cal O}_\omega^\dagger$ did.}
\begin{align}
\label{defmirror}
\begin{split}
\tO_\omega \rsz &= e^{-{\beta H \over 2}} {\cal O}_\omega^\dagger  e^{{\beta H \over 2}} \rsz ,\\
\tO_\omega {\cal O}_{\omega_1}...{\cal O}_{\omega_n} \rsz &= {\cal O}_{\omega_1}...{\cal O}_{\omega_n} \tO_\omega \rsz ,\\
[H,\tO_\omega]{\cal O}_{\omega_1}...{\cal O}_{\omega_n} \rsz &= \omega \tO_\omega {\cal O}_{\omega_1}...{\cal O}_{\omega_n} \rsz .
\end{split}
\end{align}
The last equation implies that the mirror operators are ``gravitationally dressed'' with respect to the CFT, i.e. the right boundary of the Penrose diagram. For the purposes of this paper we extend these equations to define the mirror operators, even when we include $1/N$ effects. The extension of the definition of $\tO$ to subleading orders in $1/N$ is not unique, and for the reconstruction of local bulk fields more care about this issue should be taken. However, for the thought experiments we set up later, we will take \eqref{defmirror} as the definition of the mirror operators even including $1/N$ corrections. One drawback of this choice is that it makes the hermiticity properties of the mirror operators somewhat more complicated, as explained in the paragraph below. On the other hand, this choice makes the comparison with the eternal black hole simpler, hence we will continue using it in this paper.

We now come to the hermiticity properties of the mirror operators defined in \eqref{defmirror}. In particular from this definition it follows that to leading order at large $N$ we have $\widetilde{\cal O}_{-\omega} =(\widetilde{\cal O}_{\omega})^\dagger$, but this may no longer be true at subleading orders in $1/N$. 
To see that, consider the matrix elements of these operators on two general states in the small Hilbert space, which can be written as $A\rsz$ and $B\rsz$ with $A,B \in {\cal A}$. We have
\begin{align}
\begin{split}
\lsz A^\dagger \widetilde{\cal O}_{-\omega} B \rsz  &= e^{\beta \omega \over 2}\lsz A^\dagger B {\cal O}_\omega \rsz , \\ 
\lsz A^\dagger (\widetilde{\cal O}_\omega)^\dagger B\rsz &= e^{-{\beta \omega \over 2}}	\lsz {\cal O}_\omega A^\dagger B \rsz  .
\end{split}
\end{align}
In general it is not possible to argue that these two will be the same, which suggest that in general $\widetilde{\cal O}_{-\omega} \neq (\widetilde{\cal O}_{\omega})^\dagger$ . However, if we approximate correlators on typical pure states by thermal states, an approximation we will discuss extensively in section \ref{sec:conjecture}, we find
\begin{align}
\begin{split}
\lsz A^\dagger \widetilde{\cal O}_{-\omega} B \rsz  &= e^{\beta \omega \over 2}{\rm Tr}[\rho_\beta A^\dagger B {\cal O}_\omega ] + O(1/N) , \\
\lsz A^\dagger (\widetilde{\cal O}_\omega)^\dagger B\rsz &= e^{-{\beta \omega \over 2}}	{\rm Tr}[\rho_\beta{\cal O}_\omega A^\dagger B] + O(1/N)  .
\end{split}
\end{align}
where $\rho_\beta$ is the thermal density matrix. Finally we use the KMS condition for the thermal state, which implies ${\rm Tr}[\rho_\beta A^\dagger B {\cal O}_\omega] = e^{-\beta \omega} {\rm Tr}[\rho_\beta {\cal O}_\omega A^\dagger B]$, to conclude that within the code subspace we have
\begin{equation}
 \label{hermirror}
\widetilde{\cal O}_{-\omega} = (\widetilde{\cal O}_{\omega})^\dagger + O(1/N)
\end{equation}
In particular this means, for example,  that the operator $\widetilde{\cal O}_\omega + \widetilde{\cal O}_{-\omega}$ is Hermitian up to $1/N$ corrections. We close the issue of hermiticity by saying that other extensions of the definition of mirror operators to subleading orders in $1/N$ have more manifest hermiticity properties and they may be more useful for other purposes\footnote{For example, if we define the mirror operators by using the Tomita-Takesaki equations \eqref{tomita1}-\eqref{tomita3} to all orders in $1/N$, and without making the approximation \eqref{modularh}, then it can be shown that:  if ${\cal O}$ is Hermitian, then $\widetilde{\cal O}$ is Hermitian. On the other hand, other approximations at subleading order in $1/N$ become harder with this definition.}.

Because of the restrictions in frequencies that we have imposed above, it is not meaningful to define the mirror operators for sharply localized operators ${\cal O}(t)$. As we will discuss later, in section \ref{leftcutoff}, this is related to the fact that we do not expect to be able
to reconstruct the entire left asymptotic region. We can still try to define approximately localized mirror operators $\tO(t)$ by using only the available Fourier modes, but the resulting operators will not behave as sharply localized operators. The approximation becomes better as we increase the cutoff $\omega_*$. 

We also emphasize that the equations \eqref{defmirror} are supposed to hold only inside the code subspace. This means that the operators $\tO$ do not need to commute with the operators in ${\cal A}$ in an exact operator sense, but only when their commutator is inserted inside low-point correlation functions in the small Hilbert space. This is related to the idea of black hole complementarity.

\subsubsection{Time Dependence of Mirror Operators}
\label{subsec:timedep}

Since we will be considering time-dependent perturbations of the CFT Hamiltonian, it is necessary in this context to specify the CFT time when the mirror operators are applied, or at least specify a {\it time ordering} between them and also with the normal operators. Specifying the action of the operators on the small subspace ${\cal H}_{\rsz}$, as in \eqref{defmirror}, is not sufficient to know the time when the operators act. This issue is discussed in more detail in Appendix \ref{app:timedependence}. 

We will associate the mirror operators to physical time in the CFT in such a way that when we consider time-dependent perturbations of the CFT Hamiltonian with mirror operators, then the result can be described by effective field theory in the bulk. In terms of the conjectured Penrose diagram, the requirement of having a consistent effective field theory description in the bulk requires that as the CFT time $t$ increases the corresponding mirror operators must move ``upwards'' on the left side of the diagram --- the opposite choice would not lead to a globally consistent bulk causal structure. Imposing this condition we find a one-parameter family, labeled by $T$, of useful choices\footnote{The sense in which this way of localizing the mirrors in time is useful, is that active perturbation using these mirrors can be represented geometrically by a space-time diagram and effective field theory on it. See also appendix \ref{app:timedependence}.} for how to localize the mirrors in physical time. For each choice of $T$ we define 
\be
\label{mirrortimes}
\tO_T(t) = \int_{-\omega_*}^{\omega_*} d\omega\,\, e^{-i \omega (t-T)}  \tO_\omega ,
\ee
where $t$ labels the physical CFT time at which the operator is localized. These time-dependent real-space operators are the Fourier transform of a function which has i) an upper cutoff in $\omega$ and ii) the frequencies are discretized. The discretization of frequencies does not impose a serious restriction at time-scales of $O(1)$, given that $\delta \omega$ can be very small. On the other hand the upper cutoff $\omega_*$ means that we should not be trying to resolve time with resolution smaller than $\Delta t \sim {1\over \omega_*}$. 

Time-dependent perturbations of the Hamiltonian by mirror operators have a simple bulk interpretation if the same choice of  $T$ is used for all mirror operators in the calculation. It does not matter what $T$ is, but one should not mix mirror operators with different choices of $T$, or else the bulk dual would be complicated. Notice that the choice of $T$ can also be understood in terms of representing the typical microstate in terms of the ``time-shifted eternal black hole'' \cite{Papadodimas:2015xma}, where we move one of the two boundaries by a large diffeomorphism corresponding to time translation by $T$.

We think of the operators $\tO_T(t)$ as acting at physical CFT time $t$. This means that, for given and fixed choice of  $T$, different mirror operators are time-ordered according to the obvious way with respect to the parameter $t$. i.e. a perturbation by $\tO_T(t_1)$ can affect operators $\tO_T(t_2)$ if $t_2\geq t_1$.
\begin{figure}[!t]
\begin{center}\includegraphics[width=.8\textwidth]{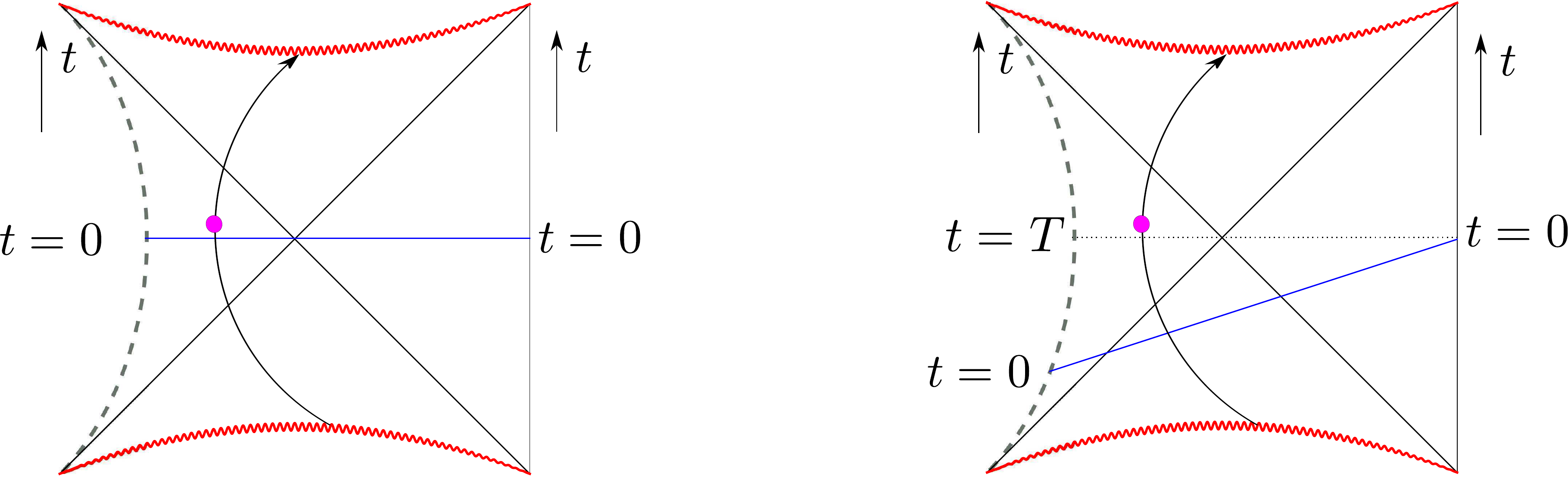}
\caption{Two different choices of localizing the mirror operators in physical time, corresponding to different identifications of physical time $t$ with the coordinate on the left region. The two diagrams  correspond to different choices of $T$, the
left diagram is for $T=0$. Notice that the bulk geometry of the left region, if it is defined relationally to the right boundary, is the same in both cases. However, the types of allowed active perturbations of the state are different.}
\label{rindsketch}
\end{center}
\end{figure}
Notice the important minus sign in the exponential in \eqref{mirrortimes}, relative to what one would naively expect for the inverse Fourier transform in the conventions of \eqref{firstfourier} and from the fact that $\tO$ seems to be associated to ${\cal O}_\omega^\dagger$ from  the first equation of \eqref{defmirror} . This minus sign can be thought of effectively as a time reversal around $t=T$. This minus sign also implies that in the Heisenberg picture, these operators obey
\be
\tO_T(t) = e^{-i H t} \tO_T(0) e^{i H t} .
\ee
The fact that this is the opposite evolution than the usual one implies that if we think of them as operators in the Schroedinger picture they are explicitly time-dependent, in such a way that effectively they behave as if they were running backwards in time i.e. in Schroedinger picture with physical time $t$ we would have the {\it explicit} time-dependence 
\be
\label{explicittime}
[\tO_T(t)]_{\rm S} = e^{-2i H t} [\tO_T(0)]_{\rm S} e^{2i Ht}.
\ee
In the rest of this paper we will be working with the choice $T=0$. For notational ease we will write $\tO_{T=0}=\tO$.

Finally, as we discussed in the previous section, and in particular equation \eqref{hermirror}, the operator $\tO_T(t)$ is not Hermitian at subleading orders in $1/N$. Hence if we want to use it as a perturbation to the CFT Hamiltonian we need to add subleading $1/N$ corrections to the operator, to ensure that the perturbed Hamiltonian is Hermitian. Provided that typical state correlators can be well approximated by thermal correlators at large $N$, a condition that we will formulate as a technical conjecture in section \ref{sec:conjecture}, these subleading corrections necessary to promote \eqref{mirrortimes} into a Hermitian operator will not alter the relevant results to leading order. For the rest of this paper whenever we discuss perturbations involving $\widetilde{\cal O}$ we will always assume that we have dressed up the operator so that it is Hermitian.

\subsection{On the Boundary of the Left Region}
\label{leftcutoff}

Since we are considering the geometry dual to a single CFT, we do not expect to be able to reconstruct the left region of the Penrose diagram all the way towards asymptotic AdS infinity. This is related to the fact that we are not able to define the mirror operators for sharply localized operators in time, or equivalently it is related to the cutoff frequency $\omega_*$ that we introduced in section \ref{subsec:defmirror}. If we think of a wavepacket in the background of an AdS Schwarzchild black hole, then if we want to localize the packet close to the boundary we need to use high frequencies $\omega$. If we have a cutoff in the allowed frequencies $|\omega|<\omega_*$ then any wavepacket constructed with this cutoff will have limited reach towards the boundary. In the limit of large $\omega_*$ we find that this translates into a cutoff in the usual $r$-coordinate in global AdS\footnote{These are coordinates where empty global AdS would have the form $ds^2= -(1+r^2) dt^2 + {1\over 1+r^2} dr^2 + r^2 d\Omega_{d-1}^2$.} as
\be
r< r_* \qquad,\qquad r_* \sim {\omega_*} .
\ee
Here we are working in units where $R_{\rm AdS}=1$. This estimate follows from the gravitational potential of AdS and from analyzing what frequencies are necessary in order to localize wavepackets around a particular region of $r$. Hence, the question of how far towards the left we can extend the geometry depends on the cutoff $\omega_*$. First we start with a conservative estimate: if we take the large $N$ limit, we can take $\omega_*$ to be as large as we like, provided that it is not $N$-dependent. This also means that the cutoff $r_*$ can be arbitrarily large, though $N$-independent. In particular this means that the left geometry can be extended to (arbitrarily) many times the Schwarzschild radius of the black hole towards the left. This is sufficient to formulate most of the thought experiments that we want to consider. 

Of course it is interesting to understand how far the left region extends. In order for the geometry to be operationally meaningful, a probe must be able to explore it. Any probe in the left region should be thought of as a spontaneous out of equilibrium excitation ``borrowing'' energy from the black hole. Hence the black hole mass provides an upper limit to the energy that these probes can have. Taking into account the redshift factor near the AdS boundary we find that this implies an ultimate upper bound $r_* \sim M $ or $r_*\sim O(N^2)$ --- but the actual bound may be much smaller. This is consistent with the fact that the mirror operators can not be defined for frequencies of order $\omega \sim O(N^2)$, since  $\langle {\cal O}_\omega^\dagger {\cal O}_\omega\rangle \sim e^{-\beta \omega}$ is almost zero, see discussion in section \ref{subsec:defmirror}. So far we have identified that the left region can be reconstructed {\it at least} up to $r_* \sim \alpha$ where $\alpha$ can be arbitrarily large but $N$ independent, and {\it at most} up to $r_* \sim O(N^2)$. The actual cutoff region must lie somewhere in-between. We have not been able to identify the more precise limit of the bulk reconstruction but this is clearly a very interesting question. We would also like to pose the following question: what happens to the space-time when we approach this cutoff region? Our conjecture is that there is no breakdown of effective field theory anywhere in the left region, but the limitation of the reconstruction arises simply from the fact that there is a restriction in the energies of the allowed probes moving in the left region. In particular,  the energies of the allowed probes are bounded and this  bound is what makes it impossible, even in principle, to probe the far-left region of the Penrose diagram and not a breakdown of bulk effective field theory. This situation is very different from that of non-typical states. For a class of such states, it was suggested that the bulk geometry has a left region bounded by some kind of end-of-the-world membrane \cite{Kourkoulou:2017zaj,Almheiri:2018ijj}, which plays the role of a hard cutoff.

\subsection{Comments on the Hamiltonian}

Let us call $M$ the ADM mass as measured in the bulk from the right side of the black hole. We argued above that the effective cutoff on the left can be pushed quite far when we are working in the large $N$ limit. Hence it is natural to define an analogue of the left ADM mass $\widetilde{M}$. The first law \cite{Bardeen:1973gs, Wald:1993nt, Iyer:1994ys, Hollands:2012sf, Jafferis:2015del} applied to the two-sided Cauchy slice $\Sigma$ up to the left cutoff implies
\be
\delta M - \delta \widetilde{M} = \delta K_{\rm bulk}^{\rm full},
\ee
where $K_{\rm bulk}^{\rm full} = \int_\Sigma * (\xi T_{\rm bulk})$, $\xi$ is the Killing vector field and $T_{\rm bulk}$ is the bulk stress tensor corresponding to EFT excitations in the left and right regions. The quantity $K_{\rm bulk
}^{\rm full}$ can naturally be split into the right and left contributions $K_{\rm bulk}^{\rm full}= K - \widetilde{K}$. Since we consider the $\tO$ operators to be gravitationally dressed with respect to the right, we have $\delta \widetilde{M} = 0$. This means that in the small Hilbert space the CFT Hamiltonian acts as
\be
\label{hamiltonian}
H =M=E_0 +  K_{\rm bulk}^{\rm full},
\ee
where $E_0$ is the energy of $\rsz$. 

We notice that according to the identification \eqref{hamiltonian} excitations which are created in the left region by right-dressed operators  have negative energy with respect to the CFT Hamiltonian. We provide a perhaps pedagogically more direct demonstration of the negative energy of excitations in the left region by considering a particular class of perturbations in appendix \ref{app:erbridge}. This negative energy is also related to the following point:  the ``physical time'', i.e. the time ordering, for the left region is taken to be pointing upwards in the Penrose diagram. On the other hand  taking commutators with the CFT Hamiltonian moves the points downwards. In other words the geometric action of $H$ coincides with the Killing vector field. This means that ``physical time evolution'' in the mirror region
is not generated by $H$ but rather by $-H$. The reason this happens is that if we think of the mirror operators as being localized in time according to the rule of the previous subsections, then these mirrors are actually explicitly time-dependent operators, as discussed around equation \eqref{explicittime}. Hence their 
physical time evolution in the Heisenberg picture is not given by ${d \widetilde{\cal O} \over dt} \stackrel{?}{=} i[H, \widetilde{\cal O}]$, but rather ${d \widetilde{\cal O} \over dt} = i[H, \widetilde{\cal O}]+ {\partial \tO \over \partial t} = - i[H,\tO]$, since the explicit time dependence in the Schr\"oedinger picture has to be taken into account.

\subsection{Perturbations of Typical States}
\label{sec:pertstates}

In this paper, we only analyze small perturbations of the quantum fields on top of the background geometry. These correspond to excitations which change the CFT energy by factors of $O(N^0)$.  

Typical states are closely related to equilibrium states, defined as states on which simple correlators are almost time-independent. In the rest of this paper we will be discussing various perturbations of a typical state $\rsz$. These perturbations can either be thought of as excited autonomous states, or as states where we actively perturb the system by turning on sources, or combinations of the two. Here we present some examples.

\subsubsection{Autonomous Excited States}
\label{sssec:autexcstate}

We use the term ``autonomous states'' to refer to quantum states where the Hamiltonian of the theory is not modified as a function of time. Hence, the entire history of the state is given by time evolution with respect to $H_{\rm CFT}$ and we are computing correlators on that state. This has to be contrasted with ``actively-perturbed states'', where we modify the CFT Hamiltonian for some period of time.
\begin{figure}[!t]
\begin{center}\includegraphics[width=.6\textwidth]{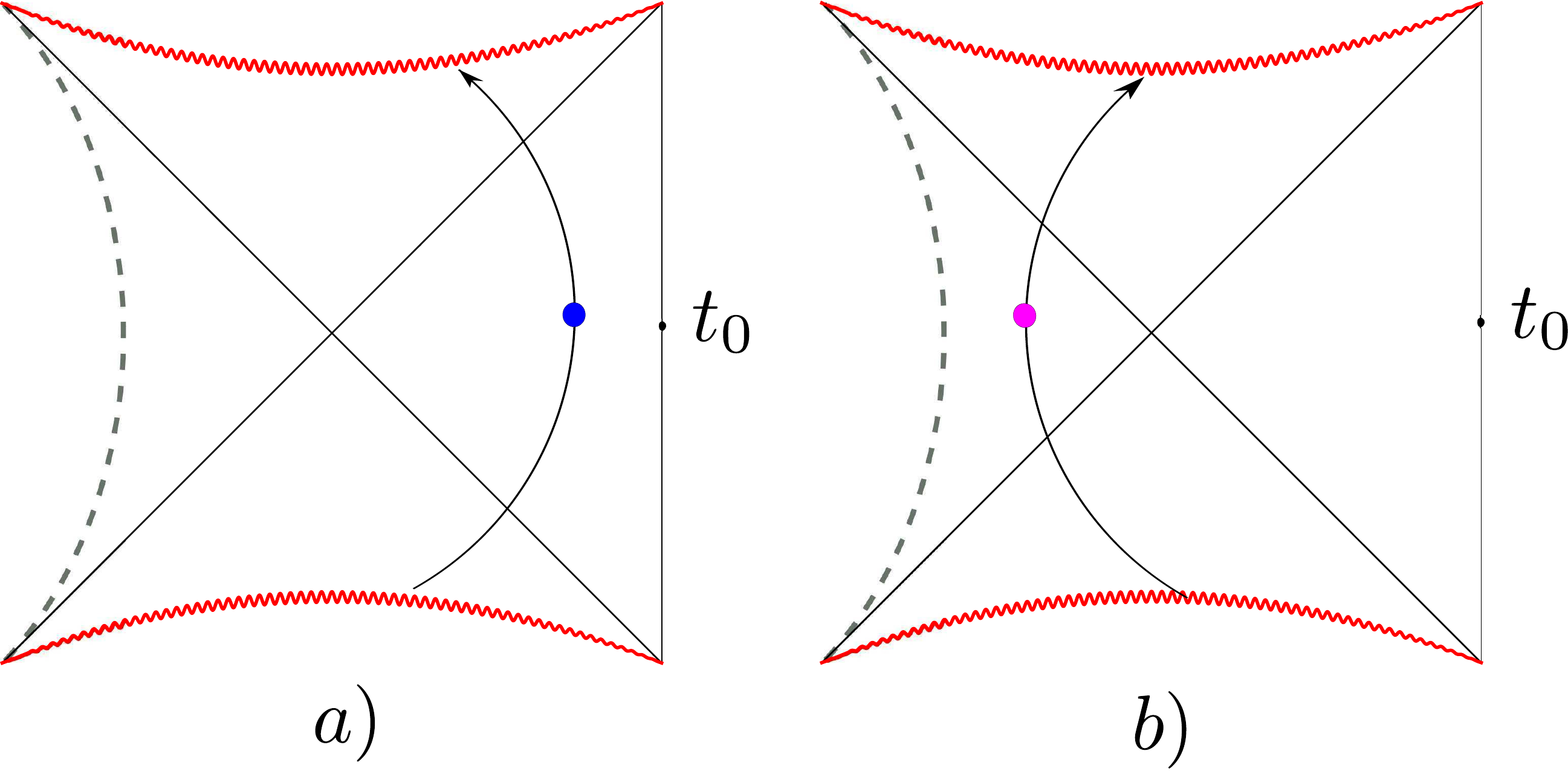}
\caption{Autonomous non-equilibrium states of the form: a) $U({\cal O}(t_0))\rsz$  and b)$U(\tO(t_0))\rsz$. }
\label{autonomous}
\end{center}
\end{figure}

In AdS/CFT for autonomous states we do not turn on any sources on the boundary. Thus states are given as initial conditions and evolve with a time independent hamiltonian after that, for example
\begin{equation}
\rs=U({\cal O}(t_0))\rsz .
\end{equation}
where $\rsz$ is a typical state. The state $\rs$ can be thought of as a state which was prepared to undergo a spontaneous fluctuation out of equilibrium at around $t=t_0$. The unitary could be something of the form $e^{i g {\cal O}(t_0)}$, appropriately smeared. For $t\ll t_0$ the state looks like an equilibrium state. At around $t=t_0$ an excitation seems to be emitted from the past horizon, coming from the white hole region, reaches a maximum distance in AdS and falls back into the future horizon. The difference of energy between $\rs$ and $\rsz$ is
\begin{align}
\label{energy1}\begin{split}
\Delta E = \ls H \rs - \lsz H \rsz & = \lsz U^\dagger [H,U]\rsz = {\rm Tr}[\rho_\beta U^\dagger [H,U]] +O(1/S) \\
& ={\rm Tr}[U \rho_\beta U^\dagger H]- {\rm Tr}[\rho_\beta H] +O(1/S) ,
\end{split}\end{align}
where $\rho_\beta = {e^{-\beta H} \over Z}$ and we used the approximation of a typical state by a thermal ensemble. This way of organizing the computation aims at keeping the error terms at $O(1/S)$. If we ignore the $1/S$ corrections, and use the positivity of the relative entropy, we find that $\Delta E \geq0$. To see that, we consider $S_{\rm rel}(\rho|\sigma)$ for $\sigma=\rho_\beta$ and $\rho= U \rho_\beta U^\dagger$. We have
$S_{\rm rel} = \Delta K - \Delta S \geq 0$. For these two density matrices we have $\Delta S =0$ hence $\Delta K ={\rm Tr}[U \rho_\beta U^\dagger H]- {\rm Tr}[\rho_\beta H] \geq 0 $, or $\Delta E \geq 0$.

Another example is
\begin{equation}
\rs=U(\tO(t_0))\rsz .
\label{insidepert}
\end{equation} 
These are states which  are prepared to undergo a spontaneous fluctuation out of equilibrium at around $t=t_0$, but now in the space of mirrors. The two types of autonomous non-equilibrium states that we have already discussed are schematically depicted in figure \ref{autonomous}.

The states \eqref{insidepert} can also be written as
\begin{equation}
\rs= W \rsz .
\end{equation}
where
\be
W \equiv e^{-{\beta H\over 2}} U({\cal O}(t_0)) e^{\beta H \over 2} .
\ee
Notice that while $W$ is not a unitary, we have $\lsz W^\dagger W \rsz = 1 +O(1/S)$, see \cite{Papadodimas:2017qit} for more details.
If we now estimate the leading order change of the energy we find
\begin{align}
\label{energy2}
\begin{split}
\Delta E & = \ls H \rs - \lsz H \rsz
= \lsz W^\dagger [H,W] \rsz + O(1/S) \\
& = {\rm Tr}[\rho_\beta W^\dagger [H,W] ]+ O(1/S)
=
-{\rm Tr}[U^\dagger  \rho_\beta U H] + {\rm Tr}[\rho_\beta H] +O(1/S) .
\end{split}
\end{align}
It is interesting that we now find $\Delta E\leq 0$. This is consistent with the discussion of the previous subsection, where we argued that placing excitations in the left region lowers the energy of the CFT. This may seem a little surprising, as we are arguing that the fixed operator $e^{-{\beta H\over 2}} U({\cal O}(t_0)) e^{\beta H \over 2}$ can lower the energy of a {\it typical} state, which at first sight seems to be inconsistent with the fact that there are fewer states at lower energies. The resolution of this apparent puzzle was discussed in detail in \cite{Papadodimas:2017qit}. We provide a more explicit example of how this works in the SYK model in subsection \ref{SYKnoneq}.

Of course we can also consider more general perturbations involving combinations of excitations in both regions. 

\subsubsection{ Perturbations of the Hamiltonian}

\begin{figure}[!t]
\begin{center}\includegraphics[width=.8\textwidth]{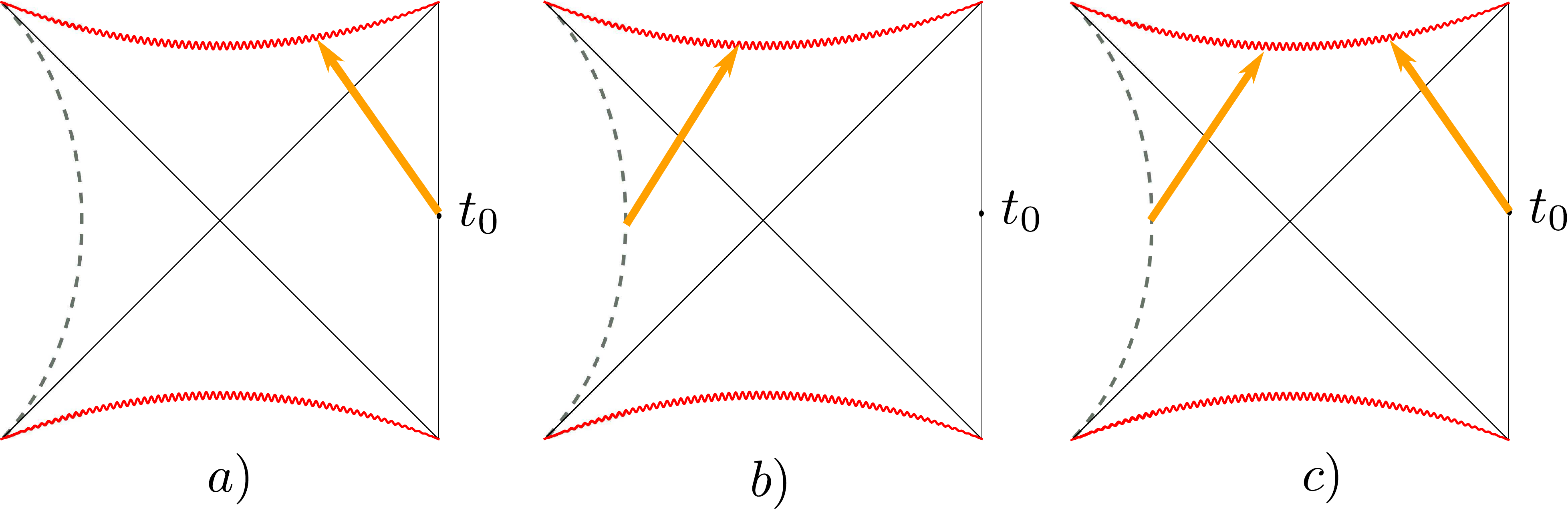}
\caption{a) A usual quench  b) A mirror quench  c) A mixed quench: the analogue of a double-trace perturbation}
\label{quenches}
\end{center}
\end{figure}

We can also consider states where the Hamiltonian is perturbed at some particular moment in time, for example
\be
\label{perthamilton}
H(t) = H_0 +  f(t) A(t) ,
\ee
where $A(t)$ is Hermitian operator localized at time $t$ and $f(t)$ is some smearing function, peaked around some particular $t_0$. The operator $A$ can be made out of the ${\cal O}$'s or the $\tO$'s or both, but it is important that all the constituents of $A$ are localized at the physical time $t$.  The most familiar type of perturbation is to take $A(t)$ to be a simple operator made out of local CFT operators ${\cal O}(t)$. This injects some energy into AdS from the boundary. In some approximation it can be described by
a shockwave falling into an AdS black hole, as shown in figure \ref{quenches}.

If $A$ is a sharply localized local operator, then in some sense the excitations are created near the boundary of AdS. We could also consider smeared $A$'s, for example using the HKLL construction, so that the excitations can be created at some finite depth in AdS. The states that are produced by the time-dependent perturbation \eqref{perthamilton} have the property that at time $t>t_0$ they look like the corresponding autonomous states (provided $A$ has been smeared),  while for $t<t_0$ they look like equilibrium states.

As we discussed before, the boundary perturbation increases the energy of the state. In a shockwave approximation the spacetime before the perturbation has mass $M$ while after the perturbation it has mass $M+\delta M$. The matching conditions across the shockwave relate $\delta M$ to the stress tensor on the shockwave. We can consider excitations created in the left region,
\be
\label{perthamiltonmirror}
H(t) = H_0 +  f(t) \widetilde{A}(t) ,
\ee
by perturbing the Hamiltonian with an operator $\widetilde{A}$ constructed out of the mirror operators $\tO$\footnote{As discussed before, we should make sure that this is a Hermitian operator by adding appropriate $1/N$ corrections.}. In the shockwave approximation we see the spacetime diagram in \ref{quenches}.
For a consistent bulk effective field theory interpretation it is important to use the specific precursors of the mirrors which localize them at the appropriate value of physical time. Notice that since the operators $\tO$ are gravitationally dressed with respect to the right, an active perturbation of the CFT Hamiltonian by $\tO$ will introduce a gravitational Wilson line in the bulk extending from the right boundary all the way into the left region where the shockwave seems to originate, see for example \cite{Donnelly:2015hta}. At order $O(1/N^2)$ these gravitational Wilson lines will backreact on the geometry and their effect has to be included as contributing to the Einstein equations. Understanding the effect of these Wilson lines on the trajectories of probes in the bulk is an interesting question, which we discuss in some more detail in subsection \ref{subsecWilson}, but we postpone a more complete analysis to future work.

It is interesting to consider the backreaction of the shockwave. Since the operators are gravitationally dressed with respect to the right, the left-mass of the spacetime does not change. On the other hand the right mass below the Wilson lines will be $M$ while above the Wilson lines $M+\delta M$. As we saw before, the effect of perturbing the CFT Hamiltonian by $\tO$ leads to $\delta M<0$, which corresponds to {\it lowering} the CFT energy.  We remind that this is not inconsistent with the 2nd law of thermodynamics, because the perturbation is state-dependent.

Finally we can consider perturbations $A(t)$ which are made out of both ${\cal O}$'s and their mirrors. It is important that these operators are localized at the same physical time. We will be interested in the particular class of perturbations of the 
schematic form
\be
\label{perthamiltonmixed}
H(t) = H_0 +  f(t) {\cal O}(t) \tO(t) .
\ee
This produces two shockwaves as indicated in the figure \ref{quenches} and as we will argue is the 1-sided analogue of the double-trace perturbations introduced in \cite{Gao:2016bin}. We will discuss this type of perturbation in more detail in section \ref{sec:doubletrace}.

\section{Traversable One-Sided Black Holes}
\label{sec:doubletrace}

In this section we will discuss in detail the state-dependent perturbation of the class \eqref{perthamiltonmixed}, which we write more precisely as
\begin{equation}
H(t) \equiv H_0+  g V(t) = H_0+ g  f(t) \, {\cal O}(t) \, \tO(t),
\end{equation}
We assume that originally the CFT is in a typical pure state $\rsz$. Here $\widetilde{\cal O}$ are the mirror operators defined by \eqref{defmirror} with respect to $\rsz$. As discussed before, we implicitly assume that these operators are supplemented by appropriate $1/N$ corrections to make the Hermitian. In the expression above we think of the operators as in the interaction picture. The function $f(t)$ is taken to be highly peaked around some time, say $t=0$. Here and hereafter, we assume that the simple operators ${\cal O}(t)$ as well as the mirror operators $\tO(t)$ have uniform support over the entire space domain
\begin{equation}
{\cal O}(t) \equiv \int d^{d-1} \, \mathbf{x} \, {\cal O}(t, \mathbf{x}) .
\end{equation}
We could also consider generalizations where the perturbation uses many light operators $\sum_{i=1}^K \, f(t) \, {\cal O}_i(t) \, \tO_i(t)$, which simplifies some computations at large $K$ \cite{Maldacena:2017axo}.

We discuss some details of the operator $V$ in subsection \ref{smearedV}. Our goal in this section is to use properties of the typical state $\ket{\Psi_0}$ and the  operator $V$ to provide evidence that typical states in the CFT correspond to the geometry proposed in figure \ref{figureintronew}. In particular, this will be evidence that typical black holes in AdS have a smooth interior and a left-exterior region with an effective cutoff, as depicted in Penrose diagram \ref{figureintronew}.  Our analysis will involve doing two thought experiments in the CFT.
\begin{itemize}
\item In {\bf Experiment 1}, we imagine sending a probe made from a mirror particle $\tphi$ at time $-t_*$, which is of the order of scrambling time $\frac{\beta} {2 \pi} \log(S)$. Then we turn on the perturbation  $V$ to the Hamiltonian at time $t=0$. Finally we compute the expectation value of a simple operator in the CFT on the resulting state. If there is a signal of the expected form, that would imply that the mirror particle escaped the horizon.  This experiment is depicted in figure \ref{fig:exp1figure} and described in subsection \ref{subsec:exp}.
\item  In {\bf Experiment 2}, we throw in a particle $\phi$ into the black hole at time $t=-t_*$. The perturbation $V$ is then turned on at time $t=0$. We then compute the response of the signal $\phi$ on a mirror operator $\tphi$ at time $t=t_*$. A non-vanishing response implies the ability of a boundary observer to reconstruct a message thrown into the black hole using the mirror particles. This experiment is reminiscent of the Hayden-Preskill protocol \cite{Hayden:2007cs}. We will elaborate on this connection in section \ref{hp_protocol}. This experiment is depicted in figure \ref{fig:exp2figure} and described in subsection \ref{experiment2}.
\end{itemize}
 Both experiments involve calculating out-of-time-ordered correlators in the typical state $\ket{\Psi_0}$. In fact, using the defining properties of the mirror operators \eqref{defmirror}, we will show that these correlators are approximately equal to left-right correlators with the same structure in the two-sided black hole geometry perturbed by a double-trace operator. 
 We will discuss the errors appearing in this approximation, which are small in the large $N$ limit. The left-right correlators in the two-sided black hole geometry were analyzed in \cite{Gao:2016bin} and \cite{Maldacena:2017axo}. We will thus review their calculation first.

\subsection{Double-Trace Perturbation of the Two-Sided Black Hole}
\label{subsec:dtreview}

We will now review some aspects of the works \cite{Gao:2016bin, Maldacena:2017axo}, where it was argued that a time-dependent double-trace perturbation to the two-sided black hole makes the wormhole connecting the two sides traversable. The eternal two-sided black hole is dual to the thermofield double state (TFD) which is
\begin{equation}
\ket{\Psi_\text{TFD}}=\frac{1}{Z(\beta)^{1/2}} \, \sum_{E} e^{-\frac{\beta E} {2}} \ket{E}_L \ket{E}_R,
\end{equation}
where the sum is over all energy eigenstates, and $\beta$ is the inverse temperature of the black hole. 

No information can be transferred between the left and the right CFT, since the operators on the left and right commute $[{\cal O}_R, {\cal O}_L]=0$. 
 However, coupling the two CFTs with a double-trace perturbation as
 \begin{equation}
H(t)  = H_0+ g \, f(t) \, {\cal O}_R(t) \, {\cal O}_L(t),
\label{tfdpert}
\end{equation}
allows for transfer of information between the two CFTs. Here, ${\cal O}_R, {\cal O}_L$ are again integrated over all of space. In \cite{Maldacena:2017axo} the perturbation was written as a sum over many fields.
In principle we would compute the effect of this perturbation in the interaction picture by a time-ordered exponential
 \begin{equation}
 \mathcal{U}_g = {\cal T} e^{i g  \int dt' f(t') {\cal O}_L(t') \, {\cal O}_R(t') },
 \end{equation}
The transfer of information between the CFTs can be diagnosed using the correlator
\begin{equation}
\mathcal{C} \equiv \bra{\Psi_\text{TFD}} e^{-i\epsilon\phi_L(-t_*)} \, \mathcal{U}_g^\dagger  \,  \phi_R(t) \mathcal{U}_g  \, e^{i\epsilon\phi_L(-t_*)} \ket{\Psi_\text{TFD}},
\label{correlatortrans}
\end{equation}
which is the one-point function of a field $\phi_R(t)$ in the right exterior region sourced by a field $\phi_L(-t_*)$ in the left exterior region, in the presence of the double-trace perturbation. Here $t_*$ is of the order of the scrambling time
${\beta \over 2\pi } \log S$. For appropriate sign of $g$ it was shown in \cite{Gao:2016bin,Maldacena:2017axo} that 
 \begin{equation}
 \mathcal{C} \neq 0 ,
 \end{equation}
indicating information transfer.
 
In the bulk, the double-trace perturbation can be thought of as inserting $\mathcal{O}(N^0)$ energy into the bulk which propagates almost lightlike and thus represents two shockwaves falling into the black hole, from both of the boundaries. One can also find a post-perturbation geometry that is smoothly glued along these shockwaves. These shockwaves backreact on the eternal black hole geometry  such that the IR of the geometry is changed \cite{Gao:2016bin}. In particular, for an appropriate sign of $g$, the quantum stress energy tensor of these shockwaves violates the averaged null energy condition, and thus allows for the wormhole to become traversable. The traversability can be seen by a non-zero commutator of two matter fields in the left and the right exterior region, captured for example by the correlator \eqref{correlatortrans}, thus allowing for transfer of information between the two boundaries.

It is instructive to discuss the quantum stress energy tensor in some more detail. For a scalar bulk field with mass $m$ in the right exterior region, it is given by
\begin{equation}
\label{eq:bulkTab}
T_{\mu \nu} (x) = \partial_\mu \, \phi(x) \, \partial_\nu \, \phi (x) -  \frac{1}{2}  \, g_{\mu \nu} \, \partial^\rho \, \phi(x) \, \partial_\rho \,  \phi (x) -  \frac{1}{2}  \, g_{\mu \nu} \,  m^2 \, \phi^2 (x).
\end{equation}
Its expectation value in the state perturbed by \eqref{tfdpert} can be calculated using the point-splitting method
\begin{equation}
\label{tfdstress}
\langle T_{\mu \nu} \rangle = \lim_{x' \rightarrow x} \,\left[ \partial_\mu \, \partial'_\nu \, G(x, x') - \frac{1}{2} \, g_{\mu \nu} \, g^{\rho \sigma} \, \partial_\rho \, \partial'_\sigma \, G(x,x') - \frac{1}{2} \, g_{\mu \nu} \, m^2 \, G(x, x')\right] ,
\end{equation}
where the short distance singularities have to be subtracted. 
One then only needs to know the corrected bulk two-point function $G(x,x')$ of field $\phi(x)$ with itself, which can be computed in perturbation theory in $g$ \cite{Gao:2016bin}
\begin{align}
\begin{split}
\label{eq:twopf}
G(x,x') &\equiv \bra{\Psi_\text{TFD}} \,  \mathcal{U}_g^\dagger \, \phi(x) \, \phi(x') \,  \mathcal{U}_g \ket{\Psi_\text{TFD}} \\
&= G_0(x, x') -ig \bigg( \int_{t_0}^{t} \, dt_1 \, f(t_1) \, \bra{\Psi_\text{TFD}}  \big[{\cal O}_R(t_1) \, {\cal O}_L(t_1), \phi(t) \big] \, \phi(t') \, \ket{\Psi_\text{TFD}}  \\
&\quad + \int_{t_0}^{t'} \, dt_2 \, f(t_2) \, \bra{\Psi_\text{TFD}}  \,   \phi(t) \, \big[{\cal O}_R(t_2) \, {\cal O}_L(t_2), \phi(t') \big] \,\ket{\Psi_\text{TFD}}  \bigg) +O(g^2),
\end{split}
\end{align}
where $t_0$ is the time before which the function $f(t)$ vanishes, $G_0$ denotes the two-point function in the absence of the perturbation and we have suppressed space coordinates. This can be further simplified using $[\phi,{\cal O}_L]=0$ since we assumed that $\phi$ is in the right region. For the calculation of the $O(g)$ term the entanglement between the two CFTs plays a crucial role. If the two CFTs were in a state very different from the TFD state, firstly there would be no wormhole in the bulk and secondly a simple double-trace perturbation would not lead to a drastic modification of the bulk two-point function.

For the wormhole to be traversable, certain no-go theorems of semi-classical gravity need to be avoided. These often use the average of the local energy, which is
\begin{equation}
\int_{-\infty}^{\infty} \, du \, \langle  T_{uu}  \rangle \equiv  \int_{u=-\infty}^{u=\infty} du \, \langle T_{\mu \nu} \, K^\mu \, K^\nu \rangle ,
\end{equation}
where the null coordinate $u$ runs along the semi-infinite null geodesic very close to the horizon and $K^\mu$ denotes a unit vector tangent to it. It was checked in \cite{Gao:2016bin} that the zeroth order term in $g$ in \eqref{tfdstress} coming from $G_0(x,x')$ integrates to zero, as expected.  At the first subleading order, i.e. at $\mathcal{O}(g)$, we already see that the averaged null energy is proportional to $g$, with the proportionality function being the null integral over derivatives of the subleading two-point function.  It was shown in \cite{Gao:2016bin} that for appropriate choice of $g$ we have
\begin{equation}
\label{eq:anec-ssbh}
\int_{-\infty}^{\infty} \, du \, \langle T_{uu} \, \rangle  < 0.
\end{equation}
This shows that the wormhole can be made traversable.

\subsection{State-Dependent Perturbations in a Single CFT}
\label{smearedV}

From now on we consider a single CFT. The typical, heavy pure state in the CFT
\be
|\Psi_0\rangle = \!\!\!\!\!\!\!\!\!\sum_{E_i \in (E_0,E_0+\delta E)} \!\!\!\!\!\!\!\!\!\!\!c_i |E_i\rangle ,
\ee
is a microstate of a typical large black hole in AdS with one asymptotic boundary. Consider the time-dependent perturbation to the CFT Hamiltonian
\begin{equation}
H(t) \equiv H_0 + g \, V(t) = H_0 + g \, \, f(t) \, {\cal O}(t) \, \tO(t) ,
\label{onesidepert}
\end{equation}
where ${\cal O}$ is a simple operator and $\tO$ is its mirror defined in subsection \ref{subsec:defmirror}.  As discussed in the beginning of section \ref{sec:doubletrace} the smearing function $f(t)$ is assumed to be highly peaked around $t=0$. The operators are uniformly smeared on the spatial sphere on which the CFT is defined.  Remember that the operators $\tO(t)$ have been defined so that they contain frequencies only up to some cut-off $\omega_*$. We  similarly define the operators ${\cal O}(t)$ to be somewhat smeared in time, so that they also contain frequencies up to $\omega_*$.  As in \cite{Maldacena:2017axo}, we can also consider perturbations involving a sum over many different pairs of operators of the form $\sum_i {\cal O}_i\tO_i$.

\subsubsection{Energy Change After the Perturbation}
\label{secenergychange}

A natural diagnostic to study after the perturbation is the change in the energy of the typical state $\ket{\Psi_0}$. The total energy of the state after the perturbation is given by
\begin{equation}
E \equiv \bra{\Psi_0}  \widehat{\mathcal{U}}_g^\dagger \, H_0 \,  \widehat{\mathcal{U}}_g \ket{\Psi_0} ,
\end{equation}
where 
\begin{equation}
\label{eqn:eqnssp}
\widehat{\mathcal{U}}_g =  {\cal T} e^{i g  \int dt' f(t') {\cal O}(t') \, \tO(t') } .
\end{equation}
The energy before the perturbation
\begin{equation}
E_0 \equiv \bra{\Psi_0}  H_0 \ket{\Psi_0} ,
\end{equation}%
is fixed in terms of the coefficients $c_i$ and the eigenvalues $E_i$ that define the typical state $\ket{\Psi_0}$. Expanding up to first order in $g$ we find
\begin{align}
\label{eqn:enexp}
\Delta E =& \int  dt' f(t') \bra{\Psi_0} \,  ig \big[H_0, {\cal O}(t') \tO(t') \big] |\Psi_0\rangle + O(g^2) .
\end{align}
It is easy to see that the first order term is zero. We consider the two-point function
\begin{equation}
G(t_1,t_2)=  \bra{\Psi_0}{\cal O}(t_1) \tO(t_2) \ket{\Psi_0} .
\end{equation}
Using equations \eqref{defmirror} and \eqref{mirrortimes}, this two-point function can be shown to be a function only of $t_1+t_2$. Then the $O(g)$ term in equation \eqref{eqn:enexp} simplifies to
\begin{equation}
\label{purefirsteen}
\Delta E_1 \sim (\partial_{t_1} - \partial_{t_2})G(t_1+t_2)|_{t_1=t_2} = 0,
\end{equation}
where we used the explicit time-dependence of the mirror operators, which gives the minus sign in front of $\partial_{t_2}$. 

It is instructive to compare this to the change in energy of the thermofield double state after a double-trace perturbation $e^{ig \, {\cal O}_L \, {\cal O}_R}$. Even the order $\mathcal{O}(g)$ change in energy is non-zero \cite{Gao:2016bin}, as one sees from the non-zero value for the correlator
\begin{align}
\begin{split}
\Delta E_1^{\rm TFD} &= - i g \, \bra{\Psi_{\text{TFD}}} \, [{\cal O}_R, H_R] \, {\cal O}_L  \, \ket{\Psi_{\text{TFD}}} \\
&= -\frac{i g}{Z(\beta)} \, {\rm Tr} \bigg( e^{-\beta H_R} \, [{\cal O}_R, H_R] \, e^{-{\beta H_R\over 2}}  \, {\cal O}_R^\dagger e^{{\beta H_R \over 2}} \, \bigg) ,
\end{split}
\end{align}
which is generally non-zero. This change in energy clearly is reflected in the change in the ADM mass of the perturbed eternal black hole solution. This raises the interesting question that in the case of a single-sided black hole, what is the bulk interpretation of the fact that the total energy  \textit{does not} change \eqref{purefirsteen} upon acting by the perturbation, equation \eqref{eqn:eqnssp}? We will discuss this question in the next subsection, where we consider the bulk properties and effects of the perturbation. 

Notice that if in the thermofield case we consider the first order variation of the {\it modular} Hamiltonian $H_R-H_L$, then it is zero to first order as $\delta H_R = \delta H_L$.  This is analogous to the one-sided case, where expectation value of the modular Hamiltonian $H-E_0$ also does not change as we found above in \eqref{purefirsteen}.

A more direct method to calculate the energy change is to write the operators in terms of spatial Fourier modes
\begin{equation}
\label{eq:local-modes}
 {\cal O}(t,\Omega) = {1\over (2\pi)^d} \, \int d \omega \sum_{lm} \bigg[e^{-i \omega t} Y_{lm}(\Omega) {\cal O}_{\omega,lm}+h.c.\bigg] ,
\end{equation}
and similarly for the mirror operators (subject to the cutoff $\omega_*$). Here $Y_{lm}$ denote spherical harmonics on ${\mathbb S}^{d-1}$. 
Since we will be working with s-waves, we will for now drop the angular momentum indices. We define the two-point function $G(\omega)$ by the equation\footnote{As discussed in the previous section, here we think of the Fourier modes ${\cal O}_\omega$ in a distributional sense.}
\begin{equation}
{1\over Z} {\rm Tr} ( e^{-\beta H} {\cal O}_{\omega}^\dagger {\cal O}_{\omega'} ) = \delta(\omega-\omega')G(\omega) .
\end{equation}
The KMS condition implies
\be
G(-\omega) = e^{\beta \omega} G(\omega) .
\ee
At large $N$ we expect that we will have similar results for the pure state:
\begin{align}
\begin{split}
\lsz {\cal O}_\omega^\dagger {\cal O}_{\omega'}\rsz &= G(\omega) \delta(\omega-\omega') + O(1/N) , \\
\lsz {\cal O}_\omega {\cal O}_{\omega'}^\dagger \rsz &= e^{\beta \omega}G(\omega) \delta(\omega-\omega') + O(1/N)  .
\end{split}
\end{align}
Finally using the definition of the mirror operators \eqref{defmirror} we can express the two-point functions between mirror Fourier modes, as well as ordinary and mirror Fourier modes in terms of the single function $G(\omega)$. Putting all this together, we can compute the first order change of the energy
\begin{align}
\begin{split}
\label{eq:OgdeltaE}
\Delta E_1 &\equiv  ig \int  dt' f(t') \bra{\Psi_0} \,   \big[H_0, {\cal O}(t') \tO(t') \big] \, \ket{\Psi_0} \\
& \propto ig  \int dt' f(t') \iint d\omega_1 d\omega_2 (\omega_1-\omega_2)  \bigg( e^{-i (\omega_1 +\omega_2)t'} - e^{i (\omega_1+\omega_2) t'
}  \bigg) e^{{\beta \omega_2\over 2}}  G(\omega_2)\, \delta(\omega_1-\omega_2)  \\
&= 0 .
\end{split}
\end{align}
Here we have ignored the bound $\omega_*$ on the frequencies for the mirror operators, which does not play a role in this calculation. The fact that  ${\cal O}$ and $\tO$ have opposite commutators with the CFT Hamiltonian plays an important role in making this energy change equal to zero.

\subsubsection{Shockwaves in One-Sided Black Hole}
\label{BHshocks}

We would now like to discuss the bulk interpretation of the state-dependent perturbation 
\begin{equation}
H \equiv H_0 + g V(t) = H_0 + g \,  f(t) \, {\cal O}(t) \, \tO(t	) .
\label{againpert}
\end{equation}
This perturbation creates shockwaves of infalling matter both in the right and left region, very similar to those in the eternal black hole. These effects are of order $O(N^0)$ and in that sense they correspond to quantum matter, rather than classical configurations of matter. If the leading order metric is normalized to be $O(N^0)$, then the backreaction of this quantum matter on the geometry modifies the metric only at order $O(1/N^2)$. We do not yet have a complete understanding of the backreacted geometry at order $O(1/N^2)$. The reason is that the operators $\widetilde{\cal O}$ have been gravitationally dressed with respect to the right. Hence the quantum matter that they create in the left exterior region of the Penrose diagram should be accompanied by appropriate gravitational Wilson lines, which extend all the way from the left region toward the CFT on the right. These gravitational Wilson lines have to be taken into account when considering the correction to the metric at the $O(1/N^2)$ order.

In order to compute the $O(N^0)$ modification of the quantum state of the fields we follow a procedure similar to that discussed in subsection \ref{subsec:dtreview}. We first compute the quantum-corrected bulk two-point function $G(x,x')$ of a scalar field $\phi(x)$ that is dual to the operator ${\cal O}$ used in the double-trace perturbation \eqref{againpert}. This leads to an equation very similar to \eqref{eq:twopf}
\begin{align}
\begin{split}
\label{eq:twopfb}
G(x,x') &\equiv \lsz \,  \widehat{\mathcal{U}}_g^\dagger \, \phi(x) \, \phi(x') \,  \widehat{\mathcal{U}}_g \rsz \\
&= G_0(x, x') -ig \bigg( \int_0^{t} \, dt_1 \, f(t_1) \,\lsz  \big[{\cal O}(t_1) \, \widetilde{\cal O}(t_1), \phi(t) \big] \, \phi(t') \,\rsz \\
&\quad + \int_{0}^{t'} \, dt_2 \, f(t_2) \,\lsz  \,   \phi(t) \, \big[{\cal O}(t_2) \, \widetilde{\cal O}(t_2), \phi(t') \big] \,\rsz  \bigg),
\end{split}
\end{align}
where $\widehat{\mathcal{U}}_g$ is given in \eqref{eqn:enexp}. From the definition of the mirror operators \eqref{defmirror}, it follows that the correction to the bulk two-point function in the typical state is the same as that in the eternal black hole \eqref{eq:twopf}. Hence, at order $O(N^0)$, we find that the scalar field $\phi(x)$ has the same bulk stress tensor as the one discussed in subsection \ref{subsec:dtreview}. This stress tensor corresponds to a shockwave falling into the black hole from the right region, as can be checked by direct calculation.

A similar calculation can be done for the left region. We consider the part of the left region which is within a few Schwarzchild radii from the bifurcation point. In that region the local bulk field $\widetilde{\phi}$ can be reconstructed, for example by an analogue of the HKLL prescription, where we will use the mirror operators $\widetilde{\cal O}$ instead of the usual operators ${\cal O}$. In that region, and in the limit of large $\omega_*$, the bulk two-points function is the same (up to the obvious left-right reflection) as the bulk two-point function in the right region. We can compute the effect of the perturbation \eqref{againpert} by following a similar analysis as in \eqref{eq:twopfb}, with the obvious replacements $\phi\leftrightarrow \widetilde{\phi}$. The final conclusion is that \eqref{againpert} produces a shock-wave like stress tensor in the left region.

All in all, we find that to order $O(N^0)$ the perturbation \eqref{eq:twopfb} creates two shockwaves of infalling matter which are similar to the two-sided case. By selecting the sign of $g$ appropriately we can make sure that these shockwaves have negative null energy. We emphasize that the existence of the shockwave on the right is completely unambiguous as it follows directly from a the algebraic properties of the $\widetilde{\cal O}$ operators and their effect on HKLL operators via the perturbation \eqref{againpert}. On the other hand the interpretation of the left shockwave relies on our conjecture about the geometry of the typical state, and that the operators $\widetilde{\cal O}$ physically describe the left region.

Also notice that while in the figures we depict the left shockwave as if it was coming from a sharply defined region of the left boundary it should be kept in mind that given that we only use frequencies $|\omega|<\omega_*$ the shockwaves are always somewhat smeared in time. 

\subsubsection{Gravitational Wilson Lines and the Backreacted Geometry}
\label{subsecWilson}

In addition to the two shockwaves, the perturbation \eqref{againpert} creates gravitational Wilson lines extending all the way to the left. This is because the operators $\widetilde{\cal O}$ are gravitationally dressed with respect to the right, in particular they do not commute with the CFT Hamiltonian. The gravitational Wilson lines are spherically symmetric. This follows from the definition of the mirror operators \eqref{defmirror}. There we have implicitly assumed that the mirrror operators are defined so that they commute with the boundary stress tensor, once its zero moved has been removed. This means that $[\tO,T'_{00}] = 0$, where $T_{00}' \equiv T_{00} - H/V$, where $V$ is that spatial volume of the sphere where the CFT lives. This is part of the definition of the mirror operators and other choices could be made which would result in non-spherically symmetric gravitational dressings of the mirror operators.

The existence of the gravitational Wilson lines is important in order to understand the vanishing energy change \eqref{purefirsteen} at first order in $g$ under the perturbation \eqref{againpert}. This perturbation
creates a negative energy shockwave in the right region. At the same time the perturbation inserts  gravitational Wilson lines due to $\tO$, which has  positive energy with respect to the CFT Hamiltonian. The Wilson lines encode the CFT energy of the left shockwave. That shockwave has negative local energy, but as it lies in the left region it has positive energy from the point of view of the CFT Hamiltonian, see for example appendix \ref{app:erbridge}. Considering both effects,  and to leading order in $g$, we find that the energy remains the same and the location of the horizon with respect to the right boundary is unchanged.

If we apply the perturbation \eqref{againpert} to a state which contains particles moving in the region behind the horizon, then we need to understand how these excitations are affected by the gravitational Wilson lines. This is equivalent to understanding how to ``glue'' the geometries, the one before the perturbation (i.e. below the Wilson lines) and the one after the perturbation (above the Wilson lines). This gluing will determine the motion of probes in the geometry. While we have not completed this analysis, the results of the following sections provide evidence that the gluing and the effect of the Wilson lines is such that the  trajectories of probes are not significantly affected when crossing the Wilson lines, in the sense that their effect is suppressed at large $N$. This is to be contrasted with the effect of the right shock-wave on the trajectory of the probe, which is $O(1)$ when the operators are separated by scrambling time.

We notice that the bulk Einstein equations are modified exactly on the Wilson line. This modification refers only to the subleading terms in $1/N^2$. The bulk equations of motion reflect the boundary dynamics. If we consider the time-dependent perturbation \eqref{againpert} the boundary equations of motion are modified for a period of time. Hence it is natural that the bulk equations may need to be supplemented by the contribution from the sources. 

Before we close this subsection we notice that these subtleties about the effect of gravitational Wilson lines, the question of gluing different geometries along spacelike slices and the modification of the bulk Einstein equations on the gluing surface is not specifically related to the mirror operators, or the conjecture about the geometry of a typical state. Similar issues arise whenever we consider perturbations of the CFT Hamiltonian by ``precursors'' and this is generally a topic which deserved further investigation.

For example, suppose we start with the CFT in the ground state $|0\rangle$ on ${\mathbb S}^{d-1}\times {\rm time}$. At time $t=0$ we act with a unitary of the form $U=e^{i g \phi(0)}$ where $\phi(0)$ is an HKLL operator in some particular gravitational gauge. Here the perturbation $U$ is a precursor, which means that while the HKLL operators are usually written as integrals over time, here we use the CFT equations of motion to localize this operator on the boundary at $t=0$. The question we want to understand is what is the bulk geometry dual to the boundary state, which suddenly switched from  $|0\rangle$ for $t<0$ to $U |0\rangle$ for $t>0$. We expect that at very early times the bulk geometry should look like empty AdS, while at very late times it will look like AdS with some particles. For intermediate times around $t=0$ the bulk interpretation is less clear. For instance, suppose we ask what is the backreacted bulk geometry. For $t<0$ the mass is zero, while for $t>0$ the mass is nonzero. We need to glue two geometries of different mass. This sudden change of mass is induced by the gravitational Wilson lines. It would be interesting to understand this toy model in more detail. It captures some of the complications that we face when trying to determine the bulk geometry in our case.

These questions are relevant only when we act with precursors, i.e. boundary operators which directly create particles deep in AdS (together with the accompanying gravitational Wilson lines). If we create the particle by switching on the source near the boundary then the geometry can be understood 
without ambiguity in terms of collapsing matter falling into AdS from infinity.

To summarize, we postpone the interesting question of understanding the bulk geometry to order $O(1/N^2)$ to further work. For now we assume that the boundary arguments presented in the following sections provide evidence that the net effect of probes going through the Wilson lines region is that their trajectory is not drastically modified.

Finally, we mention that for the  kind of typical states with narrow energy band that we are considering, it would not be straightforward to gravitationally dress the mirror operators towards the left. This is because the left dressing would 
require the algebra $[H, \tO] = 0$, which is inconsistent on such states \cite{Papadodimas:2015jra}. 

\subsection{Probing the Region Behind the Horizon}
\label{subsec:exp}

We have discussed the bulk interpretation of the state-dependent perturbation of the form ${\cal O} \tO$. In this subsection we will use this perturbation to probe the different bulk regions and study the horizon.

\subsubsection{Thought Experiment 1}
\label{experiment1}

\begin{figure}[!htb]
\begin{center}
\includegraphics[width=.6\textwidth]{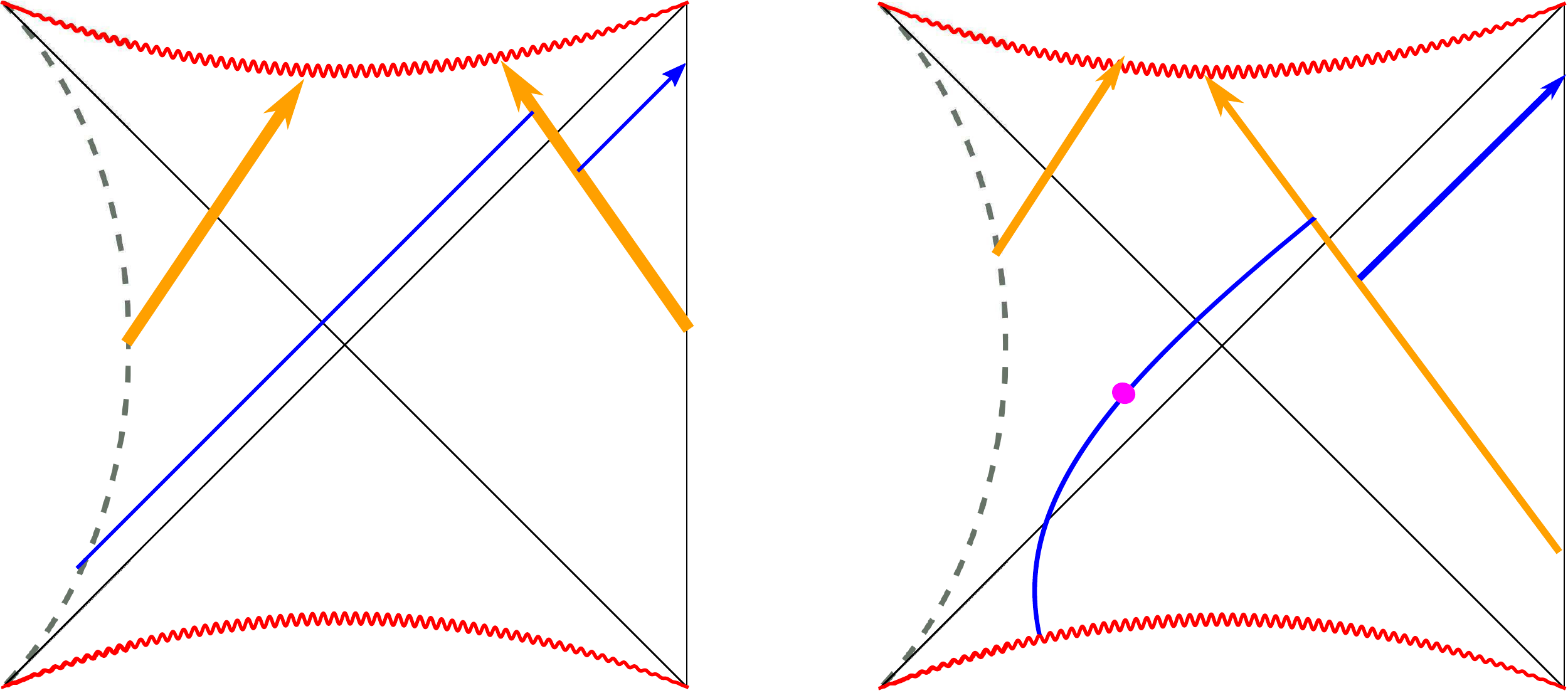}
\end{center}
\caption{Two variants of Experiment 1.}
\label{fig:exp1figure}
\end{figure}

We will now discuss the first thought experiment. In brief, this experiment is designed to probe regions behind the horizon in the conjectured Penrose diagram \ref{figureintronew}. There are two variants of this experiment, as displayed in figure \ref{fig:exp1figure}. 
Let us start with the first variant which is displayed on the left in figure \ref{fig:exp1figure}. Here, the orange lines indicate the two shockwaves and the blue line indicates a particle excitation in the left exterior region. In subsection \ref{sec:pertstates}, we argued that such excitations can be obtained by turning on time-dependent sources for the mirror operators in the CFT, say at time $t=-t_*$ where again $t_*$ is of the order of the scrambling time ${\beta \over 2\pi} \log S$. Because mirror operators commute with simple operators, such excitations cannot be detected in the CFT using simple operators. However, if we further perturb the CFT by the state-dependent perturbation $V$ \eqref{againpert}, the situation changes. We argued in the last Subsection \ref{BHshocks} that the state-dependent perturbation produces two negative energy shockwaves on either side of the horizon. The excitation in the left region interacts with the right shockwave and experiences a null shift in the right region.  After a finite proper time, it can then be detected in the CFT using a simple operator. We can thus interpret the negative energy shockwaves to have made the horizon traversable.

In order to verify that this indeed happens, we need to compute the following correlator. Turning on the mirror source corresponds to acting with $ e^{i \epsilon \tphi(-t_*)}$. Following this, we act with the perturbation at time $t=0$ using the unitary $\widehat{\cal U}_g ={\cal T} e^{ig \int dt'f(t') \, {\cal O}(t') \, \tO(t')}$. We then compute the expectation value of $\phi(t)$ on the resulting state. All in all we need
\begin{align}
\label{eq:exp1cor1}
\begin{split}
\mathcal{C}' &\equiv 
  \bra{\Psi_0} \, e^{-i \epsilon \tphi(-t_*)} \, \widehat{\cal U}_g^\dagger \, \phi(t)\, \widehat{\cal U}_g \,e^{i \epsilon \tphi(-t_*)} \ket{\Psi_0} .
\end{split}
\end{align}
This correlator is analogous to the following correlator \eqref{correlatortrans} when we do the same thought experiment in the two-sided black hole
\begin{equation}
\label{tfdcor}
\mathcal{C} \equiv \bra{\Psi_{\text{TFD}}} \,  e^{-i \epsilon {\phi}_L(-t_*)} ({\cal U}_g)^\dagger \phi_R(t) \, {\cal U}_g\,\,  e^{i \epsilon {\phi_L}(-t_*)} \ket{\Psi_{\text{TFD}}}  .
\end{equation}
where ${\cal U}_g \equiv {\cal T}e^{i g  \int dt' f(t') {\cal O}_L(t')\, {\cal O}_R(t')}$.
This was explicitly calculated in \cite{Gao:2016bin, Maldacena:2017axo} and shown to be non-zero. Instead of directly computing $\mathcal{C}'$, we will argue that it is approximately equal to $\mathcal{C}$, which is easier to calculate.

Using the defining equations for the mirror operators \ref{defmirror} repeatedly,  we can rewrite $\mathcal{C}'$ as the expectation value of a complicated string  of {\it ordinary} (i.e. non-mirror) operators\footnote{It is crucial to realize that, after the mirror operators are mapped to normal operators, the resulting correlators do not correspond to experiments that can set up by only using the normal operators. } on the state $\ket{\Psi_0}$. We call this string of operators ${\cal X}(\phi, \cal O)$, so we have
\be
\label{purecorrelator}
C' = \lsz {\cal X}(\phi,{\cal O})\rsz .
\ee
Similarly, in the case of the two-sided black hole, the action of ${\cal O}_L$ operators can be re-written in terms of the ${\cal O}_R$ operators using the properties of the TFD state
\begin{align}
\label{eq:OLTFDdef}
\begin{split}
 {\cal O}_{L,\omega} &|\Psi_{\rm TFD}\rangle  = e^{-{\beta \htfd \over 2}} {\cal O}_{R,\omega}^\dagger e^{{\beta \htfd \over 2}} |\Psi_{\rm TFD}\rangle , \\  
 {\cal O}_{L,\omega} {\cal O}_{R,\omega_1}...{\cal O}_{R,\omega_n} & |\Psi_{\rm TFD}\rangle = 
{\cal O}_{R,\omega_1}...{\cal O}_{R,\omega_n}  {\cal O}_{L,\omega}  |\Psi_{\rm TFD}\rangle  ,\\
[\htfd,{\cal O}_{L,\omega}]{\cal O}_{R,\omega_1}...{\cal O}_{R,\omega_n}  & |\Psi_{\rm TFD} \rangle= \omega {\cal O}_{L,\omega}{\cal O}_{R,\omega_1}...{\cal O}_{R,\omega_n}   |\Psi_{\rm TFD} \rangle .
\end{split}
\end{align}
where $\htfd \equiv H_R - H_L$. Notice that these equations are completely similar to equations \ref{defmirror} if we identify ${\cal O}_{L,\omega}
\leftrightarrow \tO_{\omega}, {\cal O}_{R,\omega} \leftrightarrow {\cal O}_\omega , \htfd
\leftrightarrow H$.

Using the equations \eqref{eq:OLTFDdef}, we can now repeat the same process in correlator $\mathcal{C}$, by replacing ${\cal O}_L, \phi_L$ in terms of right CFT operators. In this way we get {\it exactly the same string} ${\cal X}(\phi_R,{\cal O}_R)$, now expressed in terms of ${\cal O}_R$. This string is a function only of operators in the right CFT, and hence we can compute it by first tracing out the left CFT. Let us drop the $R$ subscript for economy. The correlator $\mathcal{C}$ then becomes a thermal correlator in the right CFT
\begin{align}
\label{eq:comparisoncor}
\mathcal{C} &= {1\over Z} {\rm Tr} \big( e^{-\beta H}{\cal X}(\phi,{\cal O} ) \big) ,
\end{align}
here ${\cal X}(\phi,{\cal O})$ is exactly the same string as the one in \eqref{purecorrelator}. We know that the correlator $\mathcal{C}$ contains a signal corresponding to the probe traversing the horizon. If the correlator $\mathcal{C}'$ is close to $\mathcal{C}$ then the same signal will be present in the one-sided black hole perturbed by the state-dependent operator \eqref{againpert}, which will be evidence that a particle was extracted from the left region of our conjectured geometry. This brings us to the main conclusion:

{\quote The conjecture that the bulk geometry of a typical state is described by the Penrose diagram discussed in section \ref{sec:mainconjecture} and that it responds to
perturbations in the way predicted by effective field theory on this diagram, requires as a necessary condition that the correlators $C,C'$ are the same at large $N$. This is essential to hold even when the time separations of the operators are taken to be of the order of scrambling time.

Thus, we have identified a technical condition for CFT correlators, necessary for the smoothness of the horizon of a typical state. We discuss this condition in more detail in section \ref{sec:conjecture}. We also provide some preliminary evidence in favor of its validity.}

\subsubsection*{A variant setup}

Now we come to the second variant of experiment 1, depicted in the right part of figure \ref{fig:exp1figure}. Here, we do not use a time-dependent source for the particle in the left region. Instead of starting with the typical state $\ket{\Psi_0}$ and acting with the operator $e^{i \epsilon \tphi(-t_*)}$, we start in the state
\begin{equation}
 \ket{\Psi_I} \equiv e^{i \epsilon \tphi(-t_*)} \, \ket{\Psi_0} .
\end{equation}
This is an {\it autonomous} non-equilibrium state, owing to the fact that $[H, \tphi] \neq 0$. As such, this state is not typical under the Haar measure, but it is an autonomous state in the full CFT Hilbert space nonetheless. Detailed discussion of such states can be found in \cite{Papadodimas:2017qit}. The experiment then consists of acting on such a non-equilibrium state by the unitary of the state-dependent perturbation. As before, one then aims to detect the mirror excitation inherent to this state by using a simple operator $\phi$. The entire experiment can be encoded in the correlator
\begin{equation}
 \mathcal{C}'' = \bra{\Psi_I} \, \widehat{\cal U}_g^\dagger \, \phi(t) \, \widehat{\cal U}_g^\dagger \, \ket{\Psi_I},
\end{equation}
where $V$ is as before given by equation \eqref{againpert}. The value of this correlator is closely related to that of $\mathcal{C}'$ in \eqref{eq:exp1cor1}. It can then be compared to similar correlators in the thermofield setup where the left side of the eternal black hole is in some autonomous non-equilibrium state.

\subsubsection{Thought Experiment 2}
\label{experiment2}

\begin{figure}[!htb]
\begin{center}
\includegraphics[width=.30\textwidth]{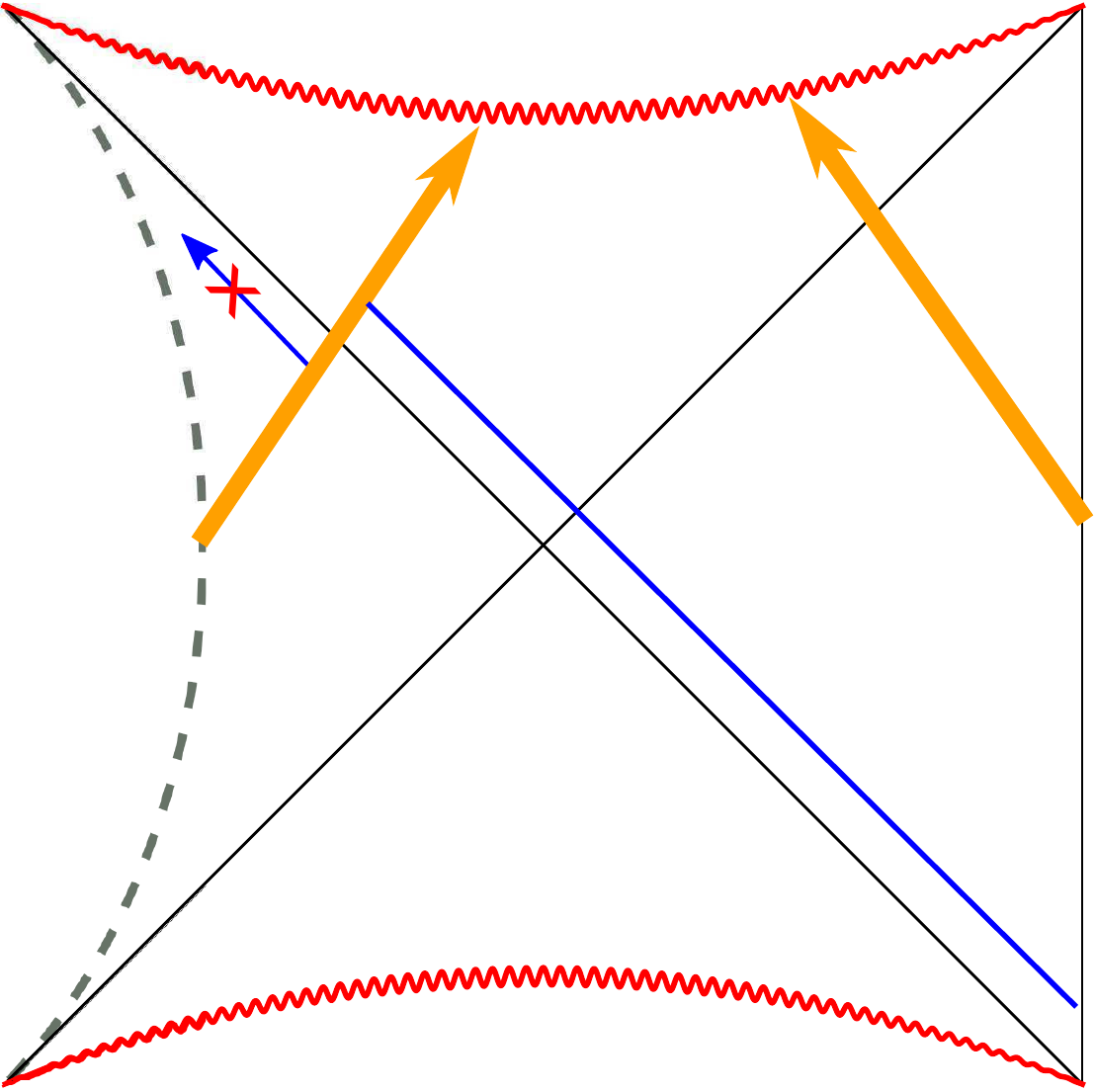}
\end{center}
\caption{Experiment 2}
\label{fig:exp2figure}
\end{figure}

We now study a second thought experiment to probe the region behind the horizon, displayed in figure \ref{fig:exp2figure}.
In this experiment, we start with the typical pure state $\ket{\Psi_0}$. Then we act on this state by a unitary of a simple operator $e^{i \epsilon \phi(-t_*)}$, where $t_*$ denotes a timescale of the order of scrambling time and $\phi$ a field near the boundary of the CFT. This creates a particle excitation in the right exterior region of the bulk, outside the black hole horizon. We have depicted this using the blue ray in figure \ref{fig:exp2figure}. The CFT state then becomes
\begin{equation}
 \ket{\Psi_E} \equiv e^{i \epsilon \phi(-t_*)} \, \ket{\Psi_0} .
 \label{te2}
\end{equation}
The particle depicted in the figure falls towards the black hole and eventually crosses the black hole horizon to go into the interior region. Our goal now is to somehow reconstruct the state of this particle. There are many way to do this, since in principle the information of the particle is always present in the CFT. Here we will describe a protocol which uses the state-dependent perturbations \eqref{againpert} and in a particular extrapolation it realizes the Hayden-Preskill protocol as formulated by \cite{Maldacena:2017axo}. This will be the subject of section \ref{hp_protocol}.

The protocol is as follows: after throwing the particle in the black hole \eqref{te2} and waiting for scrambling time, we act on the above state at time $t=0$ by the state-dependent perturbation \eqref{againpert}. This perturbation creates the shockwaves that we discussed. As shown in the figure \ref{fig:exp2figure} the shockwaves deflect the particle which moves into the left region. There it can be detected by measuring a mirror operator of the form $\tphi(t_*)$. We thus need to calculate the correlator
\begin{equation}
 \label{eq:corr_exp2}
\mathcal{C}_{2}' \equiv \bra{\Psi_0} \, e^{-i\epsilon \phi(-t_*)} \,\widehat{\cal U}_g^\dagger \, \tphi(t_*) \, \widehat{\cal U}_g\, e^{i \epsilon \phi(t_*)} \, \ket{\Psi_0} .
\end{equation}
Using similar steps as before we can reduce this correlator to a correlator of ordinary single trace operators and compare it to the corresponding correlator in the TFD state.

\subsection*{Summary}

In this section we described some thought experiments, which indirectly probe the region behind the black hole horizon. 
We showed that our conjecture for the geometry presented in section \ref{sec:mainconjecture} requires as a necessary condition that certain CFT correlators on typical pure states are close to thermal correlators. Assuming that the correlators are indeed the same at large $N$, we find that the typical black hole microstate responds to perturbations as if it contained the part of the extended Penrose diagram presented in section \ref{sec:mainconjecture}.


\section{The SYK Model as an Example} 
\label{sec:SYK}

We will exemplify some of the previous statements in the context of the SYK model. The SYK model is a toy model of holography, and although it is not expected to have an Einstein bulk dual, it still captures some important features of the bulk theory. 

\subsection{Brief Review of the SYK model}

The Sachdev-Ye-Kitaev (SYK) model \cite{kitaevtalks, Polchinski:2016syk, Maldacena:2016syk} is a  one-dimensional quantum mechanics model containing $N$ species of Majorana fermions $\psi_i$, $i=1, 2, ..., N$. The fermions satisfy  $\{\psi_i,\psi_j\}=\delta_{ij}$. In general, the fermions in the SYK model have $q$-body random interactions such that the Hamiltonian is
\begin{equation}
H=(i)^{q/2} \, \sum_{1 \le i_1 < i_2 < \cdots < i_q\leq N}  J_{i_1 \, i_2 \, \cdots i_q} \, \psi_{i_1} \, \psi_{i_2} \, \cdots \psi_{i_q} \, ,
\end{equation}
where the coupling constants $J_{i_1 \, i_2 \, \cdots i_q}$ are all chosen randomly from a Gaussian distribution with mean zero and variance
\begin{equation}
 \langle J^2_{i_1 \, i_2 \, \cdots i_q} \rangle = \frac{2^{q-1} \, (q-1)! \,  \mathcal{J}^2} {q \, N^{q-1}} \, .
\end{equation}
The parameter ${\cal J}$ has dimensions of energy and sets the scale of the problem.  The variance of the coupling $ J_{i_1 \, i_2 \, \cdots i_q} $ is chosen to depend explicitly on $N$ so that the model has interesting properties in the large $N$ limit. When $q=2 \, \text{mod} \,  4$, the factor of $(i)^{q/2}$ upfront is necessary to make the Hamiltonian Hermitian. The model becomes conformal at low energies i.e. when the frequencies are very small compared to ${\cal J}$. The conformal limit of this model has been studied in detail in \cite{Polchinski:2016syk, Maldacena:2016syk}.

\begin{figure}[!htb]
\begin{center}
\includegraphics[scale=0.3]{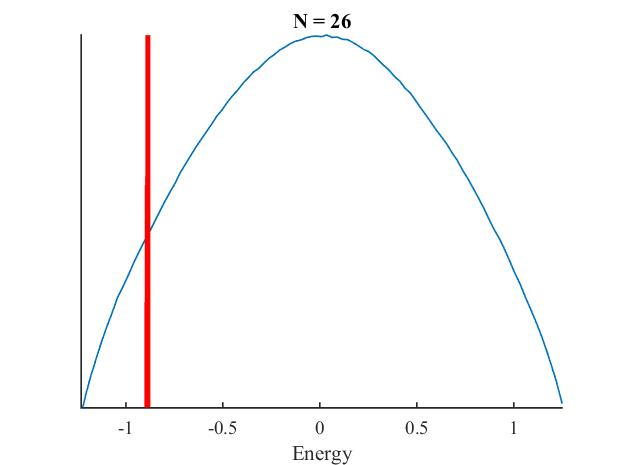}
\end{center}
\caption{Distribution of energy eigenvalues of the SYK model, the red strip shows the energy eigenstates selected for a typical state.}
\label{fig:ev_distr}
\end{figure}

It is easier to compute correlation functions in the SYK model after taking an ensemble average over the couplings $J_{i_1 \, i_2 \, \cdots i_q}$. However, we will assume a particular realization of the coupling constant $J_{i_1 \, i_2 \, \cdots \, i_q}$ to obtain a unitary model. This is not a problem because the SYK model is self-averaging: to leading order at large $N$ correlators are the same if we choose a specific realization of the coupling constants $J_{i_1 \, i_2 \, \cdots \, i_q}$ or perform a disorder-average over it.  Thus, at large $N$ we are in principle able to compare correlators calculated in a particular realization (say numerically) with the ones estimated (eg. analytically) using the disorder-average. For finite $N$, the Hamiltonian is a finite-dimensional matrix with size $2^{N/2} \times 2^{N/2}$ and has $2^{N/2}$ energy eigenvalues. It is relatively easy to find these eigenvalues and corresponding eigenstates by direct numerical diagonalization for reasonably large $N$. In figure \ref{fig:ev_distr}, we display distribution of energy eigenvalues for $N=26$. 

\subsubsection{Equilibrium and Non-Equilibrium States in the SYK Model}
\label{SYKnoneq}

The finite size of this model (at finite $N$) makes the SYK-model a good tool to numerically test various statements about typical state and the perturbations discussed earlier in section \ref{sec:pertstates}. We, moreover, have greater analytic control over some aspects of the model, which allows as to do more explicit CFT calculations. Sometimes it is easier to consider a set of spin operators
\begin{equation}
S_k \equiv 2i\psi_{2k-1}\psi_{2k}\quad, \qquad \qquad k=1,\dots,N/2 ,
\end{equation}
to further simplify calculations. These operators are bosonic and are therefore more in line with earlier discussions. We will assume that we have a particular realization of the SYK $(q=4)$ couplings $J_{ijkl}$. This means we have a well defined quantum system with a Hilbert space and unitary time-evolution. Nevertheless we will use results from disorder averaging as a mathematical technique, which allows us to estimate certain correlators for the model with a particular realization of $J_{ijkl}$, as the disorder in the SYK model is self-averaging.

The Hamiltonian has $2^{N/2}$ energy eigenstates $|E_i\rangle$, which can can be found by any diagonalization method at finite $N$. The interesting critical behavior of SYK takes place at the low-energy regime of this spectrum. We will define typical pure states in the SYK model by writing down pure states of the form
\be
\label{typicalsyk}
\rsz \; = \! \! \! \! \! \! \! \! \sum_{E_i \in (E_0,E_0+\delta E)} \! \! \! \! \! \! \! \! c_i |E_i\rangle  ,
\ee
where $|E_i\rangle$ are the exact SYK eigenstates (for a particular realization), and we select an energy window $(E_0,E_0 +\delta E)$ centered around some energy $E_0$\footnote{$E_0$ should not be confused with the energy of the ground state of the SYK model.} and with width $\delta E$. We assume $E_0$ is in the low energy regime, where the SYK model is strongly coupled and $E_0 - E_{\text{gs}} \sim a N$ where $E_{\text{gs}}$ denotes the ground state energy in SYK model and $a$ is a small number ($a<<J$) which does not scale with $N$. From basic thermodynamics and using the partition function $Z(\beta) = \sum_i e^{-\beta E_i}$ we can relate $E_0$ to $\beta$. We want to be in the regime where $\beta {\cal J}\gg 1$. We take the spread $\delta E$ to scale like $O(N^0)$ which implies that it is very small compared to $E_0 - E_{\text{gs}}$. At the same time we take $\delta E$ is large enough, so that we have exponentially many (in $N$) states contributing to equation \eqref{typicalsyk}.

\begin{figure}[!hb]
\begin{center}
\includegraphics[width=1\textwidth]{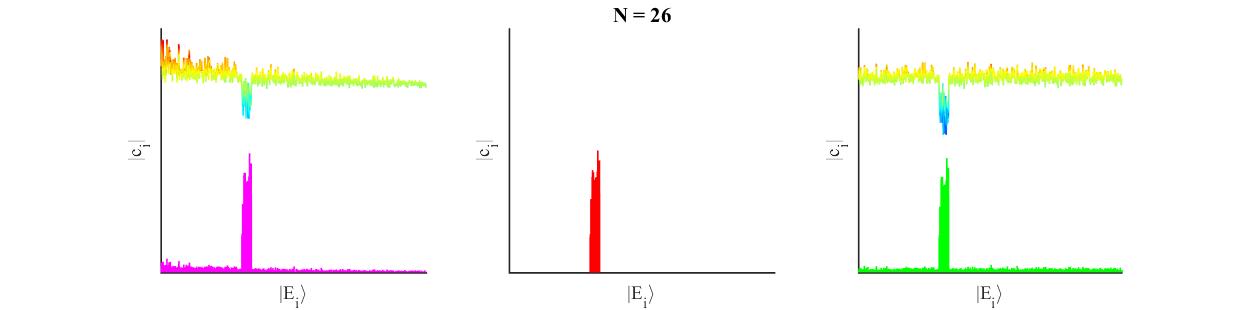}
\caption{Distribution of $|c_i|$ in non-equilibrium state of the form $U(\widetilde{S}_i)\rsz$ (magenta), typical equilibrium state $\rsz$ (red), and non-equilibrium state of the form $U(S_i)\rsz$ (green). The line above the bar plot shows, in heat map colors, which eigenstates are excited, and which ones are suppressed because of the perturbation. Blue eigenstates are suppressed, while eigenstates with other colors are excited with small (green), medium (orange), or large (red)
 magnitude.}
\label{disten}
\end{center}
\end{figure}

Let us now consider some examples of exciting an equilibrium state in the SYK model. 
Usual excited states can be written as
\be
\label{excitedsyk}
e^{i \epsilon S_i (t_0)} \rsz .
\ee
The analogue of states with excitations behind the horizon can be written as
\be
\label{excited2syk}
e^{i\epsilon \widetilde{S}_i (t_0)} \rsz = e^{-{\beta H \over 2}} e^{i \epsilon S_i(t_0)} e^{{\beta H \over 2}} \rsz .
\ee
Adding excitations behind the horizon lowers the energy of the state as shown in \eqref{energy2}. Hence, in states of the form \eqref{excited2syk} the amplitudes of lower energy eigenstates are amplified relative to $\rsz$, but coefficients of higher energy eigenstates are also turned on therefore ``borrowing'' that part of the Hilbert space. 
This explains why there is no paradox that the fixed, {\it invertible}  operator $e^{-{\beta H \over 2}} e^{i \epsilon S_i(t_0)} e^{{\beta H \over 2}}$ lowers the {\it expectation value} of the energy of a typical state, as discussed in more detail in \cite{Papadodimas:2017qit}. In figure \ref{disten} we can explicitly see this effect in the case of the SYK model.

\subsection{Mirror Operators in the SYK Model}

Having defined a typical pure state in the SYK model in equation \eqref{typicalsyk}, we now define mirror operators. The first step is to define a ``small'' algebra ${\cal A}$ of operators in the SYK model. From the AdS/CFT point of view it would seem more natural to consider only ``gauge invariant operators'' like $\psi_i \, \partial^{2m+1} \, \psi_i$. The number of such operators at a given conformal dimension does not scale as $N\rightarrow \infty$, as expected for CFTs with weakly coupled (but possibly highly curved) AdS bulk duals \cite{Heemskerk:2009pn,Fitzpatrick:2010zm,ElShowk:2011ag}.

However, it is interesting to consider the possibility of defining the mirrors for the non-gauge invariant operators like the individual fermions $\psi_i$ or the spin operators $S_k$, as many interesting statements about the SYK models can be made
directly for the fundamental operators. Thus, we will define the small algebra $\mathcal{A}$ as a span of the low-frequency components of the operators $\psi_i$ and their small products. A typical pure state cannot be annihilated by these operators, hence the construction of the mirror operators can go through. Notice that while the Hamiltonian is quartic in the fermions, it is not part of the algebra ${\cal A}$ because we have imposed the condition that only the low frequency components of the fermions are in ${\cal A}$, see discussion in subsection \ref{subsec:defmirror}, and to reconstruct $H$ we would need the fermions sharply localized at a given moment in time.

We will present the mirror construction for the spin operators $S_k$, since their bosonic nature makes the presentation simpler. Generalization to the fermions $\psi_i$ is straightforward. The operator $S_k(0)$ can be represented as a $2^{N/2} \times 2^{N/2}$ matrix, by writing the fermions as gamma matrices in the standard basis of Clifford algebra. One can also write it as a matrix in the energy eigenbasis
\begin{equation}
(S_k)_{ij} = \bra{E_i} S_k \ket{E_j} \, .
\end{equation}
Now consider the time evolution of this operator $S_k(t) = e^{i H t} S_k(0) e^{-i Ht}$ where $H$ is the SYK Hamiltonian. This defines for us the time-dependent Heisenberg operator. Its exact Fourier modes are
\begin{equation}
\label{eq:exact_mode_syk}
(S_k)^{\rm exact}_\omega \equiv \int_{-\infty}^{\infty} dt \, e^{i \omega t} \, e^{i H t} S_k(0) \, e^{-i Ht} \, .
\end{equation}
Using the definition of $S_k$, this can be written as a product of a sum of delta functions of the form $\delta(\omega-E_i+E_j)$. As we discussed in subsection \ref{subsec:defmirror}, to smoothen these delta functions out, we define the coarse-grained Fourier modes, as a function of frequencies in the bin $(\omega, \omega + \delta \omega)$. We take $\delta \omega$ to be of order $K \omega_{\rm gap}$ where $\omega_{\rm gap}$ is the typical gap between energy eigenstates
and $K \gg 1$. The coarse-grained modes then become
\begin{align} 
\label{coarseSYK} 
(S_k)_\omega  &= \frac{1}{\sqrt{\delta \omega} }\int_{\omega	} ^{\omega+\delta\omega} d\omega'\,(S_k)_{\omega'}^{\rm exact} \, .
\end{align} 
An algorithm to find the mirror operators explicitly can be constructed as follows \cite{Papadodimas:2013}. First we construct a set of vectors which will span the small Hilbert. We can use the coarse grained spin operators $S_{k,\omega}$ to do this. The set of vectors is constructed as
\begin{align}
\begin{split}
\ket{1}&=\ket{\Psi_0} ,\\
\ket{2}&=S_{1,\omega_1}\ket{\Psi_0} ,\\
\ket{3}&=S_{1, \omega_2}\ket{\Psi_0} ,\\
&\quad\vdots \\
\ket{n}&=S_{2, \omega_1}\ket{\Psi_0} ,\\
\ket{n+1}&= S_{2, \omega_2}\ket{\Psi_0} ,\\
&\quad\vdots \\
\ket{l}&=S_{2, \omega_2} S_{1, \omega_1}\ket{\Psi_0} ,\\
&\quad\vdots
\end{split}
\label{vectors}
\end{align}
This set of states is not orthogonal but we make sure that we truncate the set so that the states are linearly independent. This truncation is the same as the truncation in the definition of the small algebra ${\cal A}$ in subsection \ref{subsec:defmirror}. In the large $N$ limit the set of vectors \eqref{vectors} can be taken to be very large, and the effects of the truncation become unimportant for our purposes, see \cite{Papadodimas:2013} for a discussion of the truncation in a more general context. 

To simplify the notation, we denote these states as
\be
|I\rangle \equiv {\cal O}_I \rsz ,
\ee
where ${\cal O}_I$ is a combination of the spin operators introduced above. We define $G_{IJ}\equiv\langle I |J \rangle$ as a metric between the states. Since we demanded that the states are linearly independent, $G_{IJ}$ is an invertible matrix and
we denote its inverse by $G^{IJ}$. We also define
\be
B_{IJ,k \omega} \equiv \bra{I} \widetilde{S}_{k,\omega}\ket{J} ,
\ee
which can be computed using the equations for the mirror operators \eqref{defmirror}. We therefore have
\be
B_{IJ,k \omega}= \lsz O_I^\dagger O_J e^{-{\beta H \over 2}} S_{k,\omega} e^{{\beta H \over 2}}\rsz .
\ee
This is a matrix element involving only the ordinary operators, hence it may be in principle computed.
Finally we can represent the mirror operators explicitly as
\begin{equation}
\widetilde{S}_{k,\omega} = G^{IJ}B_{JK,k\omega}G^{KL} \ket{I}\bra{L} .
\end{equation}
While this allows us to explicitly construct the mirror operators, for example in Mathematica, it is much more economic to directly apply the equations from section \ref{subsec:defmirror} to transform the mirror operators to normal operators inside correlators.

\subsection{Comments on the Kourkoulou-Maldacena States}
\label{spins_mirrors}

We would now like to discuss some connections of the results of this paper to the work of Kourkoulou and Maldacena \cite{Kourkoulou:2017zaj}. That paper considered a class of pure states in the SYK model, denoted by $\ket{B_s(\beta)}$. To construct these states, we first consider a basis of states $\ket{B_s}$, which are eigenstates of the spin operators 
\begin{equation}
S_k \equiv 2i\psi_{2k-1}\psi_{2k} \, ,  \qquad k=1,\dots,N/2 \, .
\end{equation}
These states are defined such that
\begin{equation}
\label{eq:syk_spins}
S_k \, \ket{B_s} = s_k \ket{B_s} \, ,  \qquad  s=1,2, \cdots, 2^{N/2} \, ,
\end{equation}
where $s$ denotes the collection $s_k = \pm 1$. The states $\{ \ket{B_s} \}$ form a complete basis of states, each with mean energy around zero. Evolving the states $\ket{B_s}$ in Euclidean time by %
\begin{equation}
\label{eq:thermal_spins}
\ket{B_s(\beta)} \equiv e^{-{\beta \over 2} H } \, \ket{B_s}  \, ,
\end{equation}
we obtain states with an energy more comparable with the typical state of energy corresponding to inverse temperature $\beta$. This defines the states $\ket{B_s(\beta)}$ introduced in \cite{Kourkoulou:2017zaj}. Notice that as written in \eqref{eq:thermal_spins} the states are not yet unit-normalized.

The ensemble of states $ \{ \ket{B_s(\beta)} \}$ has the special property that it is equivalent to the thermal ensemble, in the sense
\begin{equation}
\sum_{s=1}^{2^{N/2}} \bra{B_s(\beta)} \psi \cdots \psi \ket{B_s(\beta)} = {\rm Tr} \big[ e^{-\beta H} \psi \cdots \psi  \big] .
\label{kmthermal}
\end{equation}
In the large $N$ limit the SYK model has an approximate $O(N)$ symmetry. A subgroup of this symmetry, identified as the ``flip group'' in \cite{Kourkoulou:2017zaj}, implies that certain flip invariant correlators  (for example diagonal two-point functions of the fermions) are the same at large $N$ 
on all $\ket{B_s(\beta)}$ states, and from \eqref{kmthermal}, equal to the thermal correlators. On the other hand there are non flip invariant correlators (for example non-diagonal correlators) which are different from the thermal correlators. In particular they are time-dependent, indicating that
the states $\ket{B_s(\beta)}$ are non-equilibrium states, and in particular {\it atypical} states.  Under time evolution the states thermalize and they equilibrate at late times.

While the states $\ket{B_s(\beta)}$ are atypical, and hence rather different from the typical states which are the focus of this paper, we will argue that the states  $\ket{B_s(\beta)}$ can be approximately constructed by the following procedure.
We start with a typical state of energy $E_0$ in the SYK model and we perform a measurement of many mirror operators $\widetilde{S}_k$. This produces a state which resembles $\ket{B_s(\beta)}$, with a parameter $\beta$ related to $E_0$ by the thermodynamics of the SYK model.

To see that we first consider what happens if we perform a measurement of a single mirror-spin operator. It is easier to phrase the discussion in position space, so we introduce the (smeared) position space mirror spins\footnote{As discussed before, we may need to add small $1/N$ corrections to make sure that the operators are Hermitian.}
\be
\widetilde{S}_{k}(t) = \int_{-\omega_*}^{\omega_*} \widetilde{S}_{k,\omega} e^{-i \omega t}.
\ee
As we discussed in section \ref{sec:mainconjecture} these should not be thought of as operators sharply localized in time, but in the limit where $\omega_*$ is taken to be large they do start to behave like local operators. In that limit the eigenvalues of $\widetilde{S}_{k}(t)$ are approximately $\pm 1$. If we measure $\widetilde{S}_k(0)$ we will get either $s_k=1$ or $s_k=-1$, and the state will be projected to 
\be
{1 +s_k \widetilde{S}_k(0) \over 2} \rsz .
\ee
Using the definition of the mirror operators we can rewrite this as
\begin{align}
 \label{sykbstate}
\begin{split}
 e^{-{\beta H \over 2}} \left({1 +s_k S_k(0) \over 2}\right)e^{\beta H \over 2} \rsz &=  e^{-{\beta H \over 2}} \left({1 +s_k S_k(0) \over 2}\right)e^{\beta H \over 2} \sum_{s'} c_{s'} \ket{B_{s'}(\beta)} ,\\
&=  e^{-{\beta H \over 2}} \left({1 +s_k S_k(0) \over 2}\right) \sum_{s'} c_{s'} \ket{B_{s'}} ,\\
&= \sum_{s'} c_{s'} \ket{B_{\alpha}(\beta)}\delta_{s_k',s_k} .
\end{split}
\end{align}
where the last delta function restricts the sum over $s'$ to strings where the $k$-th spin is required to be $s_k$.

Now, if we simultaneously measure $n$ mirror-spin operators, and get eigenvalues $s_{i_1},...,s_{i_n}$, we will project the state to
\be
\left({1 +s_{i_1} \widetilde{S}_{i_1}(0) \over 2}\right)...\left({1 +s_{i_n} \widetilde{S}_{i_n}(0) \over 2}\right) \rsz ,
\ee
which, using the definition of the mirror operators, can also be written as
$$
\sum_{s'} c_{s'} \ket{B_{s'}(\beta)}\delta_{s_{i_1}',s_{i_1}}...\delta_{s_{i_n}',s_{i_n}}
$$
We notice that as we increase $n$ we fix more and more spins, and if we could extrapolate to $n={N\over 2}$ we would get precisely the $\ket{B_s(\beta)}$ states
of \cite{Kourkoulou:2017zaj} as the result of measuring all mirror-spins on a typical pure state.

This is not an exact statement, for two reasons. First,  in defining the mirror operators we had to introduce the cutoff $\omega_*$ in the frequencies. Second,  we would obtain a single thermal spin state if we projected for all $N/2$  spins,  however, this would likely go beyond the small algebra and the mirror operator construction becomes difficult to control before this point.

In \cite{Kourkoulou:2017zaj}, Kourkoulou and Maldacena considered state-dependent perturbations of the Hamiltonian of the form $\delta H = g \sum s_k S_k$ on the state $|B_{s}(\beta)\rangle$, where it was shown that these perturbations lead to an extension of the time evolution from the Rindler patch of AdS$_2$ to the Poincare patch, thereby gaining access to information from behind the horizon. This effect is present even when we sum over a limited number of spins $K < N/2$, as was shown in \cite{Brustein:2018fkr}. From the arguments above it follows that in such states it is equivalent to write the perturbation as
\be
\delta H = g\sum_k S_k \widetilde{S}_k ,
\ee
which highlights some similarity with the state-dependent perturbations of the form ${\cal O} \widetilde{\cal O}$ that we have been considering in this paper. The Kourkoulou-Maldacena perturbation by 
$\delta H= g\sum_k s_k S_k$ on the state  $|B_{s}(\beta)\rangle$ can be thought as the ``quantum-teleportation-version'' of the $
\delta H = g\sum_k S_k \widetilde{S}_k$ perturbation on a typical pure state $\rsz$: first we measure $\widetilde{S}_k$ which, as we argued above, transforms
the state into $|B_s(\beta)\rangle$. Then, on this state we apply a unitary which depends on the results of the measurement and corresponds to the perturbation $\delta H= g\sum_k s_k S_k$.

\subsection{Information Behind the Horizon in SYK}
\label{exp_syk}

In subsection \ref{subsec:exp}, we discussed two thought experiments to probe the geometry dual to a typical state in a holographic CFT, especially the geometry hidden behind the horizon.
We will now illustrate these experiments in the SYK model. Because the details of the two experiments are very similar, we will discuss only the first experiment \ref{experiment1}. 

We want to consider the analogue of the correlator \eqref{eq:exp1cor1} in the SYK model. Hence the correlator we will consider is
\begin{equation}
\label{eq:SYKexp}
 \mathcal{C}' \equiv   \bra{\Psi_0}  \, e^{-i \epsilon \widetilde{S}_1(-t_*)} \, e^{-i g S_2 \widetilde{S}_2}\, S_1(t)\, e^{i g S_2 \widetilde{S}_2} \,e^{i \epsilon \widetilde{S}_1(-t_*)}  \ket{\Psi_0} \, ,
\end{equation}
where $S_2\widetilde{S}_2$ acts $t=0$, but we suppress this time label for convenience and $t_*$ is scrambling time. We are interested in the term linear in $\epsilon$ which is
\begin{align}
\begin{split}
\mathcal{C}' &=   i \epsilon \bra{\Psi_0}   \big[ e^{-i g S_2 \widetilde{S}_2}\, S_1(t)\, e^{i g S_2 \widetilde{S}_2} , \widetilde{S}_1(-t_*) \big]  \ket{\Psi_0} \\
&= -2\epsilon \, \text{Im} \bigg[\bra{\Psi_0}  e^{-i g S_2 \widetilde{S}_2}\, S_1(t)\, e^{i g S_2 \widetilde{S}_2} \cdot \widetilde{S}_1(-t_*)  \ket{\Psi_0} \bigg] ,
\end{split}
\end{align}
where we have rewritten the commutator as the imaginary part of the correlator. The action of the rightmost mirror operator $\widetilde{S}_1(-t_*)$ on the state $\ket{\Psi_0}$ can be replaced by that of $S_1(t_* + i \beta/2)$. Further, the exponential operator can be shown to be
\begin{equation}
\label{eq:simpleexp}
e^{i g S_2 \widetilde{S}_2}  = \cos(g) I +i \sin(g) S_2 \widetilde{S}_2 .
\end{equation}
Using this\footnote{The operator $\widetilde{S}_2$ in equation \eqref{eq:simpleexp} commutes with $S_1(t_*+i \beta/2)$. The easiest way to see this is to go to frequency space and observe that the imaginary argument of $S_1$ just gives a multiplicative factor, a $c$-number. } and the defining equations of the mirror operators the correlator $\mathcal{C}'$ becomes
\begin{align}
\begin{split}
\label{eq:sykcorr}
\mathcal{C}' &= -2\epsilon \cos^2(g) \, \text{Im} \big[ \bra{\Psi_0} S_1(t) \, S_1(t_*+i \beta/2) \ket{\Psi_0} \big] \\
&-  2\epsilon \cos(g) \sin(g) \, \text{Im} \bigg[ i \bra{\Psi_0} \big[S_1(t), S_2(0) \big] \, S_1(t_*+i \beta/2) \, S_2(i \beta/2) \ket{\Psi_0}  \bigg] \\
&-  2\epsilon \sin^2(g) \, \text{Im} \bigg[  \bra{\Psi_0} S_2(0) S_1(t) S_2(0)  \, S_1(t_*+i \beta/2) \,  \ket{\Psi_0}  \bigg]  .
\end{split}
\end{align}
To calculate the first term in equation \eqref{eq:sykcorr} one starts by writing the spin operators in terms of the fundamental fermions $S_j = 2i \psi_{2j-1} \psi_{2j}$. The two-point function in equation \eqref{eq:sykcorr} then becomes a four-point function of four different fermions. In the large $N$ limit of interest, the two-point function is then approximately equal to product of two thermal two-point functions of the fermions. This function can be shown to be real and hence does not contribute to the correlator $\mathcal{C}'$. 

We now focus on calculating the second line in equation \eqref{eq:sykcorr}. We will first do the calculation in a thermal correlator\footnote{This calculation is similar to a traversable wormhole calculation in the thermofield double state.} and later discuss how this approximations the computation in a typical state. So we are interested in the real part of 
\begin{equation}
\label{eq:2ndterm1}
\mathcal{C}' \sim  \langle \big[S_1(t), S_2(0) \big] \, S_1(t_*+i \beta/2) \, S_2(i \beta/2) \rangle_\beta  \, .
\end{equation}
we will not keep track of overall $O(N^0)$ real multiplicative constants as we are interested in whether there is a signal or not. We will keep track of the phase for the correlators as we need the real part of this correlator. This correlator is a fermion eight-point function
\begin{equation}
\label{eq:t2prime}
\mathcal{C}' \sim \bigg\langle \big[ \psi_1(\tau_1) \psi_2(\tau_2), \psi_3(\tau_3) \psi_4(\tau_4) \big] \, \psi_1(\tau_5) \psi_2 (\tau_6) \, \psi_3(\tau_7) \psi_4(\tau_8) \bigg\rangle_\beta ,
\end{equation}
where the time arguments are determined by equation \ref{eq:2ndterm1}. 

The easiest way to calculate the leading term is to do this calculation in the vacuum state and then apply a conformal transformation to the thermal circle and analytically continue to Lorentzian time. The Feynman graphs factorize in several pieces at different orders of $N$ \cite{Gross:2017aos}.

\begin{table}[h!]
\begin{tabular}{lc}
Products of four two-point functions.                                    & $O(N^0)$    \\
Products of a connected four-point function and two two-point functions. & $O(N^{-1})$ \\
Products of two connected four-point function.                           & $O(N^{-2})$ \\
Products of a connected six-point function and a two-point function.     & $O(N^{-2})$ \\
Connected eight-point functions.                                         & $O(N^{-3})$
\end{tabular}
\end{table}

There are more terms in general, however, terms containing an odd number of fermions are suppressed in the SYK model. The commutator further reduces this list as in some cases the order does not matter, for example in the case of the two-point functions. We, therefore, focus on the instances where the ordering does matter, such as
\begin{equation}
F=\langle \psi_1(t_*+\Delta t) \psi_3 (0) \psi_1(t_*+i \beta/2)  \psi_3(i \beta/2)\rangle_{\beta,\text{connected}},
\end{equation}
where we have defined  $\Delta t = t-t_*$ for later convenience. Observe that this is an out-of-time-order correlator. A special property of out-of-time-order correlators is that they exhibit exponential growth, a hallmark of quantum chaos. These terms can, therefore, become order $O(N^0)$ with times of the order of the scrambling time. 

The six- and eight-point functions will also have exponential growth. We can group operators that are close together in time and next to each other, which results in correlators of the form$\langle {\cal O}(\text{late}) {\cal O}(\text{early}) {\cal O}(\text{late}) {\cal O}(\text{early})\rangle$. The time ordering of such a correlator has only one exponential growing factor, which is not enough to compensate the suppression in $N$ at the same time that the four point function becomes $O(1)$. The product of two connected four point functions,  on the other hand, has two exponential growth factors, enough to compensate the $1/N^2$ suppression of these terms. Connected SYK four point functions have been calculated in \cite{Maldacena:2016syk}. They take the form
\begin{equation}
\langle \psi_1(t_*+\Delta t) \psi_3 (0) \psi_1(t_*+i \beta/2)  \psi_3(i \beta/2)\rangle_\text{chaos} \sim \frac{i} {N} \frac{e^{\frac{2\pi} {\beta} t_*}} {1+e^{- \frac{2\pi} {\beta}\Delta t}} G(\Delta t - i \beta/2) ,
\end{equation}
Here $G(t)$ is the fermion two-point function. Therefore, only the square of this four point function contributes as we need the real part of equation \eqref{eq:2ndterm1}. Thus we obtain the following result for equation \eqref{eq:SYKexp}
\begin{equation}
\mathcal{C}' \sim  \frac{1} {N^2} e^{2 \frac{2\pi} {\beta} t_*} \left(\frac{e^{ \frac{2\pi} {\beta} \Delta t}} {1+e^{\frac{2\pi} {\beta} \Delta t}}\right)^2 G(\Delta t - i \beta/2)^2 \, .
\end{equation}
We can see the signal provided that it was sent at an early time of the order of the scrambling time. 

The same steps can be taken for the third line in equation \eqref{eq:sykcorr}, however, the result is small compared to this one.

Although this correlator was calculated in the thermal state, we expect that at large $N$ we will have  the same result for the typical state. We will discuss the general reasons for this in section \ref{sec:conjecture}. Numerical evidence for this statement
is discussed in subsection \ref{subsec:numsyk22}.  We can also argue that this is true as follows. We first write the typical state as a superposition of the spin states, as discussed in section \ref{spins_mirrors}
\begin{equation}
\ket{\Psi_0} = \sum_s c_s \ket{B_s(\beta)} \, .
\end{equation}
Using this expansion for the bra and ket in the correlators that we want to compute, we get diagonal and off-diagonal terms with respect to $s$. For the diagonal terms, we have correlators in the thermal spin states which are indeed close to thermal correlators \cite{Kourkoulou:2017zaj}. Notice that here we assume that the flip symmetry remains a good approximation even for time scales of order of scrambling time, and in particular the leading large $N$ result for flip invariant correlators is the same in the $|B_s(\beta)\rangle$ states as the thermal states even at scrambling time. 

The off-diagonal terms between different thermal spin states constitute the major difference between the typical state vs the thermal correlator. Such a cross term is of the form
\begin{equation}
\bra{B_{s'}(\beta/2)} A \ket{B_s(\beta/2)} ,
\end{equation}
where $A$ is a combination of fermions given by equation \eqref{eq:sykcorr}. It is important to note that each fermion appears an even number of times in this combination. We can relate one spin state to another by flipping a spin, for example $\ket{B_{s'}} \sim \psi_{2k-1}  \ket{B_s}$, this flips the $k$'th spin. We can use this to rewrite the cross term as
\begin{equation}
\bra{B_{s}(\beta/2)}e^{\beta H/2} \Gamma e^{-\beta H/2} A \ket{B_s(\beta/2)} ,
\end{equation}
where $\Gamma$ is the combination of fermions needed to flip the string of spins $s'$ into the string $s$. In $\Gamma$ we only use the fermions with an odd label, and each of the used fermion is only used once. Therefore, the combination $e^{\beta H/2} \Gamma e^{-\beta H/2} A$ contains some fermions which appear an odd number of times and correlators of this form are suppressed in the SYK model\footnote{The spin states do have non-zero correlators with an odd number of specific fermions, however, in this case the fermions must pair up for one of the spin operators, which is not the case for the cross terms.}. The number of cross terms cannot compensate for this suppression as they come with random phases. The sum over the cross terms is, therefore, suppressed compared to the diagonal terms, and we are justified in using the thermal correlators as an approximation for the typical state.

The results are consistent with the calculations done in \cite{Maldacena:2017axo}, including the effect that the correlators becomes non-zero directly after the perturbation (though only at $O(1/N)$). This is closely related to what would be ``stringy'' corrections in other theories.

\subsection{Numerical Comparison}
\label{subsec:numsyk22}

\begin{figure}[!htb]
\begin{center}
\includegraphics[width=.5\textwidth]{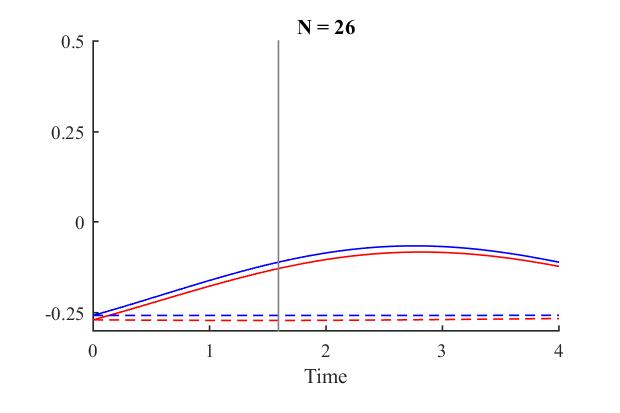}
\end{center}
\caption{Comparing the traversable wormhole correlator \eqref{eq:SYKexp} in the thermal state (blue) and typical state (red). The dotted line is the signal without the probe, which is the response one obtains from just the double-trace perturbation and should disappear in the large $N$ limit, in this case a sum over five pairs of operators was used in the perturbation to limit this effect. The vertical line denotes the scrambling time and is the time around which the probe is focused.}
\label{fig:teleSYK}
\end{figure}

The correlators \eqref{eq:SYKexp} relevant for our thought experiment, can also be calculated numerically, with the results shown in figure \ref{fig:teleSYK}. It is, unfortunately, not possible to directly compare to the analytic results, as for the values of $N$ that we can practically analyze numerically the behaviors usually associated with early time, exponential growth, and saturation overlap. It is, however, useful to check whether the the results for the typical state and the thermal state are close, and indeed they are. More about the numerics in the SYK model can be found in appendix \ref{App:SYK}.

\section{A Conjecture about Quantum Chaos in Pure and Thermal States}
\label{sec:conjecture}

In this paper we made a proposal for the interior geometry of a typical black hole microstate. In section \ref{sec:doubletrace} we related this proposed interior geometry, to  a {\it necessary} condition for CFT correlators:
\begin{quote}
 {\bf Conjecture:} 
 In a holographic large $N$ CFT, correlators\footnote{Here we are talking about Wightman correlators, in particular they need not be time-ordered.} in typical pure states are close to thermal correlators, even if the time separations $|t_i-t_j|$ are of the order of scrambling time $\beta \log S$.
\end{quote}
\begin{equation}
\label{conjectureA}
\ls {\cal O}_1(t_1,x_1)...{\cal O}_n (t_n, x_n) \rs_{\rm low-pass} = {1\over Z} {\rm Tr}[e^{-\beta H} {\cal O}_1(t_1,x_1)...{\cal O}_n (t_n, x_n)]_{\rm low-pass} + {\rm small \,\, error} ,
\end{equation}
where ``small error'' goes to zero as $N\rightarrow \infty$. Moreover, the correlators remain close to each other even after analytic continuation to imaginary time, within a strip of at most $t_E = \pm\beta/2$. The subscript ``low-pass'' means that
we only keep the low frequency components $|\omega|<\omega_*$, where $\omega_*$ is kept fixed as $N\rightarrow \infty$.

Strictly speaking we are interested in this conjecture for holographic CFTs with Einstein gravity duals, however the conjecture may apply to more general theories. For example, we have some partial evidence that it holds in the SYK model whose gravitational dual is not precisely geometric \cite{Heemskerk:2009pn,Fitzpatrick:2010zm,ElShowk:2011ag}.

In the rest of this section, we first make some general comments about the conjecture \eqref{conjectureA} and discuss why it is non-trivial. Then we explain that the conjecture can be simplified by replacing the typical pure state by the microcanonical density matrix. We also discuss some general aspects of comparing canonical and microcanonical ensembles in statistical mechanics. Finally we provide some evidence supporting \eqref{conjectureA}.
The conjecture \eqref{conjectureA} can be considered as a purely-CFT conjecture, which is independent of the discussion about the bulk interpretation and the interior of a typical black hole microstate.

\subsection{General Comments on the Conjecture}

The subscript ``low-pass'' in \eqref{conjectureA} means that we only keep the Fourier modes in the correlator with $|\omega|< \omega_*$, where $\omega_*$ is a large frequency that we keep fixed as $N\rightarrow \infty$. 
For example, if we have a function of a single variable, we define the low-pass filtered combination by
\be
\label{lowpass}
[f]_{\rm low-pass} (t) \equiv {1\over 2\pi}\int_{-\omega^*}^{\omega^*}e^{-i \omega t} d\omega \int_{-\infty}^{+\infty} dt' e^{i \omega t'} f(t') = {1\over \pi} \int_{-\infty}^{+\infty} dt' {\sin[\omega_* (t-t')] \over t-t'} f(t') .
\ee
We can similarly define the low-pass filtered correlators depending on more time-arguments.
The low-pass filtering is motivated by the fact that, as
discussed in sections \ref{sec:mainconjecture} and \ref{sec:doubletrace}, 
for the purpose of probing the black hole interior we do not need high frequencies. Moreover, the restriction of frequencies allows us to avoid certain technical problems discussed below.

First, in \eqref{conjectureA} we have not been  precise about the nature of convergence between the two correlation functions. For example, even if the correlators converge to each other point-wise in the coordinates, their derivatives
may not converge. For instance, suppose that the two correlators differ by a ``noise term'' with very small amplitude but very high frequency  of the form
\be
{1\over N} e^{-i N t} .
\label{noiseterm}
\ee
Here $t$ denotes schematically some combination of the times in \eqref{conjectureA}. While point-wise in $t$ this noise term goes to zero as $N\rightarrow \infty$, we notice that if we compute the time-derivatives of the correlators, then they will generally diverge as $N\rightarrow \infty$.

Second, a related aspect of this problem is that upon analytic continuation to imaginary time, small differences can get amplified. Consider for example the possible high-frequency noise term \eqref{noiseterm}. Typically we would 
like to compare correlators on a strip of Euclidean width of order $\beta$. We notice that upon
analytic continuation $t \rightarrow t + i \epsilon$, the term blows up exponentially for fixed $\epsilon>0$ and $N\rightarrow \infty$.

Both of these problems are avoided by considering the low-pass-filtered correlators, where only frequencies $|\omega|< \omega_*$ are kept.
For example, we avoid problematic terms such as the high frequency, small amplitude noise term \eqref{noiseterm}. For the same reason, upon analytic continuation a noise term can be amplified at most by a factor of $e^{\beta \omega_*}$. Hence
if the amplitude of the noise term in real time goes to zero as $N\rightarrow \infty$, then the same will be true in the complexified time-domain that we are interested in.

We now discuss another aspect of the conjecture \eqref{conjectureA}: one may think that the proximity of correlators between typical pure states and the thermal state is an obvious result in statistical mechanics. However condition \eqref{conjectureA}  is non-trivial for the following reasons:

\noindent (1) There is no general proof about the proximity of all expectation values in typical pure states and the thermal ensemble. The proximity can be established only if some additional assumptions are made about the nature of the observable that we 
are considering. The Eigenstate Thermalization Hypothesis (ETH) \cite{Srednicki1999approach} is an example of such an assumption. The observables in \eqref{conjectureA} involve products of operators at large time separations, so it is not automatically obvious that they obey these ETH-like assumptions.

\noindent (2) Relatedly, the general expectation in statistical mechanics is that correlators on typical pure states are close to thermal correlators up to $1/S$ corrections.  When considering out of time order correlators in large $N$ gauge theories, certain terms which begin at early times as being of order 
${1 \over N}$, grow as we increase the time separation and at scrambling time they become $O(N^0)$ and mix with leading terms. The conjecture \eqref{conjectureA} states that this effect is the same in the typical pure state and the canonical ensemble.

\noindent (3) Condition \eqref{conjectureA} requires that, as we take $N\rightarrow \infty$, the observable whose expectation value we are considering also changes, because the scrambling time explicitly depends on $N$.

\subsubsection{Replacing Typical Pure States by Microcanonical Mixed State}
\label{subsec:puremicro}

A second observation is that on very general grounds it can be shown that correlators on typical pure states are very close to those in the micrononical ensemble $\rho_m$, centered around the appropriate energy window \cite{lloyd}. In particular
\be
\label{micropure}
\ls A \rs = {\rm Tr}[\rho_m A] + O(e^{-S}) .
\ee
Contrary to the approximation between typical pure state and the thermal ensemble, the approximation between a typical pure state and the microcanonical ensemble is much more robust
and we do not expect it to break-down even at late times. The reason is that we can derive rigorous bounds on the variance of expectation values of observables among different pure states, and these bounds depend only on the norm of the observable. 
For example, it is easy to show \cite{lloyd} that for {\it any} Hermitian observable $A$ we have
\begin{align}
\begin{split}
\int [d\mu] (\langle \Psi|A|\Psi\rangle - \Tr[\rho_m A])^2 &= {1\over {\cal N}+1} \left(\Tr[\rho_m AP_m A] - \Tr[\rho_m A]^2\right) , \\
&\leq {1\over {\cal N}+1} \left(\Tr[\rho_m A^2] - \Tr[\rho_m A]^2\right) ,
\end{split}
\end{align}
where $[d\mu]$ is the usual Haar measure over pure states in the relevant energy window and ${\cal N} \sim e^{S}$ denotes the dimensionality of that Hilbert space. $P_m$ is the projector on that window and $\rho_m={P_m \over {\cal N}}$. The variance on the RHS is exponentially suppressed in $S$, times  a combination of the norms of the operators $A$ and $A^2$. For us $A = {\cal O}(t_1,x_1)...{\cal O}(t_n,x_n)$. If we work
with local operators which have been smeared in such a way that they are bounded, then separating the constituent operators ${\cal O}$ in time cannot increase the norm of the product $A$, even if the time separation is very large. This follows from two obvious observations. First since time 
evolution is unitary we have that for any bounded operator
\be
|| e^{i H t} {\cal O} e^{-i H t} || = ||{\cal O}|| .
\ee
Second, for any product of bounded operators we have
\be
|| {\cal O}(t_1,x_1)...{\cal O}(t_n,x_n)|| \leq ||{\cal O}(t_1,x_1)||...||{\cal O}(t_n, x_n)|| .
\ee
This means that if we work with bounded operators the deviation between typical pure state and microcanonical ensemble cannot grow as a function of time, and hence we can show that
\be
\label{conjectureB}
\ls {\cal O}_1(t_1,x_1)...{\cal O}_n (t_n, x_n) \rs_{\rm low-pass} = {\rm Tr}[\rho_m {\cal O}_1(t_1,x_1)...{\cal O}_n (t_n, x_n)]_{\rm low-pass} + {\rm small \,\, error}
\ee
for {\it all} time separations. Then, using the  property \eqref{conjectureB} we can simplify the original conjecture \eqref{conjectureA} to the following equivalent and simplified form
\vskip20pt

\begin{quote}
 {\bf Simplified form of the Conjecture:} 
 In a holographic large $N$ CFT correlators in the microcanonical mixed state of energy $E \approx -\partial_\beta \log Z$ are close to thermal correlators, even if the time separations $|t_i-t_j|$ are of the order of scrambling time $\beta \log S$.
\end{quote}
\begin{equation}
\label{microcan}
\Tr[ \rho_m {\cal O}_1(t_1,x_1)...{\cal O}_n (t_n, x_n)]_{\rm low-pass} = {1\over Z} {\rm Tr}[e^{-\beta H} {\cal O}_1(t_1,x_1)...{\cal O}_n (t_n, x_n)]_{\rm low-pass} + {\rm small \,\, error} ,
\end{equation}
where ``small error'' goes to zero as $N\rightarrow \infty$. Moreover the correlators remain close to each other after analytic continuation to imaginary time, in an appropriate domain, i.e. within a strip of at most $t_E = \pm\beta/2$. Again we emphasize that the comparison of the two ensembles is not trivial, for
the reasons 1-3 mentioned earlier.

\subsubsection*{Comments}

\noindent (a) First we make some clarifying remarks about the order of limits in this conjecture. The precise meaning of \eqref{microcan} is: we consider a holographic CFT with a central charge of order $N^2$. We consider a fixed number $n$ of smeared,
bounded operators approximately localized around space-time points $(t_i=a_i+b_i \log N, x_i)$, with $a_i,b_i,x_i$ fixed. Here we take all time arguments to be {\it real}. We consider a fixed inverse
temperature\footnote{The temperature must be chosen such that the system is in a black hole phase.} $\beta$ and the corresponding canonical ensemble $\rho_\beta={e^{-\beta H} \over Z}$. We consider the microcanonical ensemble defined by the energy window $(E_0\pm \Delta E)$ where
$E_0 = -\partial_\beta \log Z$ and $\Delta E$ fixed. Then we take the $N\rightarrow \infty$ limit without changing any of the ``fixed parameters''. Then the claim is that
\begin{equation}
 \lim_{N\rightarrow \infty} \left| \Tr[\rho_\beta {\cal O}_1(t_1,x_1)...{\cal O}_n (t_n, x_n)]_{\rm low-pass} - \Tr[\rho_m {\cal O}_1(t_1,x_1)...{\cal O}_n (t_n, x_n)]_{\rm low-pass}\right| =0 ,
\end{equation}
where the subscript {\it low pass} indicates that we remove frequencies $|\omega|> \omega_*$ as in \eqref{lowpass}, and $\omega_*$ is kept fixed as $N\rightarrow \infty$.

\noindent (b) For $b_i=0$ then this conjecture reduces to the standard approximation of correlators between canonical and microcanonical ensemble. The non-trivial aspect of the conjecture is when $b_i\neq 0$. In that case we want to make the statement that 
these ``chaos-enhanced $1/N$ corrections'' are the same in the pure and thermal states. Technically having $b_i\neq 0$ makes a difference since it means that as $N\rightarrow \infty$ we also change the observables that we are considering (by moving the operators in time).

\noindent (c) We assume that there the operators ${\cal O}_i$ have been smeared in an appropriate way so that they are bounded operators. In particular this will also regulate possible light-cone singularities.

\noindent (d) The conjecture formulated as \eqref{microcan} is perhaps more conservative than it could be. The correlators could be close to each other even for longer time scales. This, however, is not necessary for the purpose of the thought experiments discussed in this paper. We have also not been precise about the explicit bounds for the error terms. For the purposes of our paper it is sufficient that the error goes to zero as $N\rightarrow \infty$.

\noindent (e) Finally, even though the conjecture \eqref{microcan} is formulated for real time arguments,  the proximity of the low-pass-filtered correlators after analytic continuation to imaginary time within a strip of width $\beta$ is guaranteed by the fact that the two correlators do not contain frequencies higher than some fixed frequency $\omega_*$. In the context of black hole physics this has been discussed in earlier works, for example \cite{Balasubramanian:2007qv}, where it was suggested that perhaps the analytically continued correlators contain information about the details of the black hole microstate. However, for the purpose of probing the black hole interior using the thought experiments discussed in this paper, we seem to be sensitive to the analytic continuation of the low-pass filtered correlators and as we conjecture above, those do not vary significantly among different microstates. In special situations where one can use supersymmetry to obtain greater control, like LLM and LM geometries, the distinction between thermal and typical pure states was discussed in \cite{Balasubramanian:2005mg} and \cite{Alday:2006nd} respectively. See \cite{Balasubramanian:2008da} for a summary of these analyses and \cite{Raju:2018xue} for a discussion of
the distinguishability of typical pure states in connection to the fuzzball proposal.

\subsubsection{Comments on Comparing Canonical to Microcanonical Ensembles}

In general the proximity of the two ensembles is based on the following intuition. The two density matrices of the ensembles are
\be
\rho_\beta = \sum_i {e^{-\beta E_i} \over Z} |E_i\rangle \langle E_i| \qquad,\qquad \rho_m = \sum_{E_i \in (E_0 \pm \Delta E)} {1\over {\cal N}}|E_i\rangle \langle E_i| .
\ee
Consider an observable $A$, which in our case would be of the form $A = {\cal O}_1(t_1,x_1)...{\cal O}_n (t_n, x_n) $. The expectation value of $A$ in either ensemble receives contributions {\it only} from the diagonal matrix elements of $A$ in the energy eigenstates
\be
f(E_i)= \langle E_i| A |E_i\rangle .
\ee
First of all we will assume, in the spirit of the ETH, that $f(E_i)$ can be approximated by ``reasonably smooth`` function $f(E)$. We also assume that the discrete set of states can be described by a smooth density of 
states $\rho(E)$. Then we have
\begin{align}
{\rm Tr}[\rho_\beta A] &= \int_0^\infty dE {1\over Z} \rho(E) e^{-\beta E} f(E) , \label{canonical}\\
{\rm Tr}[\rho_m A] &= \int_{E_0-\Delta E}^{E_0 + \Delta E} dE {1\over {\cal N}} \rho(E)  f(E) . \label{microcanonical}
\end{align}
where ${\cal N}=e^S$. The usual argument in statistical mechanics is that for systems with many degrees of freedom $\rho(E)$ increases fast with energy, while $e^{-\beta E}$ decreases fast. Hence the product $\rho(E) e^{-\beta E}$ is sharply peaked at a given window of energies which depends on the temperature.
If we further assume that $f(E)$ is relatively smooth and slowly varying, then by selecting the window $(E_0\pm \Delta E)$ of the microcanonical to coincide with the window where  $\rho(E) e^{-\beta E}$ peaks, we can establish the approximation between canonical and microcanonical. 

This leads to a saddle point approximation, which relies on taking the thermodynamic limit. Usually in AdS/CFT this is achieved by taking $N\rightarrow \infty$. 
Moreover, for the saddle point method to work, it is important that $f(E)$ is a slowly varying function of $E$. For observables corresponding to small products of operators with time-separations of $O(1)$, we find that at large $N$ the function $f(E)$ actually depends on $E$ only via the temperature. This means
that for such observables ${d f(E) \over dE} \sim O(1/N^2)$ and the saddle point method is reliable, establishing the equivalence of \eqref{canonical} and \eqref{microcanonical}. The question is whether the same property of slow variation of $f(E)$ is true for observables like those in \eqref{microcan} where some of the time separations scale like $\log(N)$.

To emphasize that the slow variation of $f(E)$ is important for the equivalence of the ensembles, let us mention the following argument. When comparing two density matrices $\rho_1,\rho_2$ it is useful to consider the trace distance
\be
D(\rho_1,\rho_2) \equiv \frac{1}{2} \, {\rm Tr} \big( |\rho_1-\rho_2| \big) \, ,
\ee
where $|X|$ is defined as $|X| \equiv \sqrt{X^\dagger \, X}$. This characterizes how different the two quantum states are. More precisely for any bounded observable $A$ we have
\be
|{\rm Tr} \big( \rho_1 A \big) - {\rm Tr} \big( \rho_2 A \big) |\leq ||A|| \, D(\rho_1,\rho_2) .
\ee
The trace distance is positive and is bounded from above by $1$, and if it is close to $1$ then the two density matrices are in principle ''fully distinguishable``. When comparing the canonical and microcanonical ensembles the trace distance becomes
\be
D(\rho_\beta,\rho_m) = \int_0^{E_0-\delta E} dE {\rho(E) e^{-\beta E}\over Z} + \int_{E_0+\delta E}^\infty dE {\rho(E) e^{-\beta E}\over Z} + \int_{E_0-\delta E}^{E_0+\delta E} dE \rho(E) \left| {e^{-\beta E}\over Z} -{1 \over {\cal N}} \right| .
\ee
It is easy to check that when comparing the canonical ensemble to the microcanonical ensemble for relevant systems\footnote{For example this can be checked for 2d CFTs where $\rho(E)$ is given by the Cardy formula.} then no matter
how the window $(E_0\pm \Delta E)$ is selected, the trace distance between the two ensembles is extremely close to 1. This means that there will always exist some bounded observable $A$ with the property that its expectation value is maximally different in the two ensembles. Such an observable would have the property that $f(E)$ would have to be very rapidly varying with $E$.

All this shows that the conjectures \eqref{conjectureA} and \eqref{microcan} are essentially conjectures about the slow variation of diagonal matrix elements $f(E)$ of $A$, where $A= {\cal O}_1(t_1,x_1)...{\cal O}_n (t_n, x_n) $, when the time differences $|t_i-t_j|$ are of the order of scrambling time.

\subsection{Evidence for the Conjecture}

We showed that the original conjecture \eqref{conjectureA} can be simplified to the form \eqref{microcan}. In this subsection we provide some evidence for it. The evidence we provide does not constitute a proof. However, the statements are should be considered heuristic support for the conjecture.

A class of systems where this conjecture could be investigated more precisely might be large $c$ 2d CFTs with a sparse spectrum. In those theories the relevant correlators can be estimated by a  conformal block decomposition \cite{Fitzpatrick:2015zha}, where it is expected that the effects at scrambling time are related to the Virasoro identity block. It is, however, also important to check that other blocks do not interfere with those effects, see for example \cite{Chang:2018nzm}. This is an approach which may be worth investigating in future work.

\subsubsection{Slow Change with respect to Energy}
\label{sssec:slow_change}

We now discuss an intuitive argument suggesting the slow variation with energy of the diagonal matrix elements of the relevant observables supporting the conjecture formulated in this section. We generally expect that in AdS/CFT thermal correlators of local operators
separated by short time scales depend on the energy only via the temperature, hence the diagonal matrix elements obey\footnote{Here we consider ---for example--- the ${\cal N}=4$ SYM where in the high temperature phase $E \propto N^2 R^3 T^4$, where $R$ is the size of the sphere on which the CFT lives.}, $\frac{df(E)} {dE} \sim O(1/N^2)$, where we used the notation $f(E_i)= \langle E_i| A |E_i\rangle$. This justifies the use of the saddle point methods in comparing canonical to microcanonical. When we want to separate the operators by scrambling time, then as we take the large $N$ limit we need to {\it tune} the operator $A$ in an $N$-dependent fashion, since scrambling time for a fixed
temperature depends on $N$ via $\log N$. Hence when we apply the saddle point method at large $N$ we need to take into account that the operator $A$ itself will depend on $N$, we denote this as $A_{t_S}$.

\begin{figure}[ht]
\begin{center}
\includegraphics[width=.5\textwidth]{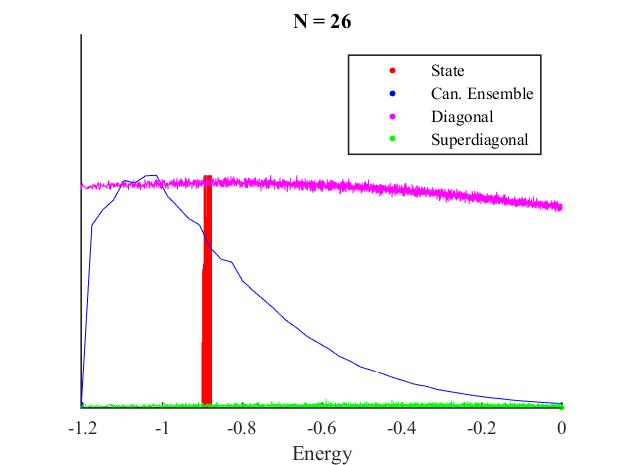}
\caption{The diagonal (magenta) matrix elements of $\{\psi_1(0),\psi(t_*)\}^2$, where $t_*$ is the scrambling time, vary slowly in the SYK model. They, moreover, dominate the elements just above the diagonal (green). The regions important for the typical state (red) and the canonical ensemble (blue, $\rho(E)e^{-\beta E}$) are shown as well. The operators were not passed through a low pass filter.}
\label{SYKenergies}
\end{center}
\end{figure}

The bulk computation in the eternal black hole suggests that the thermal expectation value of $A_{t_S}$
depends on the mass of the black hole only via the temperature, in the sense that
\be
I \equiv {d\over d\beta} \Tr \big(  \rho_\beta A_{t_S} \big)= O(1),
\ee
where we need to differentiate wrt $\beta$ both the thermal density matrix and the observable which depends on $t_S$. We can write these two contributions as
\be
\label{despest}
I = \Tr \bigg( {d \rho_\beta \over d\beta} A_{t_S} \bigg) + \Tr \bigg(\rho_\beta {d A_{t_S} \over dt} \bigg) {dt_S \over d\beta}.
\ee
This equation is rather schematic as the observable $A$ can have many time arguments of the order of the scrambling time. We expect that the correlator $\Tr[\rho_\beta {d A_{t_S} \over dt}]$ is at most
$O(N^0)$, as the growth of correlators is bounded by the Lyaponov exponent \cite{Maldacena:2015waa}. On the other hand we have ${dt_S \over d\beta} = O(\log N)$, hence the second term in \eqref{despest} grows at most like $\log N$. Since the sum of the two terms in \eqref{despest} is $O(N^0)$ we conclude that the first term
\be
\label{despest2}
\Tr \bigg( {d \rho_\beta \over d\beta} A_{t_S} \bigg) = O(\log N).
\ee
From this we can estimate the energy dependence of the diagonal matrix elements of the observable $A_{t_S}$, leading to
\begin{equation}
{d  f(E)\over dE}\sim O\left(\frac{\log N} {N^2}\right).
\end{equation}
We, therefore, see an enhancement of the error term of the saddle point method, but the error term is still suppressed in $N$. This justifies the use of the saddle point method and the approximation of the canonical and microcanonical ensemble for observables at scrambling time. This slow variation with respect to temperature can also be seen numerically in figure \ref{SYKenergies}.

\subsubsection{Connection to ETH}

The Eigenstate Thermalization Hypotheses (ETH) \cite{Srednicki1999approach} proposed that the matrix elements of an operator $V_{ij} = \bra{E_i} V \ket{E_j}$ have the form
\begin{equation}
V_{ij} = f_V(E) \delta_{ij} + e^{-S(E)/2}g_V(E,\omega)R^V_{ij} ,
\end{equation}
where $E_i = E - \omega/2, \; E_j = E +\omega/2$. The functions $f_V,g_V$ are assumed to be slowly varying and $R^V_{ij}$ are some almost random phases. We will discuss to what extent the product of operators obeying the ETH does also obey the ETH. We are interested in timescales of order of scrambling time  $t_S={\beta \over 2\pi}\log S$. For concreteness we consider, for example, the correlator $\langle [W(t),V(0)]^2\rangle$. The matrix elements for $W(t)$ can be written in a similar fashion
\begin{equation}
W(t)_{ij} = f_W(E) \delta_{ij} + e^{-i\omega t}e^{-S(E)/2}g_W(E,\omega)R^W_{ij} .
\end{equation}
To simplify the discussion we can assume that the operators have vanishing one-point functions, i.e. $f_V=f_W=0$ (or more generally we would have to consider connected correlators). We define
\be
C \equiv W(t) V(0) ,
\ee
which will help us obtain the out-of-time order correlator mentioned above. We have
\begin{align}
\label{ceth}
C=\sum_k \bigg[ &e^{-i(E_k-E_i)t} \, e^{-\{ S((E_k+E_i)/2)+S((E_k+E_j)/2) \}/2}\\ 
&\times g_W\left(\frac{E_i+E_k} {2},E_k-E_i\right)g_V\left(\frac{E_j+E_k} {2},E_j-E_k\right) R^W_{ik}R^V_{kj} \bigg]. \nonumber
\end{align}
We also need the other ordering, which we will call $C'$. It has a similar expression with some of the indices interchanged. To obtain an expression for the out-of-time ordered correlator we need terms of the form $C^2, C'^2, CC', \text{and} \, C'C$. The phases are almost random but they are assumed to have some correlation \cite{Marolf:2013dba}
\begin{equation}
R^V_{ij} R^W	_{kl} = h^{VW}(E_i, E_l) \delta_{il} \delta_{kj} + {\rm erratic} ,
\label{coreth}
\end{equation}
where $h^{VW}(E_i, E_l)$ is some smooth function. For short time scales this ansatz leads to the conclusion that the combination $[W(t),V(0)]^2$ obeys the ETH, if $W$ and $V$ do.

We will now comment on the opposite regime. For very large time scales, the time-dependent phases in $C$ and $C'$ fluctuate rapidly and 
the correlations \eqref{coreth} are washed out when $h^{VW}$ is inserted in \eqref{ceth}. In the terms $C^2$ and $C'^2$, the matrix product implies that we need to fix two indices (say $k, \ell$). As a result we are left with a sum over only one index. This sum is of order $O(e^S)$ but it cannot compensate the $O(e^{-2S})$ suppression. Thus we can neglect these terms. In the cross terms $CC'$ and $C'C$, the phases can be summed over in a coherent way and they then compensate the $O(e^{-2S})$ suppression. We will thus only look at these. If we redefine $E_k=E_i+\omega$ and $E_j=E_i+\omega_1+\omega_2$, and turn the sums into integrals, we get
\begin{equation}
CC' = \int d\omega_1 d\omega_2 e^{\beta \omega_1/2+\beta \omega_2/2}|g_W(E_i+\omega_1/2,\omega_1)|^2|g_V(E_i+\omega_1+\omega_2/2,\omega_2)|^2 ,
\end{equation}
where we used that the entropy $S$ varies slowly and that moderate values of $\omega$ dominate. ETH assumes that $g_V$ and $g_W$ are slowly varying functions in the first argument. We can, therefore, use a Taylor expansion to obtain
\begin{align}
CC'+C'C &= 2\langle V^2 \rangle \langle W^2 \rangle + \alpha_V \partial_\beta  \langle W^2 \rangle + \alpha_W \partial_\beta \langle V^2 \rangle , \\
\alpha_V &= 2 \int d\omega e^{\beta \omega /2} \partial_E |g_V(E,\omega)|^2 .
\end{align}
Generally speaking one expects that $\alpha_V \ll \langle V^2 \rangle$. This expectation can be motivated by analyzing the two-point function, which has the form 
\begin{equation}
\bra{E}V(t)V(0)\ket{E} = \int d\omega \,  e^{\beta\omega/2-i\omega t} \left(|g_V(E,\omega)|^2+\frac{\omega} {2} \partial_E |g_V(E,\omega)|^2\right) \, ,
\end{equation}
upon using the ETH ansatz. As discussed in the previous section \ref{sssec:slow_change}, we expect the derivative of the first term to be very small, thus making $\alpha_V$ small. There is an independent argument to justify that $\alpha_V \ll \langle V^2 \rangle$. Notice that it is the connected diagrams which contribute to $\alpha_V$. These are suppressed in the large $N$ limit as opposed to the disconnected ones that contribute to $\langle V^2 \rangle$, leading us to the conclusion $\alpha_V \ll \langle V^2 \rangle$. This means that for very long time scales the factorized result dominates, exactly what one expects in the region of saturation after chaotic growth.

The ETH, therefore, works with both short times and very long times. It is natural to expect that ETH will also work for intermediate times, but it seems that we need to make some additional assumptions about the matrix elements of the observables in order to prove this. We postpone this for future work.

\subsubsection{Time-Order vs Out-of-Time-Order Correlators} 

We provide another approach for readers who accept that time-ordered correlators continue to factorize at time scales of the order of the scrambling time. We will argue that this implies the validity of our conjecture also for out-of-time-order correlators.

The exponential growth of parts of chaotic correlators is seen in correlators that are out-of-time-order, for example
\begin{equation}
\label{eq:tdcomm}
\langle\mathcal{O}_1(t)\mathcal{O}_2(0)\mathcal{O}_1(t)\mathcal{O}_2(0)\rangle .
\end{equation}
Factorization breaks down in such a case and some $1/N$ corrections are enhanched to $O(1)$ at time scales of the scrambling time. This is specific to out-of-time-order correlators.

The time-order and out-of-time-order correlators stated above are related by the commutator: $[\mathcal{O}_1(t),\mathcal{O}_2(0)]$. These commutators have been studied  in the case of 2d CFTs \cite{Turiaci:2016cqc}, with the assumption that we can promote coordinate transformations $\xi(x)$ to Goldstone fields $\hat{\xi}(x)$ and that we write light operators as functions of these fields. In that case, it was argued that the commutator is dominated by a single term around scrambling time
\begin{equation}
[\mathcal{O}_1(t_1),\mathcal{O}_2(t_2)] \sim \tfrac{1} {c} e^{\lambda (t_1-t_2)}\partial_{t_1}\mathcal{O}_1(t_1) \partial_{t_2} \mathcal{O}_2(t_2) .
\end{equation}
where $c$ is the central charge, and $\lambda$ is the Lyapunov exponent giving the rate of exponential growth of out-of-time-order correlators. The commutator can be used to transform the out-of-time-order correlator to a time-order correlator, and shows how the exponential growth remains present after we convert the out-of-time-order correlators to time-order correlators. 

Nevertheless, the out-of-time-order correlator is the same in the microcanonical and the canonical ensemble if it is dominated by finitely many time-ordered terms that one obtains after using commutators such as those in equation \eqref{eq:tdcomm}. In passing, we note that these commutators are not known or easily constructed in general.

\subsubsection{SYK Numerics}

\begin{figure}[!h]
\centering
\begin{subfigure}[l]{.45\textwidth}
\includegraphics[width=\textwidth]{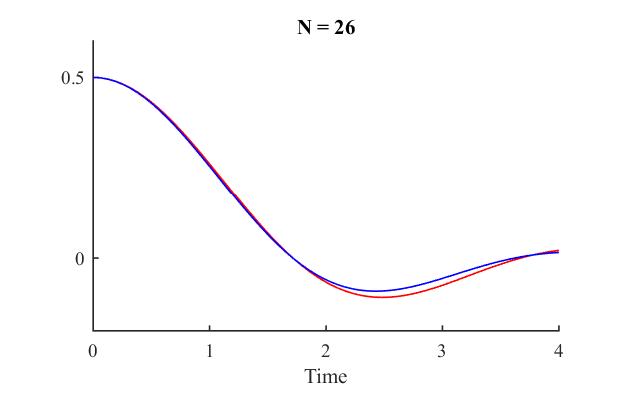}
\end{subfigure}
\begin{subfigure}[r]{.45\textwidth}
\includegraphics[width=\textwidth]{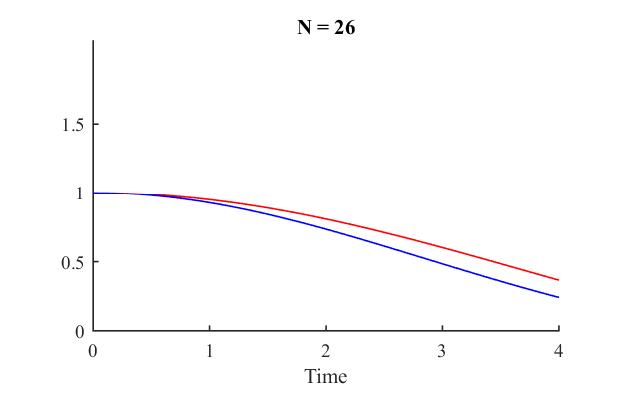}
\end{subfigure}
\caption{\label{fig:2pf}We compare the expectation value of $\langle \psi_1(t) \psi_1(0)\rangle$ (left) and $\langle S_1(t) S_1(0)\rangle$ (right) in the thermal state (blue) and pure state (red). We expect the two curves to differ by corrections suppressed by $1/N$. The observed deviation is consistent with the value of $N$ that we used.}
\end{figure}

\begin{figure}[!h]
\centering
\begin{subfigure}[l]{.45\textwidth}
\includegraphics[width=\textwidth]{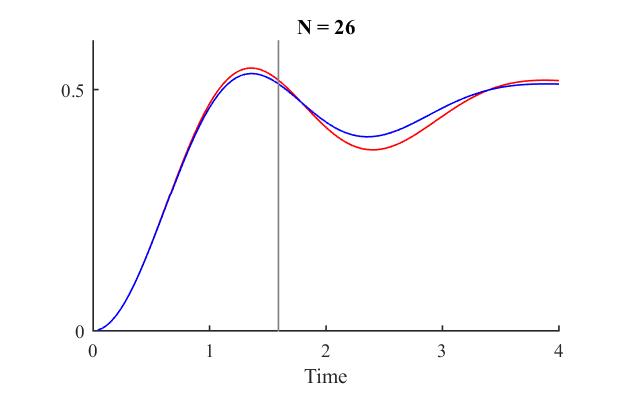}
\end{subfigure}
\begin{subfigure}[r]{.45\textwidth}
\includegraphics[width=\textwidth]{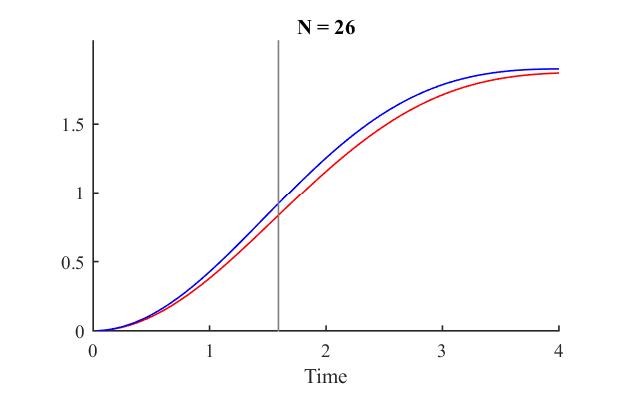}
\end{subfigure}
\caption{\label{fig:OTOC}We compare the expectation value of $\langle \{\psi_1(t),\psi_2(0)\}^2\rangle$ (left) and $-\langle[S_1(t),S_2(0)]^2\rangle$ (right) in the thermal state (blue) and pure state (red). The scrambling time is designated by the vertical line.}
\end{figure}

We can numerically check the conjecture and the various statements of this section in the SYK model. For example, in figure \ref{fig:2pf} we show the agreement between various simple correlators in the canonical ensemble and a typical pure state.
The conjecture becomes non-trivial for out-of-time-order correlators with times of the order of the scrambling times. We, therefore, look at the OTOC in figure \ref{fig:OTOC}. We can see that the pure and thermal expectation value are very close to each other up to the scrambling time, and close to each other after scrambling time. 

\begin{figure}[!h]
\centering
\begin{subfigure}[l]{.45\textwidth}
\includegraphics[width=\textwidth]{nfig_2_N26_2.jpg}
\end{subfigure}
\begin{subfigure}[r]{.45\textwidth}
\includegraphics[width=\textwidth]{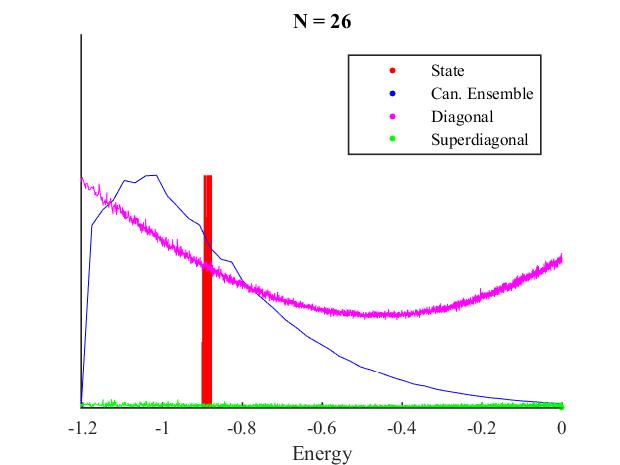}
\end{subfigure}
\caption{\label{fig:energies}We compare the diagonal (magenta) matrix elements of $\{\psi_1(t_*),\psi_2(0)\}^2$ (left) and $-[S_1(t_*),S_2(0)]^2$ (right), where $t_*$ is the scrambling time. They, moreover, dominate the elements just above the diagonal (green). The regions important for the typical state (red) and the canonical ensemble (blue, $\rho(E)e^{-\beta E}$) are shown as well. The operators were not passed through a low pass filter.}
\end{figure}

We can also check the arguments about the slowly varying energy dependence of the operators at scrambling time, see figure \ref{fig:energies}. This corresponds to the moment of the grey line in figure \ref{fig:OTOC}. 

These low $N$ calculations seem to support the conjecture. However, the $N$ used in these calculations is too low to clearly see some desired features of the correlators and it would be interesting to numerically study larger values of $N$. In appendix \ref{App:SYK} we give more details about numerics in the SYK model.


\section{Remarks on the Mirror Operators and the Hayden-Preskill Protocol}
\label{hp_protocol}

In this section we discuss how the mirror operators can be used to extract information from a black hole and highlight similarities with the Hayden-Preskill protocol. We start with a review of the original Hayden-Preskill argument and then we discuss the implementation of the protocol suggested by
\cite{Maldacena:2017axo}. After that we explain the connection with the mirror operators and the state-dependent perturbations that we have been discussing in this paper.

\subsection{The Hayden-Preskill Protocol} 

\begin{figure}[!htb]
\begin{center}
\includegraphics[scale=0.4]{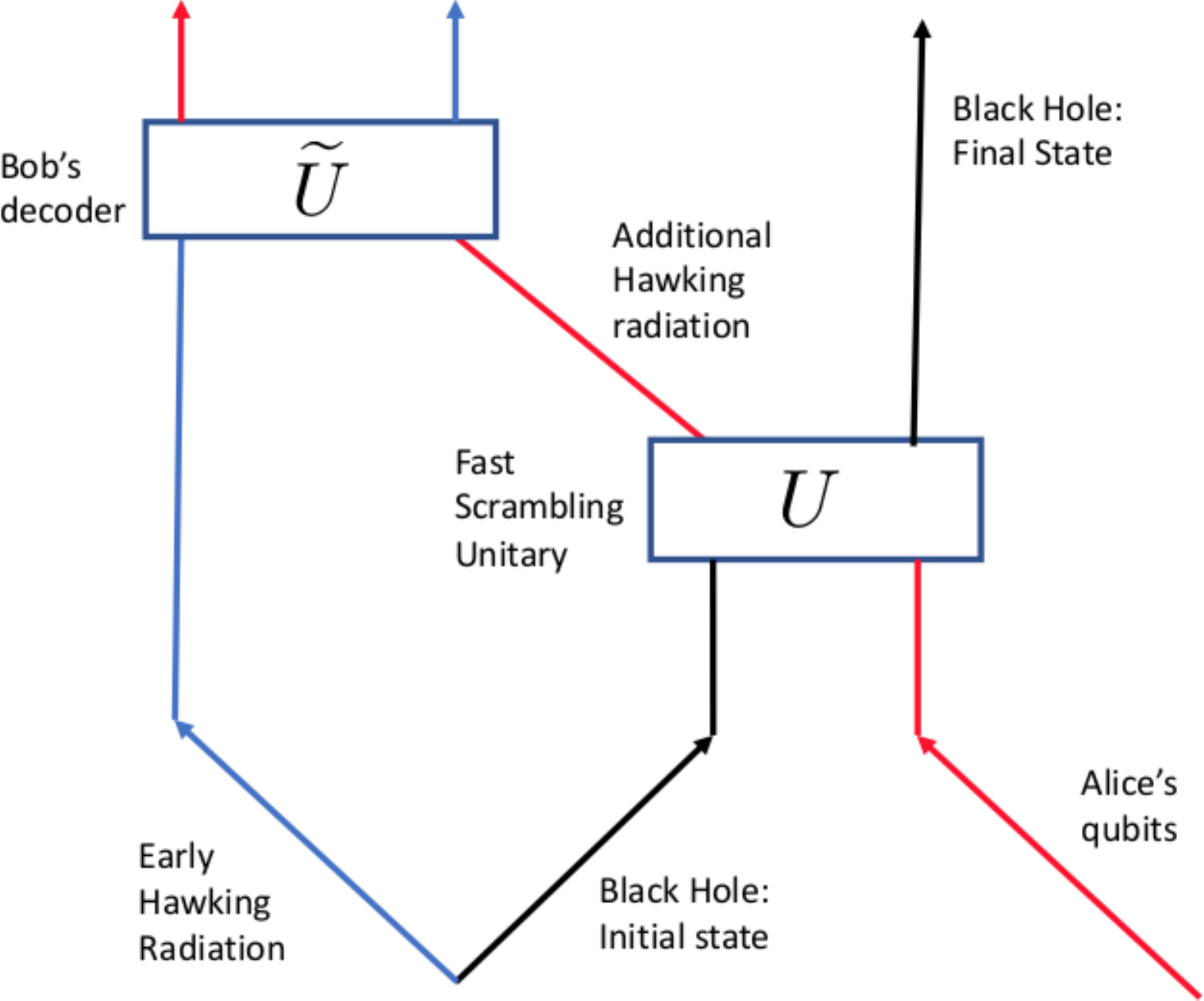}
\caption{\small Diagrammatic representation of the Hayden-Preskill experiment.}
\label{hpdiag}
\end{center}
\end{figure}

In \cite{Hayden:2007cs}, Hayden and Preskill proposed a way to extract information from old black holes i.e. those that have evaporated away at least half of their entropy. Their thought experiment is represented diagrammatically in figure \ref{hpdiag}.
Alice sends a message into a black hole after the half-point of evaporation, also called the Page time. This message is drawn as a red ray in figure \ref{hpdiag}. Bob, who wants to spy on Alice's message, has been collecting the Hawking radiation since the moment the black hole started emitting it. He has infinite resources and can study correlations between Hawking quanta exactly. He also knows the exact initial microstate of the black hole and the exact dynamics of quantum gravity governing black hole evaporation, denoted $U$ in the figure. Then the question \cite{Hayden:2007cs} asked is - 
 how many Hawking particles does Bob need to collect additionally, in order to reconstruct the message that Alice threw into the black hole? 

In figure \ref{hpdiag} time flows upwards. Bob can build a decoder using the early Hawking radiation and part of the late Hawking radiation, represented by the operation $\widetilde{U}$. We assume that the internal dynamics of the black hole $U$ is a random and rapidly-mixing unitary. Then, using the knowledge of the microstate of the black hole, \cite{Hayden:2007cs} argued that after the half-way point, Bob's decoder will need to collect only an order $\mathcal{O}(1)$ (compared to entropy) number of late Hawking quanta to be able to reconstruct Alice's message. While their argument shows that there exists in principle a quantum operation that can reconstruct the message, Hayden and Preskill did not provide a constructive algorithm to realize the decoding operation.

In \cite{Maldacena:2017axo}, it was pointed out that this quantum operation can be realized as follows:
 we imagine that Bob collects the early radiation emitted by the black hole  and he collapses it to form a second black hole. This results in two black holes which are approximately in a maximally entangled state. By applying a complicated unitary on the second black hole, Bob can bring the state of the two black holes in approximately a thermofield-like state. Then the corresponding geometry is that of the two-sided black hole with a non-traversable wormhole. We think of the original black hole as corresponding to the right side while the new black hole to the left. Alice's message can be thought of as throwing particle on the right side of the eternal black hole created by acting on the TFD state with the unitary $e^{i \phi_R(t_0)}$. The step in the Hayden-Preskill protocol of collecting a few more Hawking particles can be thought of as corresponding to the Gao-Jafferis-Wall double trace perturbation, which modifies the state by $e^{ig {\cal O}_L {\cal O}_R}$. Then Alice's message emerges on the left side in geometric form,  where it can be detected by Bob using an operator $\phi_L$. This provides us with a specific
 implementation of the Hayden-Preskill decoding.

\subsection{Information Recovery Using the Mirror Operators}

The question that was addressed in the Hayden-Preskill protocol was how to extract information which fell into the black hole,  if we only have access to the Hawking radiation. A somewhat different question is how to extract information which fell into the black hole,  if we also have access to the microscopic Hilbert space of the black hole. Of course in that case it is obvious from unitarity that information can in principle be recovered at any time, even before the Page time. 

We will present one particular protocol, using the mirror operators, which can be used to recover information from a black hole. If this protocol is applied to a large black hole in AdS, or a black hole in  flat-space  before Page time, it allows information extraction provided that one has access to the Hilbert space of the black hole. In AdS this means access to the full Hilbert space of the CFT. On the other hand if the protocol is applied to a black hole in flat space and after Page time then it becomes the analogue of the Hayden-Preskill protocol as formulated in \cite{Maldacena:2017axo} and in particular it allows information recovery purely from the Hawking radiation.

We start with a black hole in AdS dual to a microstate $\rsz$. At some time
$t \approx - t_S$  (here $t_S$ is scrambling time), we throw a qubit into the black hole. This qubit is created in the bulk by acting with the CFT operator $U_\epsilon = e^{i \epsilon \phi(t_0)}$ (appropriately smeared). We wait until the particle has been absorbed, and then we ask what is the CFT operator we need  to measure in order to extract the quantum information of the qubit. 

Of course in principle the boundary observer, who has access to the microstate of the CFT, can extract the information at any moment in time after the qubit has been injected\footnote{For example, if the boundary observer applies a time-reversal operator to the state and then evolves forward in time, then the particle will simply pop out of the black hole. Or, relatedly, we can extract the information at a later time by measuring the precursor of the operator which created the particle. We would like to thank J. Maldacena for discussions on this.}. Here we present one particular way of extracting the information, which will allow us ---in a different limit --- to make contact with the Hayden-Preskill protocol.

According to the previous discussions, one natural way to extract the information is the following: after sending in the qubit at $-t_S$, we perturb the CFT Hamiltonian at $t=0$ by an interaction of the form $U_g=e^{ig {\cal O} \tO}$, then two negative shockwaves will be produced, as discussed in section \ref{sec:doubletrace}. The infalling particle
collides with one of the shockwaves and undergoes a time-advance pushing it into the ``space of the mirrors''. It can then be measured by the mirror operator $\widetilde{\phi}(t)$ after scrambling time. All in all, the result of this measurement is captured by the correlator
\be
\label{hpcor}
\mathcal{C}'_2 \equiv \lsz \,U_\epsilon^\dagger \,U_g^\dagger \,\tphi(t)\, U_g \,U_\epsilon \,\rsz   ,
\ee
which according to the discussion of section \ref{sec:doubletrace} will allow us to measure the qubit at time $t\approx t_S$.

What we have thus is a protocol which  allows us to extract the quantum information of a particle which has crossed the horizon, in a time scale of the order of scrambling time. The protocol can be applied to black holes in flat space even before page time, but it requires  access  to the Hilbert space of the black hole. The reason is that before the Page time the black hole represents the largest part of the Hilbert space, and the mirror operators are supported on it.

Now consider a black hole in flat space, which is after its Page time. The region near the horizon is still approximately thermal, as the time-scale for evaporation is much longer than the timescales (of order scrambling time) relevant for our problem. We can still define the mirror operators for the exterior modes near the horizon. In this limit we have to work with the modular Hamiltonian of the entire system, defined by the Tomita-Takesaki construction, which will not be as simple as the CFT Hamiltonian in the case of a large AdS black hole.  After Page time the early radiation represents the largest part of the Hilbert space,  hence the mirror operators are mostly supported on the early radiation\footnote{While the mirror operators are supported on the early radiation, they can simultaneously play the role of the interior modes behind the horizon, .}. Hence, after Page time, implementing the analogue of the  state-dependent ``double trace'' protocol using the mirror operators allows us to extract the information from the cloud of Hawking radiation. In particular it is a realization of the Hayden-Preskill protocol in the form discussed in \cite{Maldacena:2017axo}.

\subsubsection*{State-dependence and the Hayden-Preskill protocol } 

The original Hayden-Preskill protocol to extract quantum information from a black hole, figure \ref{hpdiag}, can only be applied after the Page time. This is because the decoder needs to have access to the larger part of the full Hilbert space, as well as the knowledge of the microstate of the black hole. In particular, in Hayden-Preskill the decoder is state-dependent.  When we consider the realization of the protocol in two-sided black hole, as described in \cite{Maldacena:2017axo}, the decoder includes the double-trace unitary $e^{ig {\cal O}_L {\cal O}_R}$, applied {\it after}  we have brought the original state of the system in the the thermofield state, which is done by a state-dependent unitary rotation. Thus this realization of the Hayden-Preskill protocol in \cite{Maldacena:2017axo} is state-dependent as well. In the one-sided realization of the protocol that we proposed, the decoder is a function of the mirror operator $\tO$, which is also state-dependent. This is consistent with the general expectation that the decoder needs to be state-dependent. 

\subsubsection*{Comments on entanglement}

The Hayden-Preskill protocol relies on the fact that an old black hole is maximally entangled with the early radiation. This corresponds to an amount of entanglement of $O(S)$ between two tensor factors. In our case we can think of the code-subspace as made out of two tensor factors, corresponding to the algebras ${\cal A}$ and ${\cal A'}$
\begin{equation}
\label{eq:tensor_decomp}
\mathcal{H}_{\rsz} = {\cal H}_{\cal A} \otimes {\cal H}_{\cal A'}  .
\end{equation}
The state of the black hole $\rsz$ can be written as an entangled state, similar to the TFD state, in the factors ${\cal H}_{\cal A}, {\cal H}_{\cal A'}$. 
Our version of the protocol involves information transfer from the tensor factor of ${\cal A}$ to ${\cal A'}$. A natural question is what is the amount of entanglement between these two algebras. This questions depends on how exactly the algebras ${\cal A}$ are defined and how exactly we introduce a cutoff in the number of operators that can be multiplied together in ${\cal A}$. We may be able to extend the size of ${\cal A}$ to the point where the entanglement
entropy between ${\cal A}$ and ${\cal A}'$ is of order $O(S)$. The algebra ${\cal A}$ can be extended as long as the state $\rsz$ remains a separating vector i.e. it cannot be annihilated by the algebra ${\cal A}$. For example in the SYK model we noticed that we can include in ${\cal A}$ the fundamental fermions, which would result in an entanglement entropy of $O(S)$. 

Irrespective of the fact that we could in principle enlarge ${\cal A}$ to attain $O(S)$ entropy, we notice that the decoding protocol may be performed even while keeping the size of these algebras to be $O(1)$. The reason that this is not inconsistent with the $O(S)$ entanglement needed for the Hayden-Preskill protocol is that the code subspace is a very special subspace of the system where the interesting dynamics takes place. In particular it is a state-dependent subspace. For example even in the case of the eternal black hole, while the microscopic entanglement between the two CFTs is $O(S)$, the Gao-Jafferis-Wall protocol can be described within the code subspace corresponding to effective field theory excitations in the bulk which has smaller amount of entanglement.

Finally, in the extrapolation of our experiment to old black holes in flat space, the statement that the mirror operators are supported on the early radiation relies on the fact that the early radiation is maximally entangled with the remaining black hole, with an entanglement of order $O(S)$.

\subsection*{Quantum Teleportation}

The Hayden-Preskill experiment, and its two-sided analogues that we have discussed above, are related to quantum teleportation \cite{Maldacena:2017axo}. In the standard teleportation, given two maximally entanglement systems, one wishes to ``teleport" a message from one system to another. To do this, one first measures the message (some qubit) and one of the systems such that it is projected into a Bell state. Classically communicating the result of the measurement from system one to two is then used to decode the message. Note that the message need not be physically transported.

The double-trace protocol does not have this two-step measurement process. However, as discussed in \cite{Maldacena:2017axo}, the double trace protocol can be related to quantum teleportation. Notice that the the double-trace perturbation can be written as
\begin{equation}
e^{i \, g{\cal O} \, \tO  } ={1\over 2\pi g} \iint d\lambda \, d\widetilde{\lambda} \,  \,  e^{i g \lambda \, {\cal O}} \, \,   e^{i g \widetilde{\lambda}  \, \tO }\, e^{-i g \lambda \widetilde{\lambda}}  ,
\end{equation}
where we used that the operators ${\cal O}$ and $\tO$ commute. Using this identity and equation \eqref{eq:tensor_decomp}, we can calculate the reduced density matrix that corresponds to the mirror operators $\tO$ after the double-trace perturbation. This is easy to do by integrating over the $\lambda$ variables. One can then show that the result is equal to a reduced density matrix one gets after measuring ${\cal O}$ first and then acting with a unitary that depends on the outcome of the measurement on the space of $ \tO$'s. This latter operation is like the standard quantum teleportation. Thus the double-trace perturbation in our one-sided setup can be interpreted as a quantum teleportation. This allows us to extract the information $\phi(-t_*)$ by measuring the operator $\widetilde{\phi}(t_*)$.

\begin{figure}[!htb]
\begin{center}
\includegraphics[scale=0.3]{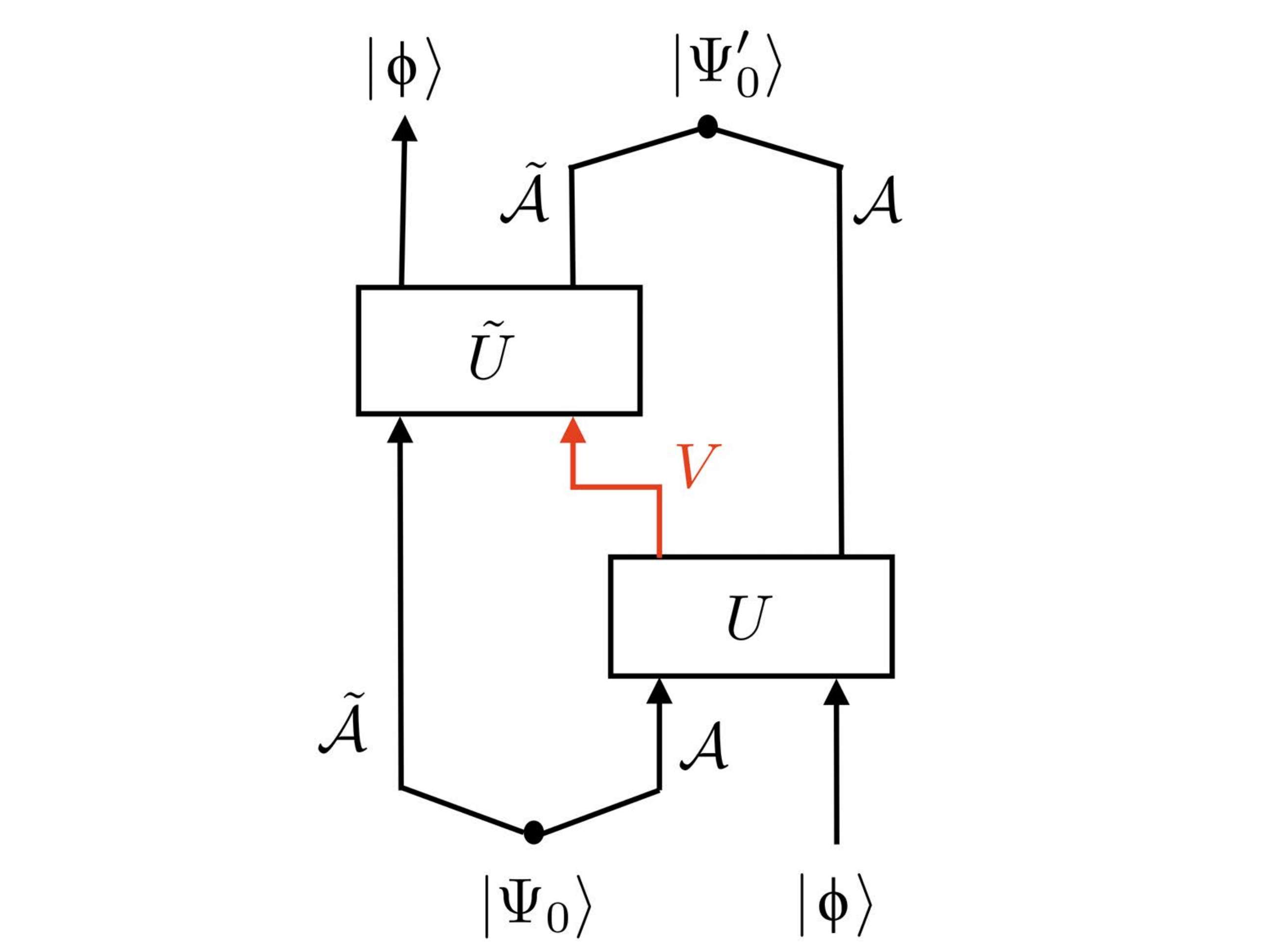}
\end{center}
\caption{\small \it An analogue of the Hayden-Preskill protocol: the code subspace approximately factorizes into a tensor product corresponding to the algebras ${\cal A}$, $\widetilde{\cal A}$. These tensor factors are entangled and provide the reservoir of EPR pairs needed to perform the teleportation. Here $U, \widetilde{U}$ is time evolution in the CFT and $V \approx {\cal O} \tO$ denotes the perturbation of the CFT Hamiltonian. The figure describes the part of the information flow that is relevant for reconstructing an infalling state $|\phi\rangle$ through measurements behind the horizon. A mirrored diagram exists for the opposite process in which a state is sent from the inside to the outside.}
\label{fig:HP_circuit_diag}
\end{figure}

We now draw a circuit diagram to represent the information transfer during the decoding protocol based on the mirror operators. It is shown in figure \ref{fig:HP_circuit_diag}.
In this circuit diagram, time flows upwards. The dot on the bottom left denotes the typical pure state $\ket{\Psi_0}$. The Hilbert space of low-energy excitations on top of this state approximately factorizes into ordinary excitations and those behind the horizon of the black hole. 
As seen in the correlator $\mathcal{C}_2$, we then create a particle $\phi$ at time $t=-t_*$. This is represented by the input of the state $\ket{\phi}$ in the bottom right part of the figure. This state is part of the Hilbert space spanned by action of algebra $\mathcal{A}$. Let $U$ denote the unitary that acts on this Hilbert space. This unitary rotates the state $\ket{\phi}$ into other states in the same Hilbert space. In physical terms, the state $\ket{\phi}$ undergoes time evolution in the CFT. The goal then is to reconstruct the state $\ket{\phi}$. The action of the double-trace operator unitary $V$ couples simple operators to their mirrors. In the circuit diagram, $\widetilde{U}$ represents the unitary that denotes time evolution in the mirror Hilbert space. This operator acts on the state $\ket{\phi}$ and mixes it with generic mirror excitations like $\ket{\widetilde{\phi}}$. One has access to the mirror excitations (in the code subspace) and in principle one can reconstruct the state $\ket{\phi}$. For this, one first evolves the state $\ket{\phi}$ with the mirror unitary $\widetilde{U}$ for a scrambling amount of time. And then one measures an appropriate mirror particle $\ket{\widetilde{\phi}}$ which the state $\ket{\phi}$ would have the most support on. This is represented by the final steps in the correlator $\mathcal{C}_2$. In the end, we are able to reconstruct the state $\ket{\phi}$.


\section{Discussion}
\label{sec:conclusions}

In this paper we discussed the bulk geometry dual to a typical CFT microstate. We proposed that the geometry corresponds to the extended AdS-Schwarzschild solution, including part of the left region. We argued that the existence of state-dependent CFT operators, the mirror operators, supports this idea.

We formulated a one-sided analogue of the Gao-Jafferis-Wall protocol to probe the geometry beyond the exterior, which shows that perturbation of the CFT Hamiltonian by state-dependent operators would allow particles in the region behind the horizon to escape the black hole and get detected in the CFT. This protocol directly relates the smoothness of typical black hole microstates to the technical conjecture that correlators in the thermal ensemble and correlators in the microcananonical ensemble are the same at leading order. The non-trivial aspect of the condition is that it must also hold at timescales of the order of the scrambling time. We looked at various aspects of this conjecture with techniques from statistical mechanics, the Eigenstate Thermalization Hypothesis, chaotic correlators, and numerics in the SYK model to provide partial evidence for the conjecture. We were, however, unable to find a full proof. It would be interesting to find a CFT proof of this conjecture, which would provide additional new evidence in favor of a smooth horizon for typical states.

An important question in understanding the black hole interior in AdS/CFT is to identify a precise CFT computation, whose result contains information about the spacetime behind the horizon. The mirror operators allow us to provide a description of the interior, however the mirror operators are {\it defined} in such a way that they reproduce a smooth interior. In order to have independent evidence it would be desirable to identify computations sensitive to the smoothness of the horizon, which involves only
ordinary state-independent CFT operators. In this work we have made a small step in this direction, by arguing that the existence of a smooth horizon for a typical state implies  certain predictions for OTOCs of ordinary single trace operators on typical states. In particular it implies that they remain close to thermal 
correlators even at scrambling time, when chaotic effects amplify $1/N$ corrections. As we mentioned, this condition is a {\it necessary condition} for a smooth horizon and it is formulated as a purely  CFT statement which can in principle be investigated without reference to the bulk dual.

This paper highlights the usefulness of considering state-dependent perturbations of the CFT Hamiltonian, which allows us to probe the region behind the horizon. Similar state-dependent perturbations, but in a class of a-typical states, have been previously considered in \cite{Kourkoulou:2017zaj, Almheiri:2018ijj, Almheiri:2018xdw, Jefferson:2018ksk, Cooper:2018cmb, Brustein:2018fkr}. 
It would be interesting to understand better the possible relevance of such perturbations for general thermal systems, not necessarily dual to black holes in AdS. The perturbations considered in this paper have the form ${\cal O} \widetilde{O}$, connecting the interior to the exterior. It would be interesting to understand
possible connections with the proposal formulated in \cite{Giddings:2017mym}.

Finally, it would be interesting to connect the results of this paper to the question of how the infalling observer can use the state-dependent operators to experience a smooth interior.

\begin{acknowledgments}
We would like to thank Dionysios Anninos, Jose Barbon, Ben Freivogel, Steve Giddings, Alessandra Gnecchi, Monica Guica, Guy Gur-Ari,  Daniel Harlow, Daniel Jafferis, Hong Liu, Juan Maldacena, Shiraz Minwalla, Lubos Motl, Suvrat Raju, Eliezer Rabinovici, Julian Sonner, Herman Verlinde, Spenta Wadia for discussions. RvB would like to thank NCCR SwissMAP of the Swiss National Science Foundation for financial support. SFL would like to acknowledge financial support from the Netherlands Organization for Scientific Research (NWO). 
\end{acknowledgments}


 \appendix
 
\section{The Exterior Geometry of Typical Black Hole Microstates}
\label{exterior}

Here we briefly review some arguments which show that at large $N$ the exterior geometry dual to a typical microstate should be the AdS-Schwarzschild geometry. The most conservative version of this statement is that, to leading order at large $N$ and for time separations that are not too large, the boundary two-point function of light single trace operators on a typical pure state is close to the two-point function that would be computed from analytic continuation starting with the geometry of the Euclidean AdS black hole. By large $N$ factorization this also implies that the leading large $N$ disconnected part of higher point functions is also the same between the typical pure state and the eternal black hole.

We will start with the assumption that  
\be
Z^{-1} \Tr[e^{-\beta H} {\cal O}(\tau,\vec{x}){\cal O}(0,\vec{0})] = \langle \phi(\tau,\vec{x}) \phi(0,\vec{0})\rangle_{\rm EBH}+O(1/N)  .
\ee 
where the RHS is the boundary-limit of the two-point function of the dual bulk field that would be computed by a Witten diagram on the Euclidean black hole geometry (the subscript EBH refers to Euclidean Black Hole).
We can not prove this statement, but we will take it as being true given that it is a basic prediction of the AdS/CFT correspondence at large $N$.

The first step is to analytically continue both sides to real time. The leading term on the RHS decays exponentially in time. For time scales of the order of scrambling time and longer, the leading term becomes comparable to the subleading term and we will not be able to control the approximations. Hence we restrict to time scales which are smaller than that. We can also restrict the correlators by 
considering only frequencies which do not scale with $N$, similar to the $\omega<\omega_*$ approximation used in the main text.
For short time scales, and filtering out high frequencies, we have that the subleading $1/N$ corrections in the Euclidean computation will remain small. Hence we conclude that
\be
Z^{-1} \Tr[e^{-\beta H} {\cal O}(t,\vec{x}){\cal O}(0,\vec{0})] = \langle \phi(t,\vec{x}) \phi(0,\vec{0})\rangle_{\rm EBH}+O(1/N)\qquad,\qquad t\ll \beta \log N
\ee 

The next step is to replace the canonical with the microcanonical ensemble. Since the RHS depends on the energy of the black hole only via the temperature, we expect that the saddle point approximation will be reliable and we conclude that
\be
\Tr[\rho_m {\cal O}(t,\vec{x}){\cal O}(0,\vec{0})] = \langle \phi(t,\vec{x}) \phi(0,\vec{0})\rangle_{\rm EBH}+O(1/N)\qquad,\qquad t\ll \beta \log N
\ee 
The next step is to replace the microcanonical with the typical pure state. Here we expect the approximation of the LHS to be good up to exponentially small corrections, as discussed in section \ref{sec:conjecture}. Hence we find
\be
\ls {\cal O}(t,\vec{x}){\cal O}(0,\vec{0})\rs = \langle \phi(t,\vec{x}) \phi(0,\vec{0})\rangle_{\rm EBH}+O(1/N)\qquad,\qquad t\ll \beta \log N
\ee
as claimed.

In general we expect that the time separation on the boundary has an inverse relation to the distance from the horizon. The argument above suggests that there cannot be any modifications to the exterior geometry (or the state of the quantum field on top of the geometry) to any $O(N^0)$ distance from the horizon. The argument above does not exclude the possibility that there are modifications at ---say--- Planckian distance from the horizon, which would be detectable by an observer hovering very near the horizon.

However, for the infalling observer these possible modifications do not affect low point functions computed at macroscopic space and time separations in the frame of the infaller. For example, if we compute a two-point function between a point in the interior and a point in the exterior, separated by a distance of horizon scale, the two point function is robust even if we introduce cutoffs in the frequencies of the modes involved, as well as on the time scales over which we have access on the boundary \cite{Papadodimas:2012aq, Kabat:2014kfa}.

The argument above does not directly apply to the connected part of higher point functions, since that is suppressed by powers of $1/N$ and hence the perturbative expansion mixes up with the corrections 
of the order of $1/N$ from comparing the ensembles. It might be possible to disentangle the two types of corrections by taking the limit where the temperature of the black hole is very large. However, we will not explore this possibility further in this paper. 

Notice that similar approximations for time scales of the order of scrambling time, and for the $O(1)$ part of higher point functions, would follow from the conjecture of section \ref{sec:conjecture} for this system, which remains currently unproven.


 \section{Time-dependence and Choice of $T$ of the Mirror Operators}
 \label{app:timedependence}

Since we are considering time-dependent perturbations, it is convenient to work in coordinate space, so we would like to define the mirror operators as a function of time. 
This suggests that we should Fourier transform $\tO_\omega$ back to coordinate time, modulo the limitations imposed by the restrictions on the frequencies discussed above. However here we encounter an interesting subtlety.

First of all we make a general observation in quantum mechanics: by specifying the matrix elements of an operator in the Heisenberg picture, as we have done in equations \eqref{defmirror}, it is not possible to conclude what is the physical time at which the given operator is localized. For instance, if we are given the matrix elements of two different operators, we do not have enough information to conclude what is the relative time-ordering of these two operators\footnote{And to analyze time-dependent perturbations of the Hamiltonian it is necessary to have a well-defined notion of time-ordering.}. Hence, assigning a particular physical time to operators requires extra information that has to be specified in addition to the Heisenberg picture matrix elements\footnote{In the Schroedinger picture the 
localization of an operator in time is rather obvious.}. 

For local operators ${\cal O}(t,x)$ in a CFT, it is natural to localize the operators in physical time which is given by the argument $t$. Notice however, that this is simply the most ``natural'' choice, but not a unique one. For example, we can imagine that the operator 
${\cal O}(t,x)$ is localized at a time $t' \neq t$. This is the notion of the ``precursor'' of an operator \cite{Polchinski:1999yd,Susskind:2013lpa}. 

In the case of local operators, the choice $t'=t$ is dictated by simplicity, as precursors are complicated operators. In the context of AdS/CFT the choice $t'=t$ for local operators, is also dictated by our desire to have a theory with a semiclassical bulk dual. If we consider a holographic CFT where the Hamiltonian is perturbed in a time-dependent fashion 
by local operators, for which the localization in physical time coincides with their argument $t$, i.e. if we avoid using perturbations by non-trivial precursors, then the dual geometry corresponds to solutions of gravity where the non-normalizable mode --- corresponding to the source of the operator --- is turned on in
a particular time-dependent fashion. However if we consider perturbations by non-trivial precursors then the bulk interpretation may go beyond effective field theory, see for example \cite{Heemskerk:2012mn} for a discussion.

This raises the question: how should we assign the mirror operators to any given physical time? In particular how should we select the time-ordering of the mirror operators? We will answer this question by imposing the following criterion: the mirror operators should be assigned a physical time in such a way that if we consider time-dependent perturbations of the Hamiltonian by mirror operators, the effect can be described by effective field theory in the bulk.

We already described in section \ref{subsec:timedep} that according to this criterion there is a one-parameter family, labeled by $T$, of useful choices. All choices of the parameter $T$ above lead to equivalent results, regarding the interpretation of the bulk geometry, if it is defined relationally with respect to the right boundary and if we do not turn on any sources for the $\tO$'s. However, since depending on the choice of $T$ the mirror operators are anchored differently on physical time, the choices of $T$ 
differ in what are the allowed time-dependent perturbations that we can do\footnote{Of course we could also act with mirror operators which do not respect a single selection of $T$. This would not be fundamentally wrong, but it would not be consistent with a simple bulk dual geometry.} --- see example at the end of this subsection. For a given desired experiment, some choices of anchoring $T$ may be more convenient than others.

We continue the discussion from section \ref{subsec:timedep} to emphasize several non-trivial aspects of the time dependence of the mirror operators together with the choice of $T$. We write the (approximate) equations for the mirror operators in position space as follows,
\begin{align}
\begin{split}
\widetilde{{\cal O}}_T (t) \ket{\Psi_0} =& \; e^{-\frac{\beta H} {2}} \mathcal{O}^\dagger(T-t) e^{\frac{\beta H} {2}} \ket{\Psi_0} , \\
\tO_T(t) A(t_1,t_2,...) \ket{\Psi_0} =& \; A(t_1,t_2,...) \tO_T(t) \ket{\Psi_0} , \\
[H,\tO_T(t)] A(t_1,t_2,...) \ket{\Psi_0} =& \;  A(t_1,t_2,..) e^{-\frac{\beta H} {2}}[H, \mathcal{O}^\dagger(T-t)] e^{\frac{\beta H} {2}} \ket{\Psi_0} .
\end{split}
\end{align}
We emphasize again that we do not want to define the mirror operators for sharply time-localized operators ${\cal O}(t)$. The meaning of the equations above is that they are correct, provided that we
convolute the operators with smearing functions in time, such that they pick out the Fourier modes with frequencies $|\omega| < \omega_*$. However, to simplify the notation we will continue writing $\tO(t)$, with the understanding that the operator has to be smeared out sufficiently. As we take $\omega_*$ to be larger, the approximation to a local operator becomes better.

\subsubsection*{Comments}

\noindent (1).  Notice that the mirror operators as defined above obey 
\be
\label{relationT}
\tO_T(t) = \tO_{T'}(t-T+T') .
\ee
Hence for any choice of $T$, we are talking about the same set of operators, but anchored differently on physical time $t$. Equation \eqref{relationT} is an example of using ``precursors'' of operators, to localize them at different moments in time. 

\noindent (2). In the bulk the choice of $T$ corresponds to whether we think of the geometry as representing either the thermofield $\tfd$ or one of its time-shifted cousins $e^{-i H_L T} \tfd$, see \cite{Papadodimas:2015xma}. Notice for instance
\be
\label{twopointtfdt}
\ls  {\cal O}(t_1) {\tO}_T (t_2)\rs = \int d\omega_1 d\omega_2 \, e^{-{\beta \omega_2 \over 2} } \ls {\cal O}_{\omega_1} {\cal O}_{\omega_2}^\dagger \rs e^{-i \omega_1 t_1 - i \omega_2 (t_2-T)} ,
\ee
given that for equilibrium states we have $\ls {\cal O}_{\omega_1} {\cal O}_{\omega_2}^\dagger \rs \propto \delta(\omega_1-\omega_2)$, we find the two-point function above is a function of $(t_1+t_2-T)$ and highly peaked around $t_1+t_2-T=0$, which is the same as the behavior of two-sided correlators 
in the time-shifted thermofield states $e^{-i T H_L} |{\rm TFD}\rangle$, where we take both times in the CFTs $t_1, t_2$ to run forward.

\noindent (3). We emphasize that if we only study correlators of operators in ``autonomous states`` in which the Hamiltonian is not time-dependent, then there is no way to distinguish between the different choices of the framing $T$.

\subsubsection*{Example} 

To illustrate the difference of possible choices of $T$, let us ask the physical question: what are the allowed perturbations of the form ${\cal O}\tO$ that we can perform at physical time $t=0$. Suppose we select $T=0$. Then as we see from the diagram we could perturb the Hamiltonian by
\be
\delta H = \delta(t) {\cal O}(0) \widetilde{{\cal O}}_{T=0}(0)  .
\ee
The operators ${\cal O}(0)$ and $ \widetilde{{\cal O}}_{T=0}(0)$ are highly entangled, and this perturbation would produce two shockwaves. However, if we make a choice of $T\neq 0$, and if we then ask what perturbation we are allowed to perform at $t=0$ then we would have
\be
\delta H =  \delta(t) {\cal O}(0) \widetilde{{\cal O}}_T(0)  .
\ee
Now we notice from \eqref{twopointtfdt} that the operators $ {\cal O}(0)$ and $\widetilde{{\cal O}}_T(0)$ are less entangled as $|T|$ increases, and the produced shockwave in the bulk becomes weaker.

Notice, however, that with a choice of $T\neq 0$, we can produce a strong shockwave by perturbing the state at time $t=T/2$ by
\be
\delta H = \delta(t-T/2) \,{\cal O}(T/2) \,\widetilde{{\cal O}}_T(T/2)  .
\ee
since again ${\cal O}_T(T/2)$ and $\widetilde{\cal O}_T(T/2)$ from \eqref{twopointtfdt} are highly entangled.


\section{Spherical Shells on an Einstein-Rosen Bridge}
\label{app:erbridge}

In this Appendix, we will study the gluing of spherical shells on an Einstein-Rosen bridge. The goal is to understand bulk energy associated to matter in the right and the left regions.
Using Einstein's equations, we will check that the mass of some matter to the left of the bifurcation horizon, if gravitationally dressed with respect to the right, is negative with respect to the right asymptotic boundary. This is similar to a particle constructed using mirror operator as discussed in Section \ref{sssec:autexcstate} using purely CFT technology. The discussion here will be purely classical but it does show some aspects that are shared with the mirror operators. We will focus on 4d asymptotically flat space-time for ease of calculation. A useful reference is \cite{Brill:1963yv}. 

We will consider the initial value problem assuming we are at a moment of time symmetry. We parametrize the induced spacelike metric of the bridge as 
\begin{equation}
\label{spatialmetric}
ds^2 = \chi(\rho)^4 (d\rho^2 + \rho^2 d\Omega_2^2)  .
\end{equation}
For time-reflection invariant initial data (no extrinsic curvature) the Hamiltonian constraint of general relativity becomes $R=0$, where $R$ is the Ricci scalar of the spatial metric. For the ansatz \eqref{spatialmetric} this becomes
$\nabla^2 \chi = 0$. The usual Einstein-Rosen bridge without matter is the solution
\begin{equation}
\chi = 1 + {a \over \rho}  .
\end{equation}
$\rho$ is not the same as the Schwarzchild radial coordinate $r$. Expanding near $\rho\rightarrow \infty$ we find $a ={GM \over 2}$. Moreover, we can multiply $\chi$ by an arbitrary constant without changing the physical metric. So more generally if we take $\chi = c_1 + {c_2 \over \rho}$ we find that the right mass is $M_R = {2 c_1 c_2 \over G}$. Notice that the left mass is also the same, which can be found by $\rho \rightarrow 1/\rho$.

We introduce a sphericall shell of matter which is momentarily at rest at $t=0$, at a location $\rho=b$. This means that the solution will be
\begin{align}
\chi &= c_1 + {c_2 \over \rho} , \qquad \rho<b , \\
\chi &= d_1 + {d_2 \over \rho} , \qquad \rho>b .
\end{align}
This depends on 5 parameters. One is trivial and eliminated by the overall rescaling. Another is eliminated by demanding continuity of $\chi$ at $b$. The others can be selected to correspond to: i) left mass $M_L$, ii) position of shell $b$  iii) $T_{00}$ of shell, which will be related to the ``kink'' of $\chi$. These will fix $M_R$. Of course one can reorganize the independent parameters differently.

Continuity of the metric at the gluing demands that:
\begin{equation}
\label{eqn:fglue}
1+ {G M_R \over 2b} = c_1 + {G M_L \over 2 c_1 b} .
\end{equation}
The last parameter $c_1$ can be expressed in terms of the stress tensor of the shell of matter. For initial data with time reflection symmetry the constraint equation in the presence of matter reads $R = 16 \pi G \rho_{\rm matter}$, where $\rho_{\rm matter} = T_{ab} n^a n^b$ and $n^a$ is the unit normal vector. In this case the unit normal is $n^a = {1\over \sqrt{f}} \delta^{a0}$, where $f=1-2 {G M_R \over r}$, expressed in terms of $\rho$.  We select an infinitely thin shell: $T_{00} =  \mu f \delta(y-y_0)$, where $y$ are locally flat radial coordinates. Converting to $\rho$ this becomes
\begin{equation}
\rho_{\rm matter} =  \mu {\delta(\rho-b) \over \chi^2|_{\rho=b}} .
\end{equation}
We must, therefore, solve 
\begin{equation}
-8{\rho \chi''+2 \chi' \over \rho \chi^5}= 16 \pi G \rho_{\rm matter}
\end{equation}
to obtain $c_1$. This can be simplified across the shell.
\begin{align}
\chi'(b^+) - \chi'(b^-) &= - 2 \pi G \mu (\chi^3)_{\rho=b}  ,\\
-{G M_R \over 2 b^2} + {G M_L \over 2 c_1 b^2} &= -2 \pi G \mu \,  \bigg(1+{GM_R \over 2b} \bigg)^3  .
\end{align}
We can use this to eliminate $c_1$ from equation \ref{eqn:fglue} to obtain an expression for $M_R$
\begin{equation}
M_R = {1 - 8 \pi^2 G^2 b^2 \mu^2- \sqrt{1+4 G^2\pi \mu M_L(1-2 G\pi \mu b)} \over 2\pi G^2\mu(2 \pi G b \mu -1)}  .
\end{equation}
Expanding to linear order we find 
\be
M_R = M_L + 4 \pi \mu (b^2 - {G^2 M_L^2\over 4}) + O(\mu^2)  .
\ee 
If we gravitationally dress the operators with respect to the right, then $M_L$ is constant. Then we see that at linear order in $\mu$ the change of the mass $M_R$ flips sign as we move the shell through the bifurcation point, which is at $b= {G M_L \over 2}$ before we add the shell. This is consistent with the commutation relations of the mirror operators with the CFT Hamiltonian in \eqref{defmirror}.


\section{Numerics in the SYK Model}
\label{App:SYK}

The numerical study of the SYK model is straightforward for low enough values of $N$. The fermions satisfy the Clifford algebra
\begin{equation}
\{\psi_i,\psi_j\} = \delta_{i,j} \, ,
\end{equation}
where we have used a slightly different normalization than usual. This allows one to represent the fermions as Euclidean gamma matrices. For $N=4$, for example, the matrices are,
\begin{align*}
\psi_1 =\frac{1} {\sqrt{2}}
\begin{bmatrix}
0 & 0 & 0 & \:\, 1\\ 
0 & 0  & \:\,1 & 0 \\ 
0 & \:\, 1 & 0 & 0\\ 
\:\,1 & 0 & 0 & 0
\end{bmatrix}\;,&
&\psi_2=\frac{i} {\sqrt{2}}
\begin{bmatrix}
0 & 0 & 0 & \:\,1\\ 
0 & 0 & -1 & 0\\ 
0 & \:\,1 & 0 & 0\\ 
-1 & 0 & 0 & 0
\end{bmatrix}\;,\\
\psi_3 =\frac{1} {\sqrt{2}}
\begin{bmatrix}
0 & 0 & 1 & 0\\ 
0 & 0 & 0 & -1\\ 
1 & 0 & 0 & 0\\ 
0 & -1 & 0 & 0
\end{bmatrix}\;,&
&\psi_4=\frac{i} {\sqrt{2}}
\begin{bmatrix}
0 & 0 & \:\,1 & 0\\ 
0 & 0 & 0 & \:\,1\\ 
-1 & 0 & 0 & 0\\ 
0 & -1  & 0 & 0
\end{bmatrix} \;.
\end{align*}
The Hamiltonian is the sum over $q$ different products of these matrices, multiplied with a Gaussian-distributed random number, with mean zero. Calculations are fastest for the minimal $q$ that still has interesting dynamics. Therefore, $q=4$ is chosen,
\begin{equation}
H = \sum_{i<j<k<l} J_{ijkl} \psi_i \psi_j \psi_k \psi_l \qquad \qquad \text{Var}(J_{ijkl})= \frac{6 J^2} {N^3} ,
\end{equation}
where $J$ sets the scale for the model. We can, therefore, set $J=1$ to simplify this further. The energy levels and energy eigenvectors can be obtain by diagonalizing the Hamiltonian. The typical state is constructed as a superposition of these vectors. 

Taking the usual Laplace transform to get the relation between energy and temperature does not work because the energy levels are not smoothed out for low $N$, see for example figure \ref{appfig:fmael}. Instead, the temperature is fixed by demanding that the average energy in the typical pure state is the same as that in the thermal ensemble. This leads to an order $O(1/N)$ error for the temperature in the large $N$ limit, but it is computationally fast. For the numerics in this paper, $\beta=5$ is chosen for the inverse temperature.

In the rest of this appendix we will show several figures that were used in the main text at different $N$. We compare the thermal correlators with the pure state correlators,
\begin{equation}
\langle A \rangle_{\text{Thermal}} = \frac{\Tr [e^{-\beta H} A]} {\Tr [e^{-\beta H}]} \;, \qquad \langle A \rangle_{\text{Pure}} = \bra{\Psi_0}A\ket{\Psi_0} .
\end{equation}

\newpage
\begin{figure}[!htb]
    \centering
    \begin{tabular}[c]{ c c c }
		\begin{subfigure}[c]{0.31\textwidth}
                \centering
                \includegraphics[width=1\textwidth]{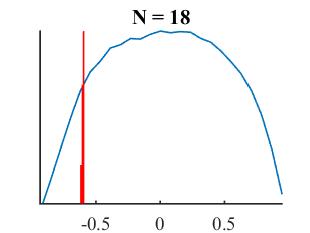}
            \end{subfigure}
		&
	\begin{subfigure}[c]{0.31\textwidth}
                \centering
                \includegraphics[width=1\textwidth]{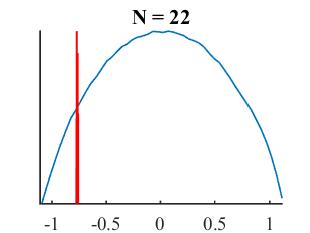}
            \end{subfigure}
		&
	\begin{subfigure}[c]{0.31\textwidth}
                \centering
                \includegraphics[width=1\textwidth]{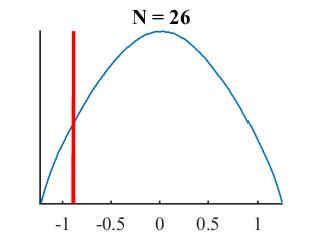}
       \end{subfigure}
\end{tabular}
    \caption{Density of states of the SYK model (blue), and the energy eigenstates excited in the equilibrium state $\rsz$ (red).}
\end{figure}
\begin{figure}[!htb]
\centering
\begin{tabular}[c]{ c }

	\begin{subfigure}[c]{1\textwidth}
                \centering
                \includegraphics[width=1\textwidth]{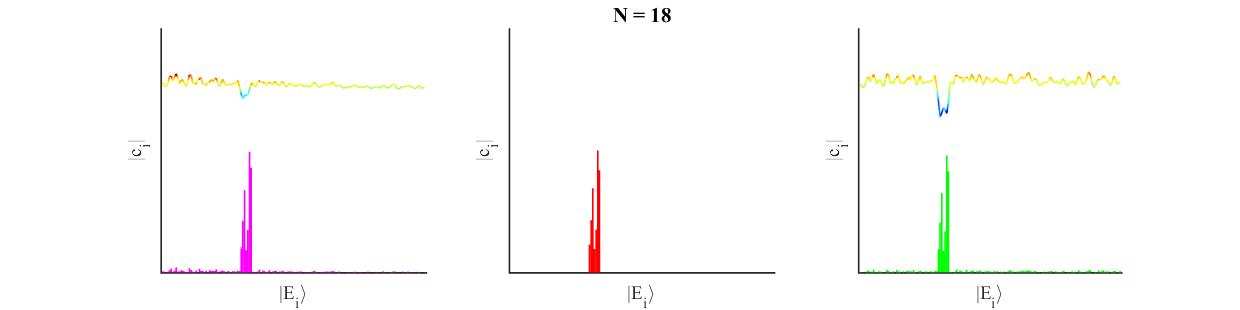}
       \end{subfigure} \\ 

	\begin{subfigure}[c]{1\textwidth}
                \centering
                \includegraphics[width=1\textwidth]{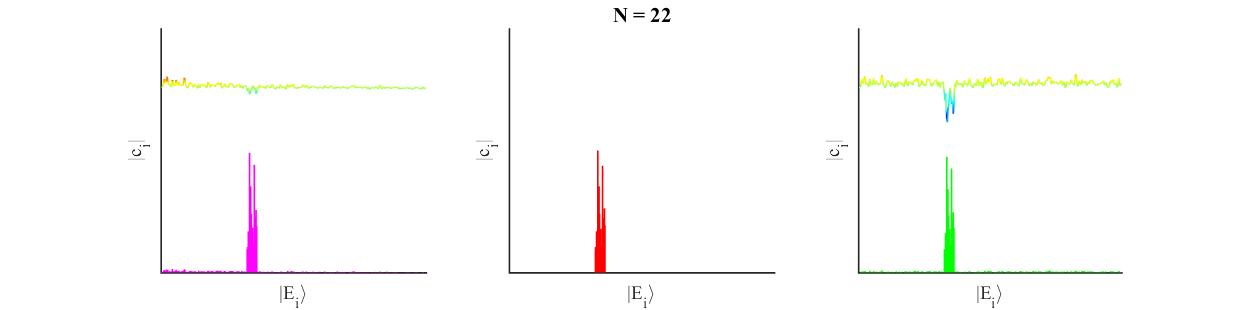}
       \end{subfigure} \\ 

	\begin{subfigure}[c]{1\textwidth}
                \centering
                \includegraphics[width=1\textwidth]{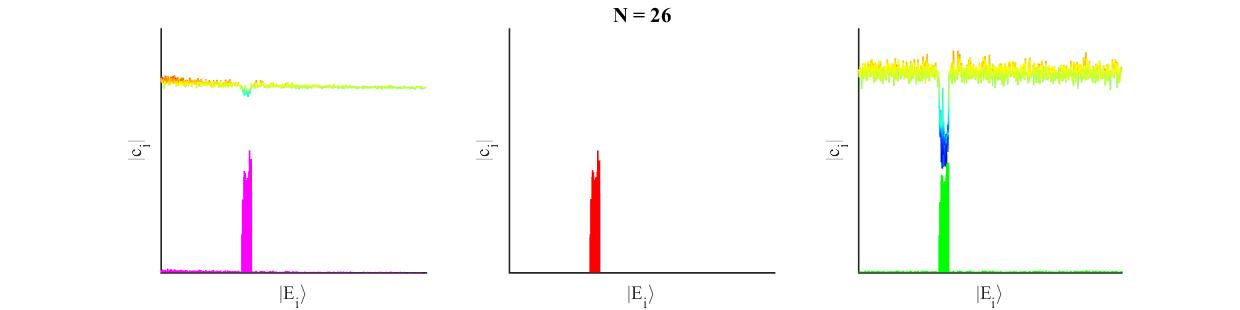}
       \end{subfigure} 

\end{tabular}

\caption{{\small Distribution of $|c_i|$ in non-equilibrium state of the form $U(\widetilde{S}_i)\rsz$ (magenta), typical equilibrium state $\rsz$ (red), and non-equilibrium state of the form $U(S_i)\rsz$ (green). The line above the bar plot shows, in heat map colors, which eigenstates are excited, and which ones are suppressed because of the perturbation. Blue eigenstates are suppressed, while eigenstates with other colors are excited with small (green), medium (orange), or large (red)
 magnitude.}}

\end{figure}

\begin{figure}[!htb]
\centering
\begin{tabular}[c]{ c }
	\begin{subfigure}[c]{1\textwidth}
                \centering
                \includegraphics[width=1\textwidth]{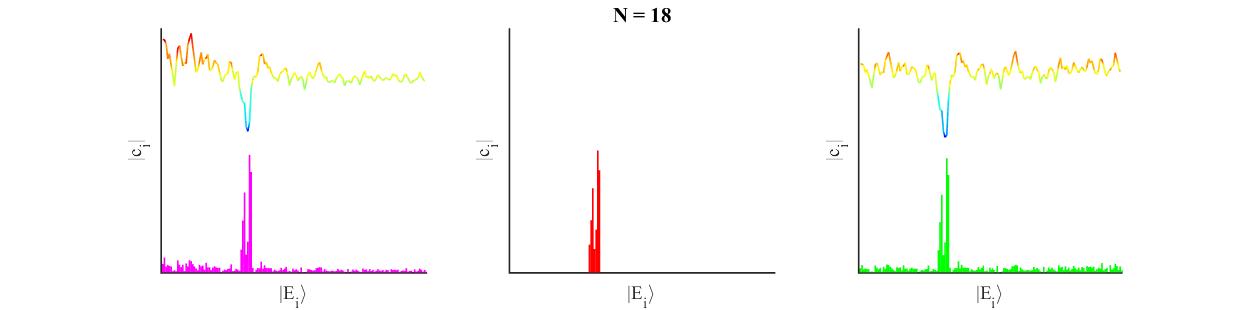}
       \end{subfigure} \\ 

	\begin{subfigure}[c]{1\textwidth}
                \centering
                \includegraphics[width=1\textwidth]{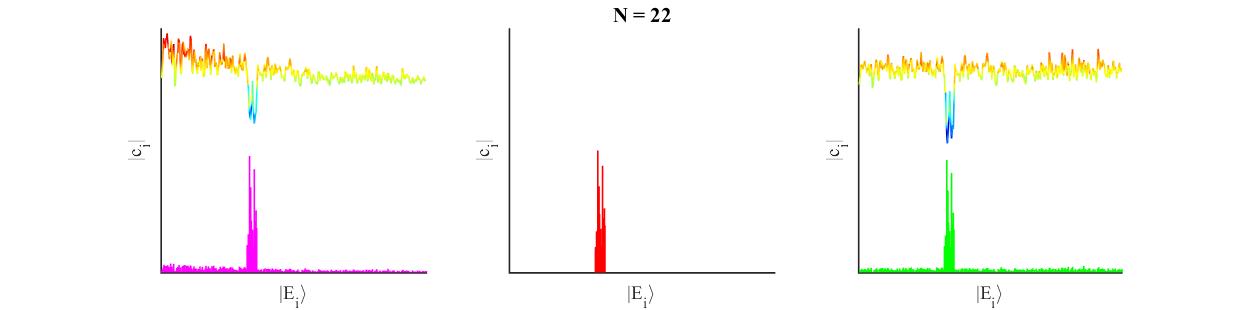}
       \end{subfigure} \\ 

	\begin{subfigure}[c]{1\textwidth}
                \centering
                \includegraphics[width=1\textwidth]{nfig_1S_N26_1.jpg}
       \end{subfigure} 
\end{tabular}

\caption{{\small Distribution of $|c_i|$ in non-equilibrium state of the form $U(\widetilde{S}_i)\rsz$ (magenta), typical equilibrium state $\rsz$ (red), and non-equilibrium state of the form $U(S_i)\rsz$ (green). The line above the bar plot shows, in heat map colors, which eigenstates are excited, and which ones are suppressed because of the perturbation. Blue eigenstates are suppressed, while eigenstates with other colors are excited with small (green), medium (orange), or large (red)
 magnitude.}}

\end{figure}

\begin{figure}[!htb]
    \centering
    \begin{tabular}[c]{ c c c }
		\begin{subfigure}[c]{0.31\textwidth}
                \centering
                \includegraphics[width=1\textwidth]{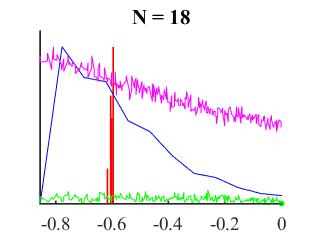}
            \end{subfigure}
		&
	\begin{subfigure}[c]{0.31\textwidth}
                \centering
                \includegraphics[width=1\textwidth]{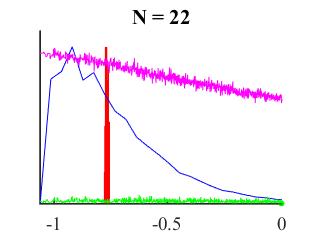}
            \end{subfigure}
		&
	\begin{subfigure}[c]{0.31\textwidth}
                \centering
                \includegraphics[width=1\textwidth]{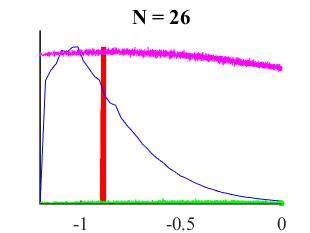}
       \end{subfigure}
\end{tabular}
    \caption{\label{appfig:fmael} The diagonal elements of $\{\psi_1(t),\psi_2(0)\}^2$ at scrambling time (magenta) dominate over the superdiagonal (green) and is slowly varying in the energy region of the thermal ensemble (blue) and the equilibrium state (red).}
\end{figure}
\begin{figure}[!htb]
    \centering
    \begin{tabular}[c]{ c c c }
		\begin{subfigure}[c]{0.31\textwidth}
                \centering
                \includegraphics[width=1\textwidth]{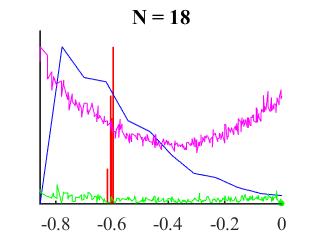}
            \end{subfigure}
		&
	\begin{subfigure}[c]{0.31\textwidth}
                \centering
                \includegraphics[width=1\textwidth]{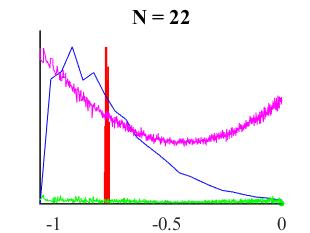}
            \end{subfigure}
		&
	\begin{subfigure}[c]{0.31\textwidth}
                \centering
                \includegraphics[width=1\textwidth]{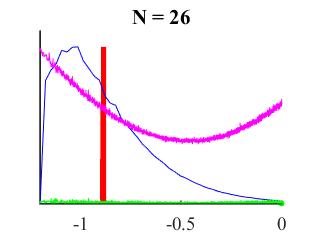}
       \end{subfigure}
\end{tabular}
    \caption{\label{appfig:fmael} The diagonal elements of $-[S_1(t),S_2(0)]^2$ at scrambling time (magenta) dominate over the superdiagonal (green) and is slowly varying in the energy region of the thermal ensemble (blue) and the equilibrium state (red).}
\end{figure}
\begin{figure}[!htb]
    \centering
    \begin{tabular}[c]{ c c c }
		\begin{subfigure}[c]{0.31\textwidth}
                \centering
                \includegraphics[width=1\textwidth]{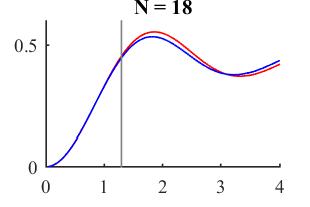}
            \end{subfigure}
		&
	\begin{subfigure}[c]{0.31\textwidth}
                \centering
                \includegraphics[width=1\textwidth]{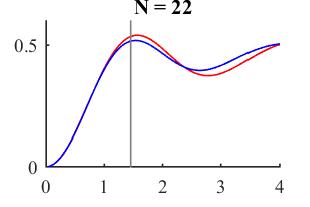}
            \end{subfigure}
		&
	\begin{subfigure}[c]{0.31\textwidth}
                \centering
                \includegraphics[width=1\textwidth]{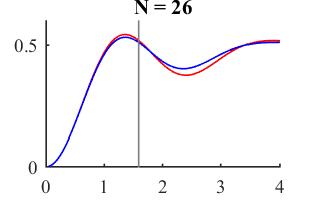}
       \end{subfigure}
\end{tabular}
    \caption{We compare the expectation value of $\langle\{\psi_1(t),\psi_2(0)\}^2\rangle$ in the thermal state (blue) and pure state (red). The scrambling time is designated by the vertical line.}
\end{figure}
\begin{figure}[!htb]
    \centering
    \begin{tabular}[c]{ c c c }
		\begin{subfigure}[c]{0.31\textwidth}
                \centering
                \includegraphics[width=1\textwidth]{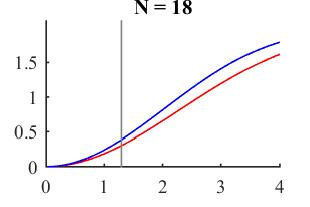}
            \end{subfigure}
		&
	\begin{subfigure}[c]{0.31\textwidth}
                \centering
                \includegraphics[width=1\textwidth]{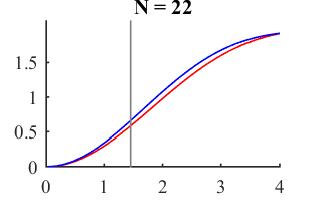}
            \end{subfigure}
		&
	\begin{subfigure}[c]{0.31\textwidth}
                \centering
                \includegraphics[width=1\textwidth]{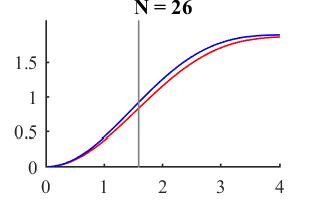}
       \end{subfigure}
\end{tabular}
    \caption{We compare the expectation value of $-\langle[S_1(t),S_2(0)]^2\rangle$ in the thermal state (blue) and pure state (red). The scrambling time is designated by the vertical line.}
\end{figure}
\begin{figure}[!htb]
    \centering
    \begin{tabular}[c]{ c c c }
		\begin{subfigure}[c]{0.31\textwidth}
                \centering
                \includegraphics[width=1\textwidth]{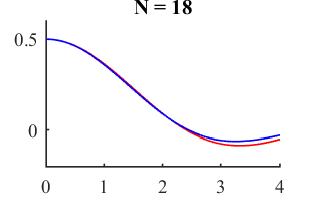}
            \end{subfigure}
		&
	\begin{subfigure}[c]{0.31\textwidth}
                \centering
                \includegraphics[width=1\textwidth]{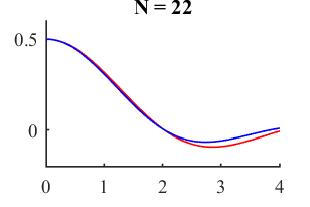}
            \end{subfigure}
		&
	\begin{subfigure}[c]{0.31\textwidth}
                \centering
                \includegraphics[width=1\textwidth]{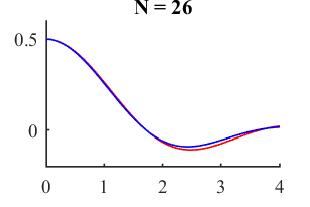}
       \end{subfigure}
\end{tabular}
    \caption{We compare the expectation value of $\langle\psi_1(t)\psi_1(0)\rangle$ in the thermal state (blue) and pure state (red).}
\end{figure}
\begin{figure}[!htb]
    \centering
    \begin{tabular}[c]{ c c c }
		\begin{subfigure}[c]{0.31\textwidth}
                \centering
                \includegraphics[width=1\textwidth]{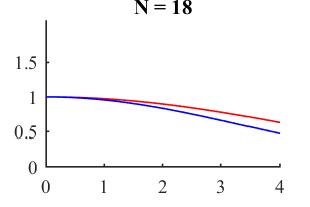}
            \end{subfigure}
		&
	\begin{subfigure}[c]{0.31\textwidth}
                \centering
                \includegraphics[width=1\textwidth]{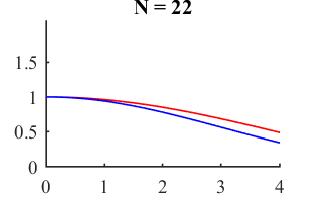}
            \end{subfigure}
		&
	\begin{subfigure}[c]{0.31\textwidth}
                \centering
                \includegraphics[width=1\textwidth]{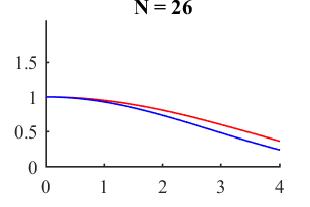}
       \end{subfigure}
\end{tabular}
    \caption{We compare the expectation value of $\langle S_1(t) S_1(0)\rangle$ in the thermal state (blue) and pure state (red).}
\end{figure}
\begin{figure}[!htb]
    \centering
    \begin{tabular}[c]{ c c c }
		\begin{subfigure}[c]{0.31\textwidth}
                \centering
                \includegraphics[width=1\textwidth]{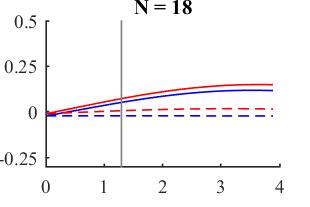}
            \end{subfigure}
		&
	\begin{subfigure}[c]{0.31\textwidth}
                \centering
                \includegraphics[width=1\textwidth]{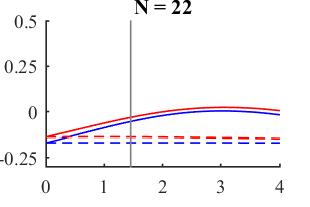}
            \end{subfigure}
		&
	\begin{subfigure}[c]{0.31\textwidth}
                \centering
                \includegraphics[width=1\textwidth]{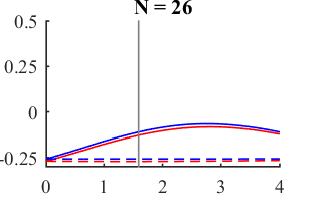}
       \end{subfigure}
\end{tabular}
        \caption{Comparing the traversable wormhole correlator \eqref{eq:SYKexp} in the thermal state (blue) and typical state (red). The dotted line is the signal without the probe, which is the response one obtains from just the state-dependent perturbation and should disappear in the large $N$ limit, in this case a sum over five pairs of operators was used in the perturbation to limit this effect. The vertical line denotes the scrambling time and is the time around which the probe is focused. }
\end{figure}

\bibliographystyle{jhep}
\bibliography{references}

\providecommand{\href}[2]{#2}\begingroup\raggedright\begin{thebibliography}{10}

\bibitem{Mathur:2009hf}
S.~D. Mathur, \emph{{The Information paradox: A Pedagogical introduction}},
  \href{https://doi.org/10.1088/0264-9381/26/22/224001}{\emph{Class.Quant.Grav.}
  {\bfseries 26} (2009) 224001}
  [\href{https://arxiv.org/abs/0909.1038}{{\ttfamily 0909.1038}}].

\bibitem{Almheiri:2012rt}
A.~Almheiri, D.~Marolf, J.~Polchinski and J.~Sully, \emph{{Black Holes:
  Complementarity or Firewalls?}},
  \href{https://doi.org/10.1007/JHEP02(2013)062}{\emph{JHEP} {\bfseries 02}
  (2013) 062} [\href{https://arxiv.org/abs/1207.3123}{{\ttfamily 1207.3123}}].

\bibitem{Almheiri:2013hfa}
A.~Almheiri, D.~Marolf, J.~Polchinski, D.~Stanford and J.~Sully, \emph{{An
  Apologia for Firewalls}},
  \href{https://doi.org/10.1007/JHEP09(2013)018}{\emph{JHEP} {\bfseries 1309}
  (2013) 018} [\href{https://arxiv.org/abs/1304.6483}{{\ttfamily 1304.6483}}].

\bibitem{Marolf:2013dba}
D.~Marolf and J.~Polchinski, \emph{{Gauge/Gravity Duality and the Black Hole
  Interior}},
  \href{https://doi.org/10.1103/PhysRevLett.111.171301}{\emph{Phys.Rev.Lett.}
  {\bfseries 111} (2013) 171301}
  [\href{https://arxiv.org/abs/1307.4706}{{\ttfamily 1307.4706}}].

\bibitem{Bousso:2013wia}
R.~Bousso, \emph{{Firewalls From Double Purity}},
  \href{https://arxiv.org/abs/1308.2665}{{\ttfamily 1308.2665}}.

\bibitem{Gao:2016bin}
P.~Gao, D.~L. Jafferis and A.~Wall, \emph{{Traversable Wormholes via a Double
  Trace Deformation}},  \href{https://arxiv.org/abs/1608.05687}{{\ttfamily
  1608.05687}}.

\bibitem{Maldacena:2017axo}
J.~Maldacena, D.~Stanford and Z.~Yang, \emph{{Diving into traversable
  wormholes}}, \href{https://doi.org/10.1002/prop.201700034}{\emph{Fortsch.
  Phys.} {\bfseries 65} (2017) 1700034}
  [\href{https://arxiv.org/abs/1704.05333}{{\ttfamily 1704.05333}}].

\bibitem{Kourkoulou:2017zaj}
I.~Kourkoulou and J.~Maldacena, \emph{{Pure states in the SYK model and
  nearly-$AdS_2$ gravity}},  \href{https://arxiv.org/abs/1707.02325}{{\ttfamily
  1707.02325}}.

\bibitem{Shenker:2013pqa}
S.~H. Shenker and D.~Stanford, \emph{{Black holes and the butterfly effect}},
  \href{https://arxiv.org/abs/1306.0622}{{\ttfamily 1306.0622}}.

\bibitem{Maldacena:2015waa}
J.~Maldacena, S.~H. Shenker and D.~Stanford, \emph{{A bound on chaos}},
  \href{https://doi.org/10.1007/JHEP08(2016)106}{\emph{JHEP} {\bfseries 08}
  (2016) 106} [\href{https://arxiv.org/abs/1503.01409}{{\ttfamily
  1503.01409}}].

\bibitem{Hayden:2007cs}
P.~Hayden and J.~Preskill, \emph{{Black holes as mirrors: Quantum information
  in random subsystems}},
  \href{https://doi.org/10.1088/1126-6708/2007/09/120}{\emph{JHEP} {\bfseries
  09} (2007) 120} [\href{https://arxiv.org/abs/0708.4025}{{\ttfamily
  0708.4025}}].

\bibitem{deBoer:2018ibj}
J.~De~Boer, S.~F. Lokhande, E.~Verlinde, R.~Van~Breukelen and K.~Papadodimas,
  \emph{{On the interior geometry of a typical black hole microstate}},
  \href{https://arxiv.org/abs/1804.10580}{{\ttfamily 1804.10580}}.

\bibitem{Almheiri:2018ijj}
A.~Almheiri, A.~Mousatov and M.~Shyani, \emph{{Escaping the Interiors of Pure
  Boundary-State Black Holes}},
  \href{https://arxiv.org/abs/1803.04434}{{\ttfamily 1803.04434}}.

\bibitem{Almheiri:2018xdw}
A.~Almheiri, \emph{{Holographic Quantum Error Correction and the Projected
  Black Hole Interior}},  \href{https://arxiv.org/abs/1810.02055}{{\ttfamily
  1810.02055}}.

\bibitem{Jefferson:2018ksk}
R.~Jefferson, \emph{{Comments on black hole interiors and modular inclusions}},
   \href{https://arxiv.org/abs/1811.08900}{{\ttfamily 1811.08900}}.

\bibitem{Cooper:2018cmb}
S.~Cooper, M.~Rozali, B.~Swingle, M.~Van~Raamsdonk, C.~Waddell and D.~Wakeham,
  \emph{{Black Hole Microstate Cosmology}},
  \href{https://arxiv.org/abs/1810.10601}{{\ttfamily 1810.10601}}.

\bibitem{Brustein:2018fkr}
R.~Brustein and Y.~Zigdon, \emph{{Revealing the interior of black holes out of
  equilibrium in the Sachdev-Ye-Kitaev model}},
  \href{https://doi.org/10.1103/PhysRevD.98.066013}{\emph{Phys. Rev.}
  {\bfseries D98} (2018) 066013}
  [\href{https://arxiv.org/abs/1804.09017}{{\ttfamily 1804.09017}}].

\bibitem{Papadodimas:2015jra}
K.~Papadodimas and S.~Raju, \emph{{Remarks on the necessity and implications of
  state-dependence in the black hole interior}},
  \href{https://doi.org/10.1103/PhysRevD.93.084049}{\emph{Phys. Rev.}
  {\bfseries D93} (2016) 084049}
  [\href{https://arxiv.org/abs/1503.08825}{{\ttfamily 1503.08825}}].

\bibitem{tHooft:1993dmi}
G.~'t~Hooft, \emph{{Dimensional reduction in quantum gravity}}, {\emph{Conf.
  Proc.} {\bfseries C930308} (1993) 284}
  [\href{https://arxiv.org/abs/gr-qc/9310026}{{\ttfamily gr-qc/9310026}}].

\bibitem{Papadodimas:2017qit}
K.~Papadodimas, \emph{{A class of non-equilibrium states and the black hole
  interior}},  \href{https://arxiv.org/abs/1708.06328}{{\ttfamily 1708.06328}}.

\bibitem{Susskind:2018pmk}
L.~Susskind, \emph{{Three Lectures on Complexity and Black Holes}},  2018,
  \href{https://arxiv.org/abs/1810.11563}{{\ttfamily 1810.11563}}.

\bibitem{Srednicki1999approach}
M.~Srednicki, \emph{The approach to thermal equilibrium in quantized chaotic
  systems}, {\emph{Journal of Physics A: Mathematical and General} {\bfseries
  32} (1999) 1163}.

\bibitem{Maldacena:2013xja}
J.~Maldacena and L.~Susskind, \emph{{Cool horizons for entangled black holes}},
   \href{https://arxiv.org/abs/1306.0533}{{\ttfamily 1306.0533}}.

\bibitem{Papadodimas:2012aq}
K.~Papadodimas and S.~Raju, \emph{{An Infalling Observer in AdS/CFT}},
  \href{https://doi.org/10.1007/JHEP10(2013)212}{\emph{JHEP} {\bfseries 1310}
  (2013) 212} [\href{https://arxiv.org/abs/1211.6767}{{\ttfamily 1211.6767}}].

\bibitem{Papadodimas:2013b}
K.~Papadodimas and S.~Raju, \emph{{The Black Hole Interior in AdS/CFT and the
  Information Paradox}},  \href{https://arxiv.org/abs/1310.6334}{{\ttfamily
  1310.6334}}.

\bibitem{Papadodimas:2013}
K.~Papadodimas and S.~Raju, \emph{{State-Dependent Bulk-Boundary Maps and Black
  Hole Complementarity}},  \href{https://arxiv.org/abs/1310.6335}{{\ttfamily
  1310.6335}}.

\bibitem{Verlinde:2012cy}
E.~Verlinde and H.~Verlinde, \emph{{Black Hole Entanglement and Quantum Error
  Correction}}, \href{https://doi.org/10.1007/JHEP10(2013)107}{\emph{JHEP}
  {\bfseries 1310} (2013) 107}
  [\href{https://arxiv.org/abs/1211.6913}{{\ttfamily 1211.6913}}].

\bibitem{Verlinde:2013vja}
E.~Verlinde and H.~Verlinde, \emph{{Black Hole Information as Topological
  Qubits}},  \href{https://arxiv.org/abs/1306.0516}{{\ttfamily 1306.0516}}.

\bibitem{Verlinde:2013qya}
E.~Verlinde and H.~Verlinde, \emph{{Behind the Horizon in AdS/CFT}},
  \href{https://arxiv.org/abs/1311.1137}{{\ttfamily 1311.1137}}.

\bibitem{Haag:1992hx}
R.~Haag, \emph{{Local quantum physics: Fields, particles, algebras, 2nd ed.}}
  Springer, 1992.

\bibitem{Papadodimas:2015xma}
K.~Papadodimas and S.~Raju, \emph{{Local Operators in the Eternal Black Hole}},
  \href{https://doi.org/10.1103/PhysRevLett.115.211601}{\emph{Phys. Rev. Lett.}
  {\bfseries 115} (2015) 211601}
  [\href{https://arxiv.org/abs/1502.06692}{{\ttfamily 1502.06692}}].

\bibitem{Bardeen:1973gs}
J.~M. Bardeen, B.~Carter and S.~W. Hawking, \emph{{The Four laws of black hole
  mechanics}}, \href{https://doi.org/10.1007/BF01645742}{\emph{Commun. Math.
  Phys.} {\bfseries 31} (1973) 161}.

\bibitem{Wald:1993nt}
R.~M. Wald, \emph{{Black hole entropy is the Noether charge}},
  \href{https://doi.org/10.1103/PhysRevD.48.R3427}{\emph{Phys. Rev.} {\bfseries
  D48} (1993) R3427} [\href{https://arxiv.org/abs/gr-qc/9307038}{{\ttfamily
  gr-qc/9307038}}].

\bibitem{Iyer:1994ys}
V.~Iyer and R.~M. Wald, \emph{{Some properties of Noether charge and a proposal
  for dynamical black hole entropy}},
  \href{https://doi.org/10.1103/PhysRevD.50.846}{\emph{Phys. Rev.} {\bfseries
  D50} (1994) 846} [\href{https://arxiv.org/abs/gr-qc/9403028}{{\ttfamily
  gr-qc/9403028}}].

\bibitem{Hollands:2012sf}
S.~Hollands and R.~M. Wald, \emph{{Stability of Black Holes and Black Branes}},
  \href{https://doi.org/10.1007/s00220-012-1638-1}{\emph{Commun. Math. Phys.}
  {\bfseries 321} (2013) 629}
  [\href{https://arxiv.org/abs/1201.0463}{{\ttfamily 1201.0463}}].

\bibitem{Jafferis:2015del}
D.~L. Jafferis, A.~Lewkowycz, J.~Maldacena and S.~J. Suh, \emph{{Relative
  entropy equals bulk relative entropy}},
  \href{https://doi.org/10.1007/JHEP06(2016)004}{\emph{JHEP} {\bfseries 06}
  (2016) 004} [\href{https://arxiv.org/abs/1512.06431}{{\ttfamily
  1512.06431}}].

\bibitem{Donnelly:2015hta}
W.~Donnelly and S.~B. Giddings, \emph{{Diffeomorphism-invariant observables and
  their nonlocal algebra}}, \href{https://doi.org/10.1103/PhysRevD.94.029903,
  10.1103/PhysRevD.93.024030}{\emph{Phys. Rev.} {\bfseries D93} (2016) 024030}
  [\href{https://arxiv.org/abs/1507.07921}{{\ttfamily 1507.07921}}].

\bibitem{kitaevtalks}
A.~Kitaev, ``{A Simple Model of Quantum Holography}.''
\newblock http://online.kitp.ucsb.edu/online/entangled15/kitaev/.

\bibitem{Polchinski:2016syk}
J.~Polchinski and V.~Rosenhaus, \emph{The spectrum in the sachdev-ye-kitaev
  model}, \href{https://doi.org/10.1007/JHEP04(2016)001}{\emph{Journal of High
  Energy Physics} {\bfseries 2016} (2016) 1}.

\bibitem{Maldacena:2016syk}
J.~Maldacena and D.~Stanford, \emph{Remarks on the sachdev-ye-kitaev model},
  \href{https://doi.org/10.1103/PhysRevD.94.106002}{\emph{Phys. Rev. D}
  {\bfseries 94} (2016) 106002}.

\bibitem{Heemskerk:2009pn}
I.~Heemskerk, J.~Penedones, J.~Polchinski and J.~Sully, \emph{{Holography from
  Conformal Field Theory}},
  \href{https://doi.org/10.1088/1126-6708/2009/10/079}{\emph{JHEP} {\bfseries
  10} (2009) 079} [\href{https://arxiv.org/abs/0907.0151}{{\ttfamily
  0907.0151}}].

\bibitem{Fitzpatrick:2010zm}
A.~L. Fitzpatrick, E.~Katz, D.~Poland and D.~Simmons-Duffin, \emph{{Effective
  Conformal Theory and the Flat-Space Limit of AdS}},
  \href{https://doi.org/10.1007/JHEP07(2011)023}{\emph{JHEP} {\bfseries 07}
  (2011) 023} [\href{https://arxiv.org/abs/1007.2412}{{\ttfamily 1007.2412}}].

\bibitem{ElShowk:2011ag}
S.~El-Showk and K.~Papadodimas, \emph{{Emergent Spacetime and Holographic
  CFTs}}, \href{https://doi.org/10.1007/JHEP10(2012)106}{\emph{JHEP} {\bfseries
  10} (2012) 106} [\href{https://arxiv.org/abs/1101.4163}{{\ttfamily
  1101.4163}}].

\bibitem{Gross:2017aos}
D.~J. Gross and V.~Rosenhaus, \emph{{All point correlation functions in SYK}},
  \href{https://doi.org/10.1007/JHEP12(2017)148}{\emph{JHEP} {\bfseries 12}
  (2017) 148} [\href{https://arxiv.org/abs/1710.08113}{{\ttfamily
  1710.08113}}].

\bibitem{lloyd}
S.~{Lloyd}, \emph{{Pure state quantum statistical mechanics and black holes}},
  {\emph{ArXiv e-prints} (2013) }
  [\href{https://arxiv.org/abs/1307.0378}{{\ttfamily 1307.0378}}].

\bibitem{Balasubramanian:2007qv}
V.~Balasubramanian, B.~Czech, V.~E. Hubeny, K.~Larjo, M.~Rangamani and
  J.~Simon, \emph{{Typicality versus thermality: An Analytic distinction}},
  \href{https://doi.org/10.1007/s10714-008-0606-8}{\emph{Gen. Rel. Grav.}
  {\bfseries 40} (2008) 1863}
  [\href{https://arxiv.org/abs/hep-th/0701122}{{\ttfamily hep-th/0701122}}].

\bibitem{Balasubramanian:2005mg}
V.~Balasubramanian, J.~de~Boer, V.~Jejjala and J.~Simon, \emph{{The Library of
  Babel: On the origin of gravitational thermodynamics}},
  \href{https://doi.org/10.1088/1126-6708/2005/12/006}{\emph{JHEP} {\bfseries
  12} (2005) 006} [\href{https://arxiv.org/abs/hep-th/0508023}{{\ttfamily
  hep-th/0508023}}].

\bibitem{Alday:2006nd}
L.~F. Alday, J.~de~Boer and I.~Messamah, \emph{{The Gravitational description
  of coarse grained microstates}},
  \href{https://doi.org/10.1088/1126-6708/2006/12/063}{\emph{JHEP} {\bfseries
  12} (2006) 063} [\href{https://arxiv.org/abs/hep-th/0607222}{{\ttfamily
  hep-th/0607222}}].

\bibitem{Balasubramanian:2008da}
V.~Balasubramanian, J.~de~Boer, S.~El-Showk and I.~Messamah, \emph{{Black Holes
  as Effective Geometries}},
  \href{https://doi.org/10.1088/0264-9381/25/21/214004}{\emph{Class. Quant.
  Grav.} {\bfseries 25} (2008) 214004}
  [\href{https://arxiv.org/abs/0811.0263}{{\ttfamily 0811.0263}}].

\bibitem{Raju:2018xue}
S.~Raju and P.~Shrivastava, \emph{{A Critique of the Fuzzball Program}},
  \href{https://arxiv.org/abs/1804.10616}{{\ttfamily 1804.10616}}.

\bibitem{Fitzpatrick:2015zha}
A.~L. Fitzpatrick, J.~Kaplan and M.~T. Walters, \emph{{Virasoro Conformal
  Blocks and Thermality from Classical Background Fields}},
  \href{https://doi.org/10.1007/JHEP11(2015)200}{\emph{JHEP} {\bfseries 11}
  (2015) 200} [\href{https://arxiv.org/abs/1501.05315}{{\ttfamily
  1501.05315}}].

\bibitem{Chang:2018nzm}
C.-M. Chang, D.~M. Ramirez and M.~Rangamani, \emph{{Spinning constraints on
  chaotic large $c$ CFTs}},  \href{https://arxiv.org/abs/1812.05585}{{\ttfamily
  1812.05585}}.

\bibitem{Turiaci:2016cqc}
G.~J. Turiaci and H.~Verlinde, \emph{On cft and quantum chaos},
  \href{https://doi.org/10.1007/JHEP12(2016)110}{\emph{Journal of High Energy
  Physics} {\bfseries 2016} (2016) 110}.

\bibitem{Giddings:2017mym}
S.~B. Giddings, \emph{{Nonviolent unitarization: basic postulates to soft
  quantum structure of black holes}},
  \href{https://doi.org/10.1007/JHEP12(2017)047}{\emph{JHEP} {\bfseries 12}
  (2017) 047} [\href{https://arxiv.org/abs/1701.08765}{{\ttfamily
  1701.08765}}].

\bibitem{Kabat:2014kfa}
D.~Kabat and G.~Lifschytz, \emph{{Finite N and the failure of bulk locality:
  Black holes in AdS/CFT}},
  \href{https://doi.org/10.1007/JHEP09(2014)077}{\emph{JHEP} {\bfseries 09}
  (2014) 077} [\href{https://arxiv.org/abs/1405.6394}{{\ttfamily 1405.6394}}].

\bibitem{Polchinski:1999yd}
J.~Polchinski, L.~Susskind and N.~Toumbas, \emph{{Negative energy,
  superluminosity and holography}},
  \href{https://doi.org/10.1103/PhysRevD.60.084006}{\emph{Phys. Rev.}
  {\bfseries D60} (1999) 084006}
  [\href{https://arxiv.org/abs/hep-th/9903228}{{\ttfamily hep-th/9903228}}].

\bibitem{Susskind:2013lpa}
L.~Susskind, \emph{{New Concepts for Old Black Holes}},
  \href{https://arxiv.org/abs/1311.3335}{{\ttfamily 1311.3335}}.

\bibitem{Heemskerk:2012mn}
I.~Heemskerk, D.~Marolf, J.~Polchinski and J.~Sully, \emph{{Bulk and
  Transhorizon Measurements in AdS/CFT}},
  \href{https://doi.org/10.1007/JHEP10(2012)165}{\emph{JHEP} {\bfseries 10}
  (2012) 165} [\href{https://arxiv.org/abs/1201.3664}{{\ttfamily 1201.3664}}].

\bibitem{Brill:1963yv}
D.~R. Brill and R.~W. Lindquist, \emph{{Interaction energy in
  geometrostatics}}, \href{https://doi.org/10.1103/PhysRev.131.471}{\emph{Phys.
  Rev.} {\bfseries 131} (1963) 471}.

\end{thebibliography}\endgroup

\end{document}